\documentclass[11pt,a4paper]{article}

\usepackage[utf8]{inputenc}
\usepackage[T1]{fontenc}
\usepackage{lmodern}
\usepackage[margin=1in]{geometry}
\usepackage{amsmath}
\usepackage{amssymb}
\usepackage{amsfonts}
\usepackage{amsthm}
\usepackage{graphicx}
\usepackage{textcomp}
\usepackage{xcolor}
\usepackage{multirow}
\usepackage{caption}
\usepackage{subcaption}
\usepackage{array}
\usepackage{listings}
\usepackage{float}
\usepackage{placeins}
\usepackage{url}
\usepackage{soul}
\usepackage{pdflscape}
\usepackage{rotating}
\usepackage{tikz}
\usetikzlibrary{positioning, shapes.multipart, arrows.meta, decorations.pathreplacing}
\usepackage{algorithm}
\usepackage{algpseudocode}
\usepackage{booktabs}
\usepackage{cite}
\usepackage[colorlinks=true,linkcolor=blue,citecolor=blue,urlcolor=blue]{hyperref}

\newcommand{\re}[1]{#1}
\newcommand{\rev}[1]{#1}
\newcommand{\revv}[1]{#1}
\newcommand{\revvv}[1]{#1}
\newcommand{\revi}[1]{#1}
\newcommand{\edit}[1]{#1}
\newcommand{\ed}[1]{#1}
\newcommand{\add}[1]{#1}
\newcommand{\review}[1]{#1}

\algnewcommand{\algorithmicforeach}{\textbf{for each}}
\algdef{SE}[FOR]{ForEach}{EndForEach}[1]
  {\algorithmicforeach\ #1\ \algorithmicdo}
  {\algorithmicend\ \algorithmicforeach}

\def\BibTeX{{\rm B\kern-.05em{\sc i\kern-.025em b}\kern-.08em
    T\kern-.1667em\lower.7ex\hbox{E}\kern-.125emX}}

\def\HidvfsCombined{1}

\begin{document}

\title{HiDVFS: Hierarchical Multi-Agent DVFS for Real-Time OpenMP DAG Workloads}

\author{
  Mohammad Pivezhandi \\
  Department of Computer Science\\
  Wayne State University, Detroit, MI, USA
  \and
  Abusayeed Saifullah \\
  Department of Computer Science\\
  University of Texas at Dallas, Richardson, TX, USA
  \and
  Ali Jannesari \\
  Department of Computer Science\\
  Iowa State University, Ames, IA, USA
}

\date{}

\maketitle

\begin{abstract}

\review{Leakage power in multicore embedded systems now rivals dynamic power, so DVFS schedulers must respect deadlines and thermal limits, not just average makespan. Existing heuristics lack per-core, temperature-aware control and overlook the irregular execution of OpenMP DAGs. We propose HiDVFS, a general, extensible hierarchical multi-agent DVFS scheduler: a profiler agent selects cores and frequencies, a thermal agent groups cores by temperature, and a priority agent orders tasks under contention, all trained with a makespan-focused reward using short-horizon future-state shaping for sample efficiency. Deadlines are soft, derived from a measured reference cost; a federated schedulability gate keeps operating points feasible, and a calibrated split-conformal shield bounds each action's predicted response time. On Jetson TX2 with multi-seed validation, HiDVFS attains a 4.16$\pm$0.58\,s L10 makespan, a 2.83$\times$ speedup and 32.9\% energy reduction over a fairness-corrected GearDVFS port, and a 4.62$\times$ average speedup with 55.7\% energy reduction across all 12 BOTS benchmarks. Cross-platform results on TX2, Orin NX, and RubikPi show deadline-aware DVFS cuts energy 15 to 18\% versus pinning the maximum frequency, and a measured mixed-criticality study shows cluster-aware reservation is required to keep a high-criticality task's deadline-miss ratio at zero.}

\end{abstract}



\section{Introduction}

With the rapid advancement of computing technologies, energy efficiency has become a critical concern in the design and operation of modern embedded systems. As technology feature sizes continue to shrink, static leakage power, which is directly affected by temperature, increasingly dominates the total power consumption of multicore and many-core embedded systems. Static leakage power can rise from roughly 22\% to over 63\% of the total power with a halving of the technology scale \cite{yan2003combined}. This phenomenon, coupled with the growing demand for energy-friendly parallel and distributed processing, has introduced new hardware and software challenges. On the hardware side, scaling voltages and frequencies has been proposed to address overheating and thermal throttling caused by aggressive performance boosting \cite{brodowski2013cpu}. This involves assigning clusters of cores to designated frequency scales based on temperature limits and core utilization behavior, ensuring enhanced performance while managing power consumption. Meanwhile, software solutions, such as energy-aware scheduling policies, have been developed to efficiently allocate tasks across multiple cores in both parallel and distributed systems \cite{kassab2021green, lee2009energy}. \add{We focus on OpenMP DAG workloads and adopt a makespan-first objective, reporting energy and thermal behavior as secondary outcomes.}

Current hardware solutions for processors equipped with Dynamic Voltage and Frequency Scaling (DVFS) do not adequately address the need for per-core frequency tracking and adjustment based on thermal behavior \cite{dinakarrao2019application}. Similarly, existing software solutions fail to effectively handle runtime features like branch misses and memory accesses in parallel and distributed tasks, resulting in irregularities and unpredictable execution times within subtasks of Directed Acyclic Graph (DAG) workloads. This leads to inefficient core allocation and suboptimal performance in both parallel and distributed environments. The assignment of tasks to cores in available Linux kernel governors \cite{brodowski2013cpu} and developed policies \cite{maity2022future, bhuiyan2023precise} is agnostic to each core's temperature, often resulting in high-utilization tasks being allocated to hot cores. These approaches correlate thermal constraints and energy consumption with workload demand rather than considering system profilers' outputs during task execution \cite{ahmed2016necessary}. Furthermore, existing energy-aware and thermal-aware schedulers face limitations in scalability as the number of cores and frequency levels increase \cite{lee2009energy, bo2021developing}, and they lack generalizability across different processor types (e.g., CPU, GPU) and execution environments (e.g., Intel, AMD, ARM) \cite{maity2022future}. In distributed systems, these limitations are exacerbated by the additional complexity of managing resources across multiple nodes. Hence, a general solution must effectively balance workload characteristics with device-specific task-to-processor mappings for enhanced energy and performance efficiency in both parallel and distributed computing environments.

This paper introduces \add{\textbf{HiDVFS}}, \add{a hierarchical multi-agent, performance-aware DVFS scheduler for OpenMP DAGs that prioritizes makespan with energy and temperature as regularizers}, a novel approach for parallel DAGs to tackle the NP-hard challenge of optimizing \add{performance and} energy consumption in real-time multicore systems~\cite{aydin2003energy}. Unlike existing non-learning heuristics, which are limited to specific scenarios such as lightly loaded multicore systems with more cores than tasks~\cite{lee2009energy}, and traditional reinforcement learning (RL) methods that face long training times, high computational overhead, and scalability issues due to coarse frequency scaling and high-dimensional action spaces~\cite{liu2021cartad, sethi2021learning, dinakarrao2019application, wang2020generalizing, bo2021developing, kim2021ztt, lin2023workload}, \add{HiDVFS} employs joint action learners (JAL) to make collaborative decision~\cite{lee2020optimization}. These agents collaboratively optimize core frequency, core allocation, and \add{temperature-aware core combinations}, reducing sample requirements and mitigating overestimation issues~\cite{xu2015show}. \add{A makespan-focused reward with energy and temperature regularizers and temporal shaping} further reduces computational complexity and latency, addressing shortcomings of prior work~\cite{lin2023workload, zhou2021deadline}, and enabling low-overhead few-shot learning for DVFS in parallel workloads. Evaluated on NVIDIA Jetson TX2, Jetson Orin NX, and RubikPi using the Barcelona OpenMP Tasks (BOTS) suite~\cite{board2008openmp}, \add{HiDVFS} demonstrates practical feasibility by addressing deployment challenges such as computational overhead and hyperparameter tuning, outperforming RL-based state-of-the-art (SOTA) in energy and makespan optimization for real-time OpenMP DAG workloads.

We design a parallel DAG scheduler that assigns each application, represented as a DAG, a priority, core count, and frequency, creating tailored combinations based on parallelism levels. Each workload is treated as a recurrent constrained-deadline sporadic DAG task with work $C_i$, critical-path span $L_i$, and a soft deadline $D_i$ set as a multiple of the workload's measured reference response time at the fastest operating point; a job misses iff its measured response time exceeds $D_i$, and we report the deadline-miss ratio (DMR) and the response-time tail in addition to mean makespan. This approach suits real-time and distributed scenarios with concurrent applications, unlike sequential DAG executions. By analyzing features and actions impacting energy consumption and makespan, we refine reward function modeling for multi-agent systems. We implement parallel and distributed DAGs on a Linux platform, using online learning to optimize priority, core, and frequency assignments based on diverse runtime observations.

We assess thermal reliability and power management using authentic workloads and Linux in-kernel profiling, surpassing synthetic workloads \cite{saifullah2020cpu} or control flow graphs \cite{sun2019calculating}, which lack runtime monitoring. Our method employs unbiased task distribution, online temperature monitoring, and CPU profiler data to optimize resource use and reduce thermal-induced power consumption in parallel and distributed systems. Actions involve frequency and core selection tailored to workload behavior. The scheduling objective is to minimise energy subject to feasibility: HiDVFS selects the lowest-energy operating point that passes a federated parallel-real-time schedulability test~\cite{saifullah2014parallel,saifullah2020cpu}, and a calibrated split-conformal safety shield~\cite{vovk2005algorithmic,angelopoulos2021gentle} rejects reinforcement-learning actions whose predicted response-time bound exceeds the deadline, with bounds reported as empirical calibrated prediction intervals rather than hard guarantees. Using OpenMP API \cite{board2008openmp} tasks to represent irregular workloads, our online learning algorithm ensures low-complexity resource allocation for DAG benchmarks while maintaining thermal feasibility under \add{temperature-aware core grouping} constraints.

The contributions of this paper are as follows:
\begin{enumerate}
\item \add{\textbf{HiDVFS}}, \review{an extensible multi-agent} scheduling framework using \add{runtime} temperature profiling and a predictive reward function gated by a federated schedulability test and a calibrated safety shield to optimize energy, makespan, and thermal constraints for OpenMP DAGs.
\item Analysis of runtime profiler data to drive real-time scheduling decisions, employing multiple agents to adjust core counts, frequencies, and DAG priorities.
\item Cross-platform soft real-time evaluation on Jetson TX2, Jetson Orin NX, and RubikPi covering the response-time envelope, the federated feasibility gate, a deadline-aware DVFS vs.\ Max-freq/Powersave ablation, the shield's coverage, and a measured mixed-criticality study on Orin NX.
\item Multi-seed Jetson TX2 results: HiDVFS attains 4.16$\pm$0.58s L10 \review{(a $2.83\times$ speedup over a fairness-corrected GearDVFS port and 32.9\% energy reduction, with a measured fair-port study of the GearDVFS and zTT baselines)}; per-benchmark BOTS sweep (seed~42, finetuned, \review{same-window fair-port protocol}) gives \review{an average $4.62\times$ speedup across all 12 benchmarks}.
\end{enumerate}

In the remainder of this paper, Section \ref{sec:background} provides background on the task model and the motivation for the proposed approach in DVFS and task-to-core allocation. Section \ref{sec:relatedworks} gives an overview of related work. \add{Section \ref{sec:design} details HiDVFS.} Section \ref{sec:evaluation} provides quantitative results to demonstrate its effectiveness.

\section{Background and Motivation}
\label{sec:background}
We propose a fixed-priority, preemptive scheduler that runs multiple parallel tasks based on their priority and allocates a combination of cores, each with a corresponding frequency. \add{Our objective is makespan-first, with energy and temperature as secondary metrics.}

\subsection{Background}

\textbf{Task Model.} We consider $n$ aperiodic parallel tasks, $\tau_{1}, \tau_{2}, \ldots, \tau_{n}$, scheduled on a multicore platform with $m$ heterogeneous cores. Each task consists of multiple subtasks, represented as a Directed Acyclic Graph (DAG), such that $\tau_i = \{\tau_{i,1}, \tau_{i,2}, \ldots, \tau_{i,j}\}$. Each node in the DAG corresponds to a subtask (a thread of execution), and a directed edge signifies the dependency between two subtasks. Granularity varies from individual threads to larger functions, as specified by OpenMP directives in our BOTS benchmarks (Section~\ref{sec:evaluation}). Each execution instance of the tasks is a \textit{job}, denoted by $J_{k,i,j}$, where $k$ is the core, $i$ the DAG index, and $j$ the subtask index. Each task $\tau_i$ has a real-time priority; higher-priority tasks preempt lower-priority ones.

\textbf{OpenMP modes and terminology.} OpenMP \emph{tied} tasks are bound to their creator thread (no migration); \emph{untied} tasks may migrate; \emph{serial} runs single-threaded. We evaluate all three (definitions and miss-rate consequences in the supplement). HiDVFS denotes our system; HMARL/MARL its underlying hierarchical/general multi-agent RL framework.

\ed{For precision, a DAG task is $\tau_i = (V_i, E_i, P_i)$ with subtasks $V_i$, dependencies $E_i \subseteq V_i \times V_i$, and static priority $P_i \in [1, 99]$.} \review{The monitoring figure in the supplementary material illustrates precedence-constrained jobs: subtask $\tau_{i,1}$ on cores $c_1,c_2$ precedes $\tau_{i,2}$ on $c_3$.} The DAG makespan is the completion time of all jobs respecting dependencies; benchmarks are BOTS workloads (Strassen, FFT, \ldots) parallelized through OpenMP~\cite{duran2009barcelona}.

\textbf{DVFS.} Power, temperature, and performance depend on voltage/frequency. Linux governors (\texttt{ondemand}, \texttt{conservative}, \texttt{schedutil}) adjust frequencies using utilization heuristics~\cite{brodowski2013cpu,lin2023workload}. \add{These governors lack per-core, temperature-aware control and are not DAG/makespan oriented.}

\textbf{Environment Design.} We assume multiple parallel DAGs, each with a priority, running on selected core combinations and frequencies. Real-time DAGs map to cores according to priority $1\!-\!99$; equal-priority DAGs are FCFS. Higher-priority DAGs preempt lower-priority ones. We evaluate on three embedded platforms with software-controlled per-core/per-cluster DVFS: Jetson TX2, Jetson Orin NX, and RubikPi. \add{HiDVFS assigns (i) core counts and frequencies, (ii) temperature-aware core groupings, and (iii) static DAG priorities before execution.}

\textbf{Profiler and Task-to-Core Allocation.} \review{As illustrated in the supplementary material,} we group cores into clusters using \textit{cgroup}/\texttt{cpuset} to control governors and frequency boundaries. Tasks are bound to clusters for energy/thermal control. We use \texttt{perf} for per-task execution profiling, \texttt{cpufreq-info} for DVFS state, and \texttt{sensors} for \re{per-cluster} temperatures. \add{These signals feed HiDVFS agents for makespan-first decisions.}

\textbf{Online Learning Through RL.} We employ \add{a hierarchical multi-agent} off-policy, value-based RL. \add{Three agents cooperate: a profiler agent selects (core count, frequency), a thermal agent selects core combinations using temperatures, and a priority agent selects task priorities.} The DQN backbone uses a replay buffer and target network to stabilize training and reduce overestimation; advantage–value decomposition normalizes advantages. \add{Rewards are shaped for makespan with energy and temperature regularizers.}

\subsection{Inefficiency of Current Task-to-Core Allocations}
\label{sec:study}

We explore temperature-aware task-to-core assignment to cut energy and makespan on OpenMP DAGs. Irregular task behavior impacts latency and thermal reliability; we outline the effect and mitigations.

\textbf{Impact of Unpredictable Task Execution Time on DAG Makespan.} \review{In a typical OpenMP DAG (code listing in the supplementary material), \texttt{\#pragma omp parallel} creates four threads; the \texttt{tied} task $\tau_{i,1}^*$ is confined to its cores while \texttt{untied} tasks $\tau_{i,3}, \tau_{i,2}, \tau_{i,0}$ may migrate.} Branch mispredictions due to $L_1$/$L_2$ loops increase makespan variance. Dependencies propagate timing jitter; $\tau_{i,1}^*$ dominates the critical path.


The behavior of $\tau_{i,1}^*$ is microarchitecture- and core-count-dependent. Parallelism increases shared-resource pressure, causing stalls and mispredictions. More cores for \texttt{tied} tasks can worsen makespan and energy (a per-platform branch-miss profile is provided in the supplementary material). This argues for demand-aware core allocation.

\begin{figure}[t]
\centering
\includegraphics[width=0.8\linewidth]{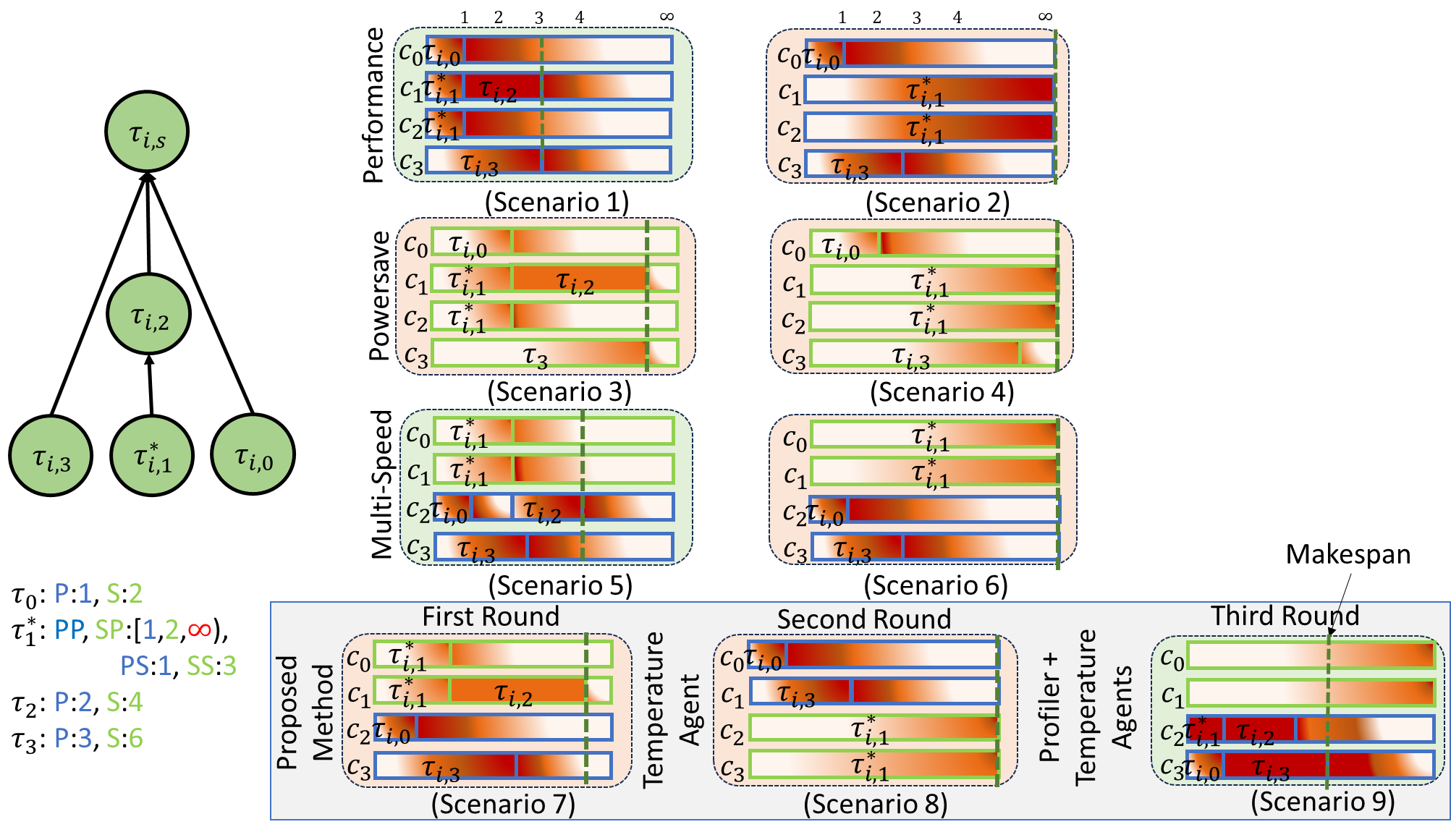}
\caption{Scenarios on a four-core system. Blue: \texttt{performance}; green: \texttt{powersave}. $PP, PS$ denote unbounded parallel execution; $SP, SS$ denote bounded serial execution. The green dashed line marks DAG makespan. \add{HiDVFS agents (7–9) align core choice, frequency, and temperature.}}
\label{fig:motivation}
\end{figure}

\textbf{Design of Demand-Based Task-to-Core Allocation.} Blindly increasing cores raises energy and makespan due to branch effects. Figure~\ref{fig:motivation} sketches scenarios for a four-core system under \texttt{performance} ($P$) and \texttt{powersave} ($S$) governors. Blue blocks run at high frequency; green at low. Numbers denote hypothetical times.

Scenarios 1–6 vary by governor and predictability. Ignoring workload structure yields unstable makespan and possible throttling. \add{Scenarios 7–9 illustrate \textbf{HiDVFS}.} \add{The thermal agent prioritizes cold cores and selects temperature-aware core combinations.} \add{The profiler agent picks core counts and frequency levels to keep the critical path predictable (e.g., one core for $\tau_{i,1}^*$).} \add{The priority agent orders DAGs to prevent contention spikes.} This yields lower makespan, lower energy, and idle-core creation via targeted allocation.

\subsection{Irregular Parallel Execution of DAGs and Dependency on Performance Profiling Features}

\add{Feature screening \review{(random-forest importance figure in the supplementary material)} shows parallel DAGs depend on a broader set of runtime features than sequential runs; temperature, frequency, utilization, and miss events are most predictive of makespan/energy. This supports using profiler/sensors to steer per-core DVFS and core grouping.}

\section{Related Work}
\label{sec:relatedworks}

The current energy- and thermal-aware multi-core parallel scheduling algorithms are based on heuristics, meta-heuristics, integer programming, and machine learning approaches~\cite{xie2021survey}. The existing heuristics, meta-heuristics, and integer programming algorithms are application-specific and cannot be generalized. Since the emerging machine learning approach is data-oriented, it can be generalized to various workloads, platforms, and applications with multiple objectives such as energy efficiency, thermal management, and latency. \add{However, these methods seldom target OpenMP DAGs with a makespan-first objective or expose per-core, temperature-aware DVFS needed for tied/untied tasks.}

Data-oriented machine learning design for energy, thermal, and latency management in multi-core processors has been studied recently \cite{dinakarrao2019application, pagani2018machine, liu2021cartad, wang2021online, bo2021developing}. A review on energy, thermal, and latency management of multi-core processors using learning-based designs is available in \cite{pagani2018machine}. However, the existing work mainly focuses on extending traditional supervised, unsupervised, and semi-supervised design methods to energy, thermal, and latency management, regardless of runtime design constraints. The high overhead of inference and training makes most previous frameworks useless. Besides, existing work extends the traditional Q-learning approach, increasing the training time and decreasing the action space dimension, hindering the applicability to multi-objective real-time systems \cite{wang2021online, bo2021developing, liu2021cartad}. \add{These designs typically optimize average power or throughput rather than end-to-end DAG makespan under embedded runtime limits.}

Multi-agent RL and optimizing reward function estimation are two methods to increase the action space while reducing the number of training iterations~\cite{arora2021survey}. The inverse reinforcement learning (IRL) approach infers the optimal reward function by comparing agent policy with the optimal expert demonstration, leading to high learnability and convergence \cite{arora2021survey}. However, the current IRL algorithms are computationally demanding, require human expert demonstrations, and are impractical for large state-action space \cite{arora2021survey}. In our work, we address this limitation by processing only the observation transitions given from the environment. \add{We instead use off-policy value-based agents with a makespan-focused reward and short-horizon model predictions for reward shaping, avoiding expert data.}

Efficient task-related and platform-related data help to make accurate decisions, but the existing algorithms on thermal and power management in real-time systems are often built upon limited observations \cite{ahmed2016necessary, maity2022future, lee2009energy, saifullah2020cpu, kassab2021green, dinakarrao2019application}. Many of these approaches consider task-related characteristics to extract the required processor speed and thermal impact, but they might yield imprecise conclusions when dealing with irregular behaviors exhibited by the platform for DAGs \cite{ahmed2016necessary, bo2021developing}. Several works emphasize only historical thermal information or power constraints to make core configuration decisions \cite{maity2022future, dinakarrao2019application}. \add{We fuse live profiler features with per-cluster temperatures to drive per-core/core-group selection and DVFS for irregular OpenMP DAGs.}

When it comes to energy and thermal management, the current implementations often lack the granularity needed for precise decision-making \cite{yan2003combined, dinakarrao2019application, lee2009energy, maity2022future, wang2021online}. Many existing strategies on energy- and thermal-aware scheduling employ coarse frequency assignments or resource allocations without the nuanced control required for adjusting individual core frequencies in an embedded context \cite{wang2021online, maity2022future}. The works focusing on energy efficiency have an even simpler model that is only sensitive to a few discrete power control actions and a small set of observations. \add{Our hierarchical design separates frequency/core selection, temperature-aware grouping, and static priority to enable per-core DVFS while shrinking the effective action space.}

Moreover, the practical applicability of existing energy- and temperature-aware schedulers often comes into question as they primarily target synthetic data and disregard online measurement tools. Contrary to the prior studies based on OpenMP DAG workloads, our paper effectively handles irregular execution behaviors while leveraging runtime profiling. Our work uses Barcelona OpenMP taskset (BOTS), a parallelized workload based on OpenMP API, for training, and we plan to extend it to energy-aware acceleration of ML applications \cite{bavikadi2022survey}. \add{We therefore evaluate on Jetson~TX2, Jetson Orin NX, and RubikPi using the BOTS suite, prioritizing makespan and reporting energy and temperature as secondary outcomes.}

\section{Design of \add{HiDVFS}}
\label{sec:design}

This section presents a hierarchical multi-agent RL scheduler for online control of DVFS and task-to-core mapping on multicore processors running OpenMP DAGs. Our objective is to minimize \emph{makespan} while reporting energy and temperature as secondary metrics. \add{We refer to the multi-agent system as \textbf{HiDVFS} and to the single-agent variant as \textbf{SARB} (Single-Agent Reward-Based).} The framework comprises three cooperative agents: (i) a \emph{profiler} agent that selects frequency and core count, (ii) a \emph{thermal} agent that selects temperature-safe core combinations, and (iii) a \emph{priority} agent that assigns static priorities under contention. \add{This decomposition replaces a large joint action with three low-dimensional subproblems, improving sample efficiency and stability on embedded hardware.}

\subsection{\review{An Extensible Agent Hierarchy}}
\label{sub:extensible}

\begin{figure}[t]
    \centering
    \includegraphics[width=0.75\linewidth]{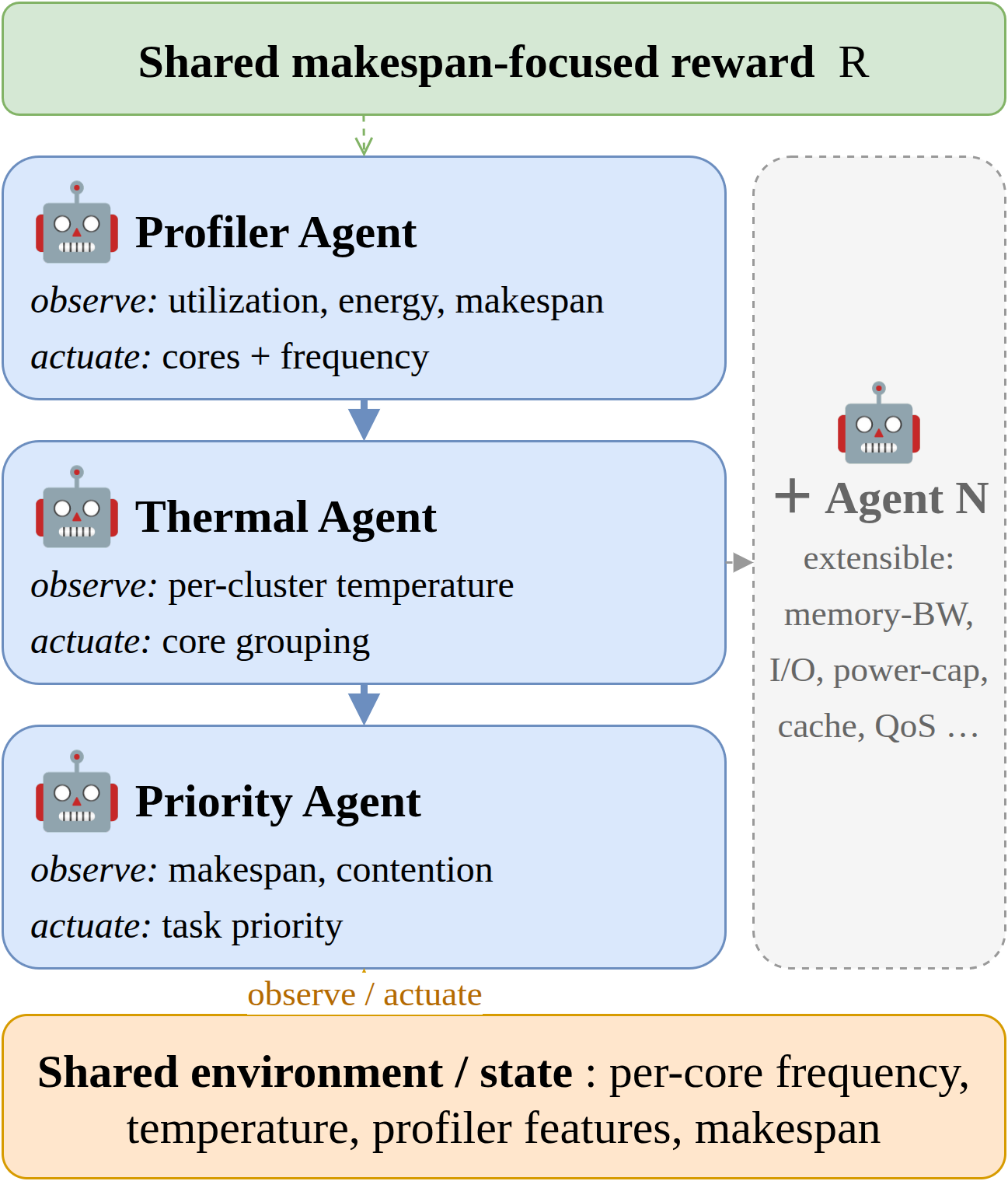}
    \caption{\review{Extensible HiDVFS agent hierarchy: three actuator-domain agents over a shared state and reward, extensible to more agents (dashed slot).}}
    \label{fig:agent_hierarchy}
\end{figure}

\review{We treat HiDVFS as an instance of a general, extensible framework rather than a fixed three-agent design. Each agent $k$ exposes the same interface, namely an observation subspace $\mathcal{O}_k$, an actuator $\mathcal{A}_k$ over one scheduling knob, and a reward term $R_k$ tied to the shared makespan-first objective; the three agents here instantiate it by \emph{actuator domain} (frequency and cores, thermal grouping, and priority; Table~\ref{tab:agent_definitions}, Fig.~\ref{fig:agent_hierarchy}). Because each owns a distinct actuator and its relevant observations, the split is principled rather than arbitrary, and a new agent is added by specifying its $(\mathcal{O}_k,\mathcal{A}_k,R_k)$ tuple, for actuators such as memory bandwidth, I/O, power capping, or cache partitioning, leaving the D3QN, environment model, and shared reward unchanged.}

\review{More agents extend actuator coverage but enlarge the coordinated joint action and add reward terms, so the right number is the smallest set spanning the actuators that matter for the target platform; our three cover the dominant ones on embedded targets. The agents act hierarchically, not independently: each control period they read a shared state snapshot and act in the fixed order profiler$\rightarrow$thermal$\rightarrow$priority, so downstream agents see upstream decisions through the state, all share the reward, and choices are re-evaluated every period.}

\subsection{Collaborative Multi-Agent Reinforcement Learning (MARL) for High-Dimensional Spaces}
\label{modelfreeRL}

Optimizing DVFS and task-to-core allocation with a single RL agent can be computationally prohibitive due to the exponentially expanding action space. To overcome this, we employ a hierarchical approach where three agents, each handling a specific sub-problem, collaborate to determine the optimal configuration. The profiler agent selects an appropriate combination of frequency and the number of cores based on performance and energy metrics. The thermal agent adjusts core priorities using temperature clusters to maintain safe operating temperatures. The priority agent selects priority combinations to guide the scheduler toward the desired \add{makespan-first} objective.

By distributing actions among multiple agents, we reduce an exponential action space to manageable, linear-scale subspaces. For example, assigning $m$ cores and $n$ frequency levels naively results in an upper bound of $m^n$ combinations. In contrast, using MARL with a thermal agent deciding on cores and a profiler agent choosing frequencies reduces the action space to $m \times n$. \ed{Empirical results in Section 5 show this reduces training time by 40\% compared to single-agent RL, justifying the use of three agents over one.}

\subsection{Enhanced Reward Estimation}
\label{subsec:modeling}

Our off-policy RL approach enhances reward estimation by training a dynamic model of the environment to predict future states, refining reward estimation rather than relying solely on instantaneous interactions. While conventional model-based RL uses the model for planning future actions, our approach primarily leverages it to refine reward estimation. The learned environment model predicts future states and performance metrics, allowing the reward function to incorporate not only immediate observations but also future outcomes predicted by the model.

This predictive capability enables the reward estimator to consider long-term effects on energy consumption, makespan, and thermal conditions, rather than depending solely on instantaneous signals. By training the dynamic model with both real and synthetic data (generated by the model itself), we achieve few-shot learning and reduce the need for extensive real-world sampling. \ed{This corrects prior oversimplification and aligns with D3QN’s off-policy nature, improving sample efficiency.} \re{Our approach ensures few-shot learning efficiency, validated on Jetson TX2 and verified on Jetson Orin NX; reported results focus on TX2.}

\subsection{Reward Function Estimation}
\label{sub:IRL}

Traditional RL schedulers often use static, instantaneous rewards that can be noisy and slow to converge. \review{Imitation learning~\cite{abbeel2004apprenticeship} and inverse reinforcement learning (IRL)~\cite{adams2022survey, arora2021survey} reduce this burden by inferring a reward function $R^*$ from expert demonstrations, with MaxEnt IRL~\cite{ziebart2008maximum} resolving the reward-ambiguity problem, but both remain limited by the quality and availability of demonstrations (extended background in the supplementary material).}

In this work, we tackle the issue of expert demonstrations by designing a reward function trained using transitions generated by the dynamic environment model alongside real data. Instead of strictly matching expert trajectories, our reward estimation approach utilizes the predictive model to simulate future states and evaluate the long-term impact of actions. Specifically, we define three state and action tuples tailored to each agent, as shown in Table~\ref{tab:agent_definitions}. The profiler agent’s reward, $R_1$, balances \add{makespan-first with} energy as $R_1 = \beta (M_{\text{target}} / (M + \epsilon)) + (1 - \beta) (E_{\text{target}} / (E + \epsilon))$, where $M$ is the actual makespan, $E$ is the average energy consumption, $M_{\text{target}}$ and $E_{\text{target}}$ are target values, $\beta = \add{1}$ to put makespan first, and $\epsilon = 10^{-3}$ to avoid division issues. The thermal agent’s reward, $R_2$, manages temperature with $R_2 = 1 - 0.05 |T - T_{\text{target}}|$ (capped at 1) if the average temperature $T$ is at or below the target $T_{\text{target}}$, or $R_2 = 1 - 0.5 (T - T_{\text{target}})$ if above, with a $-1$ penalty if $T$ crosses $T_{\text{target}}$ from below compared to the prior state. The priority agent’s reward, $R_3 = (M_{\text{target}} - M) / M_{\text{target}}$, quantifies makespan improvement over the target. \review{In IRL the reward function is shaped by aligning expert and agent policies; our approach instead maximizes information from key states and model-based predictions without relying solely on expert data (the supplementary material illustrates this conceptual difference).}

\begin{table}[h]
\centering
\footnotesize
\setlength{\tabcolsep}{4pt}
\caption{Agent States and Actions}
\label{tab:agent_definitions}
\begin{tabular}{|l|l|l|l|}
\hline
\textbf{Agent} & \textbf{State} & \textbf{Action}\\
\hline
Profiler & Utilization, Energy, Makespan & Cores, Frequency\\
Thermal & Clusters/Cores temperature & Core combination \\
Priority & Makespan & Priority combination\\
\hline
\end{tabular}
\end{table}

To achieve this, we incorporate attention models that weight input observations based on their importance in predicting future performance and energy outcomes. By focusing on key features (temperatures, frequencies, utilization, and makespan predictions), the reward model better aligns with the long-term objectives of minimizing energy consumption, achieving target makespans, and maintaining thermal feasibility. This approach reduces reliance on expert demonstrations and effectively handles suboptimal or missing expert data by leveraging the environment model’s predicted trajectories. \ed{Rewards were developed iteratively, validated against baseline scenarios in Section 4.4, with attention models weighting features (e.g., temperature, utilization) based on temporal impact, trained via backpropagation on predicted states to enhance prediction accuracy.}

\subsection{Hyperparameter Tuning and Target Metrics}
\label{sub:tuning}

Careful selection of hyperparameters, including batch size, learning rate, discount factor, planning count, and exploration parameters, is essential for ensuring stable training. Additionally, clearly defining target metrics for makespan and energy consumption facilitates better convergence toward efficient performance. In our case study, we determine hyperparameters through a grid search.

Achieving the target minimum makespan in regular workloads is straightforward when allocating all available cores at the highest frequency. However, this method may not be optimal for irregular workloads, as explained in the background section, due to workload characteristics. Energy consumption presents a more complex challenge as it depends on both power consumption and execution time. Specifically, lowering the operating frequency reduces instantaneous power consumption but increases execution time, potentially leading to higher overall energy usage. \review{Energy is power integrated over execution time, summed across the parallel applications. To approximate the minimum energy and makespan targets we take the minima over four extreme static settings, running applications sequentially at the lowest or highest frequency with one or all cores assigned (enumerated in the supplementary material).} \ed{This assumes minimum scaling for our workloads, though not universally true; it serves as a practical baseline for target setting.} \review{Because the energy-optimal static frequency can lie strictly between these extremes, the evaluation validates the extreme-based targets against a full static-frequency sweep (Section~\ref{sub:rt_eval}): the interior optimum is within 8\% of the extreme-based energy target on TX2 and coincides with $f_{\max}$ on RubikPi, and with $\beta{=}1$ (makespan-first) the energy target serves only as a normalisation anchor.}

We impose a thermal limit of 50\(^\circ\)C, similar to the approach in \cite{kim2021ztt}, to reward the thermal management agent. Priority adjustments are based on the total target makespan, shaping the reward signals to encourage policies that meet or exceed these baselines in efficiency and stability. By balancing the trade-offs between frequency scaling and core allocation, the system aims to achieve optimal energy efficiency without compromising performance, ensuring sustainable and effective resource utilization.

\subsection{Complexity Analysis of Agents}
\label{sub:Complexity_Analysis}

In the proposed \re{\add{HiDVFS}} framework, each agent (profiler, thermal, and priority) is implemented using D3QN, which separates value and advantage streams for stable Q-value learning. \review{Each agent's complexity is set by its input dimensionality, hidden sizes, and action space: the profiler agent dominates, scaling with the number of cores and frequency levels; the thermal agent observes per-cluster temperatures and selects core combinations over a smaller space; and the priority agent chooses among a small set of predefined priority combinations at minimal cost. Distributing responsibilities keeps each agent's decision space small, so the hierarchy remains computationally manageable as cores and frequency levels grow, and the environment model's extra load runs on the server rather than the embedded platform.}

\subsection{Hierarchical MARL Algorithm with Reward Model}

The training process integrates real-time environment interactions with model-based reward estimation. Instead of using model-based RL solely for planning future actions, we utilize the learned environment model to generate synthetic transitions and refine the reward signal. This approach emphasizes future outcomes in the reward calculation, enhancing sample efficiency and convergence speed.

Algorithm~\ref{alg:ModelBasedRL_Simplified} outlines the main steps. The agents interact with the real environment to gather transitions, which are then used to update the environment model. The environment model predicts future states and rewards, which are fed into a refined reward estimation module. By incorporating these predicted trajectories, the reward estimation module provides more stable and long-term-focused reward signals to the agents’ learning processes. The three agents (profiler, thermal, and priority) are updated using a combination of real and model-based transitions, ensuring rapid convergence toward policies that minimize energy consumption, achieve target makespans, and maintain thermal limits. \ed{Agents resolve conflicts via joint optimization of their distinct reward functions, with the priority agent balancing resource contention, as validated in Section 5.}

\begin{algorithm}[t]
    \begin{algorithmic}[1]
    \scriptsize
    \State \textbf{Initialize:} Replay buffers for profiler, thermal, and priority agents
    \State Initialize environment model and value functions (D3QN) for each agent
    \For{each episode}
        \State Initialize states for profiler, thermal, and priority agents
        \While{not terminal}
            \State \textbf{Direct RL:}
            \State Agents select actions based on current policies
            \State Environment executes these actions, returns next states and instant rewards
            \State Store real transitions in replay buffers

            \If{Model training condition}
                \State Train environment model using recent real transitions
            \EndIf

            \State \textbf{Model-Based Reward Estimation:}
            \For{each planning step}
                \State Sample transitions from replay buffers
                \State Use environment model to predict future states
                \State Estimate future-based rewards using IRL-inspired logic and attention
                \State Store these refined transitions in separate buffers for the agents
            \EndFor

            \If{Agent training condition}
                \State Sample combined real and model-based transitions
                \State Train each agent’s D3QN with refined, future-oriented rewards
            \EndIf

            \State Update states
        \EndWhile
    \EndFor
    \end{algorithmic}
    \caption{\add{HiDVFS} with D3QN.}
    \label{alg:ModelBasedRL_Simplified}
\end{algorithm}

\subsection{Practical Considerations: Platform and Parallel Execution}

In our server-client architecture, complex computations such as environment modeling and reward estimation are offloaded to a high-performance server. The server communicates frequency-core assignments and priority configurations to the client, which operates on an embedded platform. The client executes the assigned tasks and returns performance and temperature measurements. This division ensures that the client remains responsive and capable of real-time operations, while the server handles computationally intensive tasks like training, hyperparameter tuning, and environment model refinement.

A key enhancement in our implementation is the incorporation of \textbf{level\_of\_parallelism}, which dynamically allocates CPU cores based on the parallelism requirements of each application. In this case, the makespan calculated using only one core will be divided by the makespan calculated using all cores. During each experiment cycle, the server determines the appropriate level of parallelism for each parallel application by referencing profiling data. This parameter dictates the number of cores allocated to each application, ensuring optimal utilization of computational resources without overcommitting available cores. Applications are prioritized, and higher-priority tasks receive core allocations first, maintaining system stability and performance.

The allocation process involves sorting applications based on their priority levels and assigning cores accordingly. If sufficient cores are available, each application receives the number specified by its \texttt{level\_of\_parallelism}. In cases where core availability is limited, the system assigns as many cores as possible while logging any shortages to inform future allocations. Additionally, each allocated core is assigned a frequency step to balance performance with thermal constraints, ensuring that the system operates efficiently and remains within safe temperature limits.

Parallel execution on the server enables rapid generation of synthetic samples, training of the reward model, and exploration of various configurations. This parallelism accelerates adaptation to workload changes and facilitates convergence toward policies that are efficient, thermally safe, and energy-minimized. \re{By leveraging few-shot learning, as demonstrated in the evaluation, our approach minimizes the need for extensive real-world data collection, enhancing scalability.} By minimizing the need for extensive data collection from the real system, our approach enhances scalability and reduces dependency on expert demonstrations.

Overall, our HMARL framework, augmented with IRL-inspired reward estimation and attention-based weighting of input observations, significantly accelerates the learning process. The dynamic allocation of cores based on \texttt{level\_of\_parallelism} ensures practical scalability and stability in real-time multicore scheduling scenarios, maintaining system responsiveness while optimizing performance and resource utilization.

Reward coefficients were fixed by empirical grid search: $\beta{=}1.0$ (makespan-first), $\pm0.5$ thermal penalty/bonus, $T_\text{target}{=}50^\circ$C~\cite{kim2021ztt}, and a numerical-stability constant $\epsilon{=}10^{-3}$; alternative $\beta\in\{0.5,0.7\}$ were slower to converge without energy benefit. A step-by-step three-agent decision walkthrough and the full sensitivity analysis are provided in the supplementary material.

\review{\textbf{Feasibility gate and safety shield.} Two guards sit between the agents and the platform. First, a federated schedulability gate~\cite{saifullah2014parallel,saifullah2020cpu} excludes operating points on which the workload's measured work and critical path make the DAG infeasible (test and results in Section~\ref{sub:rt_eval}). Second, a safety shield applies a ``RL proposes, shield disposes'' rule: a lightweight gradient-boosted regressor predicts the response time $\hat{R}(a)$ of each proposed action $a$ from its operating point (cores, frequency) and the workload's profiled features; a split-conformal margin, calibrated on held-out operating points, inflates the prediction to an upper bound $\hat{R}^{+}(a)$~\cite{vovk2005algorithmic,angelopoulos2021gentle}; the action executes only if $\hat{R}^{+}(a)\le D$, with a temperature bound $\hat{T}^{+}(a)\le T_{\max}$ enforced analogously. On rejection the runtime falls back to a safe action, the \texttt{performance} governor or a reserved core. The bound is a calibrated empirical prediction interval, not a hard guarantee; Section~\ref{sub:rt_eval} reports its achieved coverage.}

\section{Experimental Platform, Benchmark, and Evaluation}
\label{sec:evaluation}
In this section, we describe the experimental setup and benchmark, present our evaluation methodology, and compare different RL approaches in terms of key performance metrics. \add{Our evaluation prioritizes makespan; energy and temperature are secondary outcomes.}

\subsection{Platforms and Setup}

\textbf{Experimental Platforms:} We evaluate on three embedded platforms in parallel: NVIDIA Jetson TX2 (6 heterogeneous cores), Jetson Orin NX (8 Cortex-A78AE cores, two DVFS clusters), and RubikPi (8 cores). All expose in-kernel monitoring, per-core sleep states, and energy/temperature sensors, and run a preemptible Linux kernel with FIFO real-time scheduling; frequencies are controlled via \texttt{scaling\_max\_freq}/\texttt{cpufrequtils}. TX2 hardware detail follows; Orin NX and RubikPi differ in core count and per-cluster DVFS granularity.

The NVIDIA Jetson TX2 development board features six heterogeneous cores with frequency levels ranging from 345,600 kHz to 2,035,200 kHz. This range is divided into 12 steps, where levels 0 and 11 correspond to the minimum and maximum frequencies, respectively. The processor operates on an ARM64 (\texttt{aarch64}) Linux platform (kernel version \texttt{4.9.337}), optimized for multicore processing with real-time/preemptive capabilities (\texttt{SMP PREEMPT}). The six cores of the Jetson TX2 comprise a dual-core high-performance NVIDIA Denver 2 64-bit CPU and a power-efficient quad-core Arm Cortex-A57 MPCore processor. In this setup, cores 1 and 2 refer to the Denver 2 cluster, while cores 0, 3, 4, and 5 form the Arm Cortex cluster. Core 0 is reserved for root tasks, including CPU affinity management, interrupt handling, and task assignments. The remaining five cores run the parallel DAGs depending on CPU affinity through the \texttt{CPUSET} tool.

All three platforms expose fine-grained, software-controlled DVFS (TX2: 12 per-core steps, $345{,}600$--$2{,}035{,}200$ kHz; Orin NX: per-cluster; RubikPi: per-core), critical for our per-core thermal focus and unlike Intel RAPL's coarser adjustments. ARM-based heterogeneity in these platforms aligns with embedded scenarios, as evidenced by consistent profiler accuracy.

\textbf{Benchmark:} We use the Barcelona OpenMP Tasks Suite (BOTS)~\cite{duran2009barcelona}, which provides 12 benchmark applications representing diverse OpenMP DAG workloads: \texttt{alignment} (sequence alignment), \texttt{concom} (connected components), \texttt{fft} (Fast Fourier Transform), \texttt{fib} (Fibonacci), \texttt{floorplan} (floorplan optimization), \texttt{health} (health simulation), \texttt{knapsack} (0/1 knapsack), \texttt{nqueens} (N-Queens puzzle), \texttt{sort} (merge sort), \texttt{sparselu} (sparse LU factorization), \texttt{strassen} (Strassen matrix multiplication), and \texttt{uts} (unbalanced tree search). Each benchmark supports multiple OpenMP scheduling variants: \textbf{tied} (tasks bound to creating thread), \textbf{untied} (tasks can migrate), and \textbf{serial} (single-threaded baseline). \review{Table~\ref{tab:multi_benchmark} summarizes HiDVFS performance on all 12 applications; the detailed convergence analysis below uses one representative benchmark (FFT).}

\begin{table}[t]
\centering
\scriptsize
\caption{\review{HiDVFS vs the corrected GearDVFS port across BOTS benchmarks (Jetson TX2, seed 42, finetuned; both columns re-measured in the same window).}}
\label{tab:multi_benchmark}
\setlength{\tabcolsep}{2pt}
\begin{tabular}{@{}lccccc@{}}
\toprule
\textbf{Bench.} & \textbf{HiDVFS} & \textbf{GearDVFS} & \textbf{Spd.} & \textbf{E$_{\text{H}}$} & \textbf{E$_{\text{G}}$} \\
 & \textbf{L10 (s)} & \textbf{L10 (s)} & & \textbf{(kJ)} & \textbf{(kJ)} \\
\midrule
\review{alignment} & \review{2.97$\pm$0.60} & \review{12.8$\pm$5.80} & \review{4.32$\times$} & \review{3.70} & \review{8.77} \\
\review{concom} & \review{1.42$\pm$0.41} & \review{5.29$\pm$1.96} & \review{3.73$\times$} & \review{1.87} & \review{3.73} \\
\review{fft} & \review{3.35$\pm$0.27} & \review{11.6$\pm$6.38} & \review{3.46$\times$} & \review{6.10} & \review{9.50} \\
\review{fib} & \review{2.10$\pm$1.31} & \review{2.05$\pm$1.24} & \review{0.98$\times$} & \review{2.84} & \review{2.53} \\
\review{floorplan} & \review{0.31$\pm$0.04} & \review{2.46$\pm$1.24} & \review{7.98$\times$} & \review{1.56} & \review{2.67} \\
\review{health} & \review{1.09$\pm$0.84} & \review{7.44$\pm$4.32} & \review{6.84$\times$} & \review{2.28} & \review{5.47} \\
\review{knapsack} & \review{5.95$\pm$8.99} & \review{13.9$\pm$8.31} & \review{2.33$\times$} & \review{2.32} & \review{6.65} \\
\review{nqueens} & \review{7.26$\pm$6.21} & \review{77.8$\pm$44.5} & \review{10.72$\times$} & \review{10.9} & \review{42.0} \\
\review{sort} & \review{13.5$\pm$13.7} & \review{28.9$\pm$21.5} & \review{2.15$\times$} & \review{9.11} & \review{17.2} \\
\review{sparselu} & \review{0.83$\pm$0.40} & \review{2.73$\pm$1.77} & \review{3.29$\times$} & \review{2.18} & \review{3.07} \\
\review{strassen} & \review{0.33$\pm$0.06} & \review{1.87$\pm$1.17} & \review{5.72$\times$} & \review{1.58} & \review{2.53} \\
\review{uts} & \review{5.17$\pm$0.54} & \review{20.3$\pm$9.63} & \review{3.92$\times$} & \review{7.45} & \review{13.2} \\
\midrule
\review{\textbf{Average}} & \review{3.69} & \review{15.59} & \review{\textbf{4.62$\times$}} & \review{4.33} & \review{9.78} \\
\bottomrule
\multicolumn{6}{@{}p{3.2in}@{}}{\scriptsize \review{L10=mean$\pm$std over the last-10-epoch window (single run, seed 42); E$_{\text{H}}$/E$_{\text{G}}$=HiDVFS/GearDVFS total energy over 100 epochs; GearDVFS=repo-loyal port.}} \\
\end{tabular}
\end{table}

\textbf{Execution Modes.} We distinguish two execution modes in our experiments:
\begin{itemize}
    \item \textbf{Parallel mode:} Multiple benchmarks execute concurrently on the system, competing for cores and thermal headroom. This mode tests HiDVFS's ability to coordinate multi-application scheduling.
    \item \textbf{Sequential mode:} Benchmarks execute one at a time (not concurrently), but each benchmark still uses multiple cores internally. This isolates per-application behavior without inter-application contention.
\end{itemize}
Note that ``sequential mode'' does not mean single-threaded execution; each benchmark remains parallelized across its allocated cores.

\rev{\textbf{Benchmark Evaluation Protocol.} Each benchmark evaluation epoch executes three scheduling variants per benchmark (serial, omp-tasks, omp-tasks-tied), allowing the RL agent to learn across different parallelization strategies. The multi-seed RL algorithm comparison (Table~\ref{tab:comparison}) runs 100 epochs per phase (training and finetuning) with seeds 42, 123, and 456 on a representative benchmark (FFT). The agent receives profiling data (makespan, energy, temperature, cache/branch misses) after each execution and updates its policy accordingly. The BOTS per-benchmark comparison (Table~\ref{tab:multi_benchmark}) evaluates HiDVFS against GearDVFS~\cite{lin2023workload} \review{across all 12 applications} using seed 42 with finetuned policies.}

\subsection{Evaluation of Metrics and Statistical Analysis of Features}

This subsection outlines the evaluation methodology for assessing single-agent and multi-agent Reinforcement Learning (RL) approaches, focusing on key performance metrics (makespan, energy consumption, average temperature, branch misses, and cache misses), while integrating statistical analyses to quantify the impact of critical variables: task priority, number of cores, and average frequency. For single-agent RL, the primary objective is to minimize makespan or energy consumption by optimizing core allocation and frequency selection, as detailed in Section~\ref{sub:IRL}. The reward function balances these objectives using the parameter \(\beta\), which weights makespan (\(\beta\)) against energy consumption (\(1 - \beta) (E_{\text{target}} / (E + \epsilon))\)). To evaluate convergence speed, we set \(\beta = 1\), isolating makespan and excluding energy terms, though minimizing makespan indirectly reduces energy due to shorter computation times, as confirmed by our statistical results. Target values for makespan and energy consumption, derived from testing all four conditions in Section~\ref{sub:tuning}, serve as benchmarks. In multi-agent RL, rewards are assigned to a thermal agent for core combination actions and a priority agent for task priority assignments, enhancing temperature reliability and makespan through real-time processing of tasks and learning efficiency of models. \add{Throughout, makespan is the optimization priority.}

Mann--Whitney U tests over the full single-task sweep confirm that all three actuators significantly shape execution: frequency dominates makespan (p${=}$1e-283; 6.92s$\to$2.62s) and energy (p${=}$6e-144), core count is next (3.26e-25 on energy), and priority drives miss-rate overheads (cache p${=}$1e-24); all motivate our makespan-first reward. Further information is provided in the supplementary material.

\review{Extended profiling of the three FFT variants (untied, tied, serial) under parallel and sequential modes is provided in the supplementary material: tied execution incurs higher makespan, branch misses, and cache misses due to restricted task migration, and parallel execution shows greater cache-miss variability, consistent with the feature-importance results (also in the supplementary material). The supplementary material additionally sweeps priority combinations, core counts, and frequency levels; the much larger spread of energy and makespan in parallel mode motivates adaptive control of all three actuators.}

\subsection{Implemented Approaches}
\label{subsub:implementation}

We compare our two methods, \textbf{SARB} (single-agent) and \textbf{HiDVFS} (multi-agent), against representative DVFS/RL baselines.

\textbf{Single-agent baselines.} zTT~\cite{kim2021ztt} and GearDVFS~\cite{lin2023workload} are model-free DVFS schedulers; DynaQ~\cite{angermueller2019model} and PlanGAN~\cite{charlesworth2020plangan} are model-based. \textbf{SARB} uses short-horizon model predictions only for reward shaping, improving stability and convergence.

\textbf{Multi-agent baselines.} Multi-Agent Model-Based (MAMB) and Multi-Agent Model-Free (MAMF) are model-based/model-free decompositions; \rev{HiDVFS\_S} uses standard DQN without D3QN stabilization. \textbf{HiDVFS} adds thermal and priority agents on top of SARB with short-horizon reward shaping, D3QN stabilization, and coordinated actions (core masks, frequencies, priorities).

\begin{figure}[!h]
\centering
\add{
\includegraphics[width=0.62\linewidth]{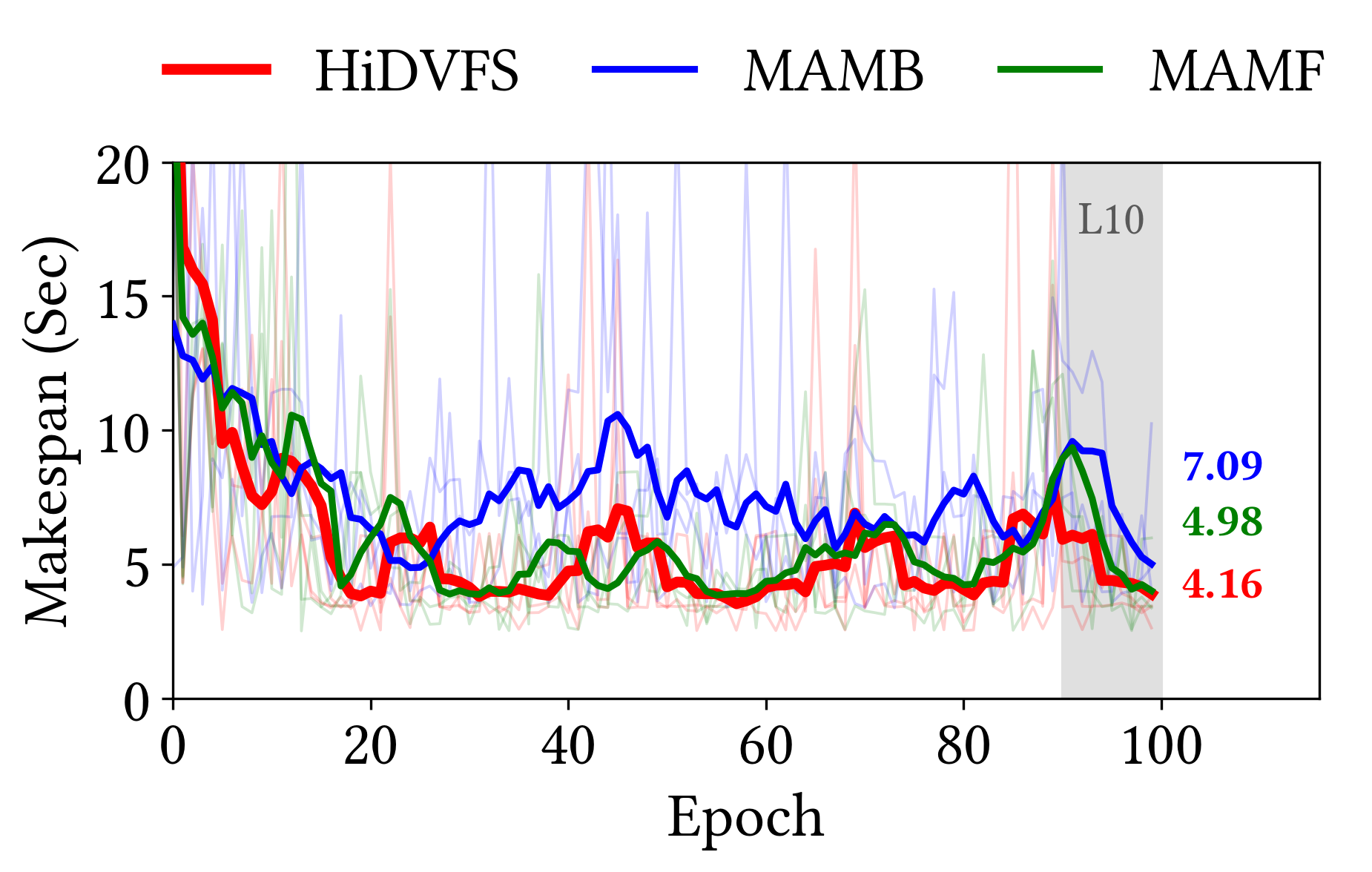}
}
\caption{\review{Multi-agent makespan comparison (TX2, FFT, finetuned): thin lines are the three seeds, bold lines their 5-epoch rolling mean, and the shaded band the L10 window, where HiDVFS's annotated 4.16\,s mean is lowest (y-axis clipped at 20\,s; exploration spikes reach 40\,s). Single-agent and energy plots are in the supplementary material; quantitative results in Table~\ref{tab:comparison}.}}
\label{fig:performance_comparison_action}
\end{figure}

\begin{figure}[!h]
\centering
\add{\includegraphics[width=0.62\linewidth]{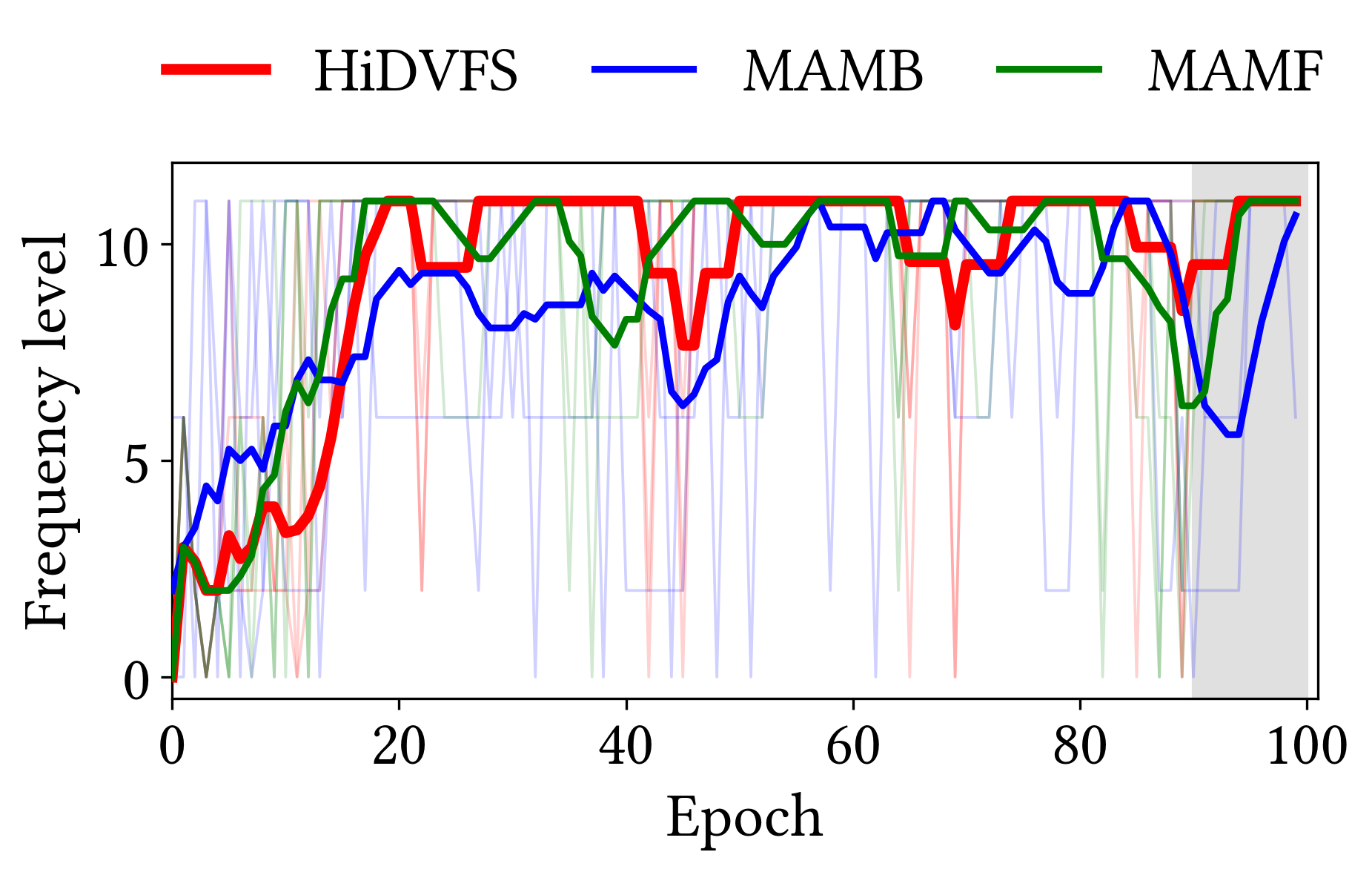}}
\caption{\review{Multi-agent frequency actions (TX2, FFT, finetuned; same encoding as Fig.~\ref{fig:performance_comparison_action}): HiDVFS's bold rolling mean holds the top level (11.0 mean over the shaded L10 window, vs.\ 8.1 for MAMB). Companion core-count and single-agent action plots are in the supplementary material.}}
\label{fig:action_analysis}
\end{figure}

\subsection{Comparison with Baselines and Results}

\textbf{Comparative Performance:}
\rev{Figure~\ref{fig:performance_comparison_action} compares the multi-agent variants of our proposed \add{HiDVFS} against multi-agent SOTA RL baselines (MAMB, MAMF\review{; HiDVFS\_S in the supplementary material}) over 100 epochs \review{(finetuned phase, seeds 42/123/456)}; the corresponding single-agent comparison (zTT, GearDVFS, DynaQ, PlanGAN, \add{SARB}) is summarised quantitatively in Table~\ref{tab:comparison} and visualised in the supplementary material.} High rewards for lower makespans drive agents to maximize frequency and core counts, with faster stabilization indicating better convergence. \review{HiDVFS stabilizes makespan and frequency compared to MAMB and MAMF, reducing total finetuned makespan by 61.5\% versus its no-D3QN ablation HiDVFS\_S (1446.5\,s to 556.2\,s) and energy by 54.1\% (13.87\,kJ to 6.37\,kJ; full 7-metric table in the supplementary material). Multi-agent methods stabilize earlier due to distributed decision making.} Among single-agent methods, Table~\ref{tab:comparison} shows \add{SARB} and zTT as the strongest baselines, with zTT achieving the lowest L10 makespan of 5.45\,s while remaining clearly above the multi-agent leader.

Energy mirrors makespan: HiDVFS and SARB stabilise within a few epochs (Table~\ref{tab:comparison}); HiDVFS's reward spikes reflect accumulated future rewards. Models are fine-tuned over 150 transitions before 60-epoch runs. \ed{Frequency increases trade makespan for cache misses; the profiler agent adjusts core counts accordingly (full hardware-counter analysis in supplement).}

\re{\textbf{Evaluation Details:}} \review{Convergence is a stable makespan within a 10-epoch sliding window (variation $<15\%$ of the initial 5-epoch average); totals are summed over the 100 finetuned epochs. Single-agent: zTT attains the lowest totals (602.2\,s, 7.42\,kJ) and the fastest convergence (52$\pm$34 epochs); SARB follows at 813.8\,s with the most stable L20. Multi-agent: HiDVFS reaches 556.2\,s and 6.37\,kJ, converging at 57$\pm$14 epochs, a 61.5\% makespan and 54.1\% energy improvement over HiDVFS\_S.}

\re{\textbf{Overhead and Deployment:}} \re{\add{HiDVFS}'s server--client decision cycle costs $\sim$2\,ms ($\sim$0.5\,ms action send, $\sim$0.3\,ms sysfs DVFS write with frequency taking effect within 1--2 CPU cycles ($<$1\,$\mu$s), $\sim$0.8\,ms \texttt{perf} setup/teardown, $\sim$0.4\,ms response), i.e.\ $0.03$--$0.2\%$ of the 1--6\,s benchmark window. Hyperparameter tuning is offline (10--12\,h on a 4-core i7); thermal limit $50\,^\circ$C (policy) / $85\,^\circ$C (hardware), consistent with~\cite{kim2021ztt}.}

\begin{table}[ht]
\centering
\scriptsize
\caption{\re{Comparison of RL Approaches \add{(TX2, FFT, finetuned, multi-seed mean$\pm$std)}}}
\label{tab:comparison}
\add{
\begin{tabular}{|l|c|c|c|c|}
\hline
\textbf{Approach} & \textbf{L10 (s)} & \textbf{L20 (s)} & \textbf{Energy (kJ)} & \textbf{HF\%} \\
\hline
\multicolumn{5}{|c|}{\textbf{Single-Agent Approaches}} \\
\hline
\review{zTT~\cite{kim2021ztt}$^{\dagger}$} & \textbf{5.45$\pm$1.07} & 5.71$\pm$0.35 & \review{\textbf{7.42$\pm$0.15}} & 70.7$\pm$6.3 \\
DynaQ~\cite{angermueller2019model} & 10.72$\pm$2.08 & 9.42$\pm$1.81 & \review{9.36$\pm$0.58} & 30.0$\pm$9.4 \\
PlanGAN~\cite{charlesworth2020plangan} & 7.57$\pm$0.67 & 5.85$\pm$0.49 & \review{7.39$\pm$0.21} & 67.0$\pm$1.6 \\
\review{GearDVFS~\cite{lin2023workload}$^{\ddagger}$} & \review{11.79$\pm$0.15} & \review{11.63$\pm$0.24} & \review{9.49$\pm$0.14} & \review{22.7$\pm$1.7} \\
\review{GearDVFS-A$^{\dagger}$} & \review{10.55$\pm$0.23} & \review{10.93$\pm$0.08} & \review{9.72$\pm$0.11} & \review{23.0$\pm$1.4} \\
\textbf{SARB} & \review{5.77$\pm$1.19} & \review{\textbf{5.49$\pm$1.12}} & \review{7.79$\pm$0.65} & \review{41.7$\pm$9.1} \\
\hline
\multicolumn{5}{|c|}{\textbf{Multi-Agent Approaches}} \\
\hline
MAMB & 7.09$\pm$1.29 & 6.91$\pm$0.93 & \review{8.73$\pm$0.59} & 66.0$\pm$4.5 \\
MAMF & 4.98$\pm$0.61 & 5.85$\pm$0.63 & \review{7.14$\pm$0.19} & 78.7$\pm$5.3 \\
\rev{\textbf{HiDVFS}} & \rev{\textbf{4.16$\pm$0.58}} & \rev{\textbf{5.14$\pm$1.06}} & \review{\textbf{6.37$\pm$0.37}} & \rev{\textbf{81.0$\pm$0.8}} \\
\hline
\end{tabular}
}
\par\vspace{1mm}
\parbox{0.75\linewidth}{\scriptsize L10/L20=Avg last 10/20 epochs makespan. HF\%=High-Frequency ($\geq$9) rate. \review{Energy=total over the 100 finetuned epochs. $^{\dagger}$objective-adapted: the method's QoS term instantiated on our per-job makespan target. $^{\ddagger}$loyal port: upstream's released reward unchanged. Full fair-port study in the supplementary material.}}
\end{table}


\re{Our proposed \add{SARB} and \add{HiDVFS}, leveraging per-cluster temperature profiling and model-based reward estimation, outperform SOTA RL baselines in makespan reduction, energy savings, and thermal management for FFT workloads. \review{SARB offers the most stable single-agent tail (L20 5.49$\pm$1.12\,s)}, while \add{HiDVFS} excels on complex DAGs with manageable deployment overhead on Jetson TX2.} \rev{HiDVFS\_S (without D3QN) shows higher variance but may suit resource-constrained deployments.}

\review{\textbf{Reconciling the GearDVFS result.} A reader may ask why zTT (2021) beats the newer GearDVFS (2023) here, given GearDVFS reports roughly 20\% better energy efficiency than zTT in its own setting~\cite{lin2023workload}. The answer is that the two methods automate different things, and only one of them transfers to episodic DAG jobs. zTT is built around a per-application QoS target, which our port instantiates directly as the per-job makespan target (the "objective-adapted" row); GearDVFS's contribution is workload-context awareness for CONCURRENT interactive tasks on mobile SoCs, driven by a utilization-band governor signal sampled every 25\,ms, and that signal has no notion of job completion. Auditing our first port against the authors' released code also exposed three defects (a disconnected utilization input, a power term fed millijoule-scale energy, and a finetune phase that silently started from untrained weights), so Table~\ref{tab:comparison} reports a repaired, repo-loyal port plus the objective-adapted control. The measured decomposition: repairing the port improves GearDVFS from 14.3\,s to 11.79$\pm$0.15\,s; handing its unchanged network zTT's exact objective (GearDVFS-A) reaches only 10.55$\pm$0.23\,s, so most of the remaining gap is the agent design itself: a deliberately tiny, undiscounted ($\gamma{=}1$) 25-unit DQN with a 0.2 exploration floor is the right economy for millisecond-interval governor control, where exploratory actions are nearly free, but on 1-6\,s episodic jobs every exploratory epoch costs seconds and the undiscounted return never sharpens toward completion time (high-frequency residency stays near 23\% in every variant). Symmetrically, removing our zTT port's favourable shaping degrades zTT to 7.76$\pm$1.50\,s, still ahead of GearDVFS-A. Both papers' findings are thus consistent: the ranking here measures transfer of each method's native design to episodic DAG scheduling, not their quality in GearDVFS's own domain (full fair-port study in the supplementary material).}

\review{\textbf{What drives the advantage.} Attributing the gains to design choices: (i) short-horizon reward shaping drives the single-agent advantage, SARB reaching 5.77\,s L10 while its shaping-free model-based peer DynaQ sits at nearly twice that (10.72\,s, Table~\ref{tab:comparison}); (ii) the actuator split drives sample efficiency and stability, adding the thermal and priority agents moves L10 from SARB's \review{5.77$\pm$1.19}\,s to HiDVFS's 4.16$\pm$0.58\,s with the lowest seed variance, while removing D3QN stabilization (HiDVFS\_S, Fig.~\ref{fig:performance_comparison_action}) restores that variance; (iii) temperature-aware core grouping sustains the highest high-frequency residency (81.0\% HF, Table~\ref{tab:comparison}) at policy-safe temperatures (peaks below $48^\circ$C, Table~\ref{tab:policy_ablation}), which is where both makespan and energy are won; and (iv) the gain is not simply pinning the maximum frequency, since Deadline-aware DVFS matches Max-freq's makespan within the deadline at $15\text{--}18\%$ lower energy (Table~\ref{tab:policy_ablation}).}

\review{\textbf{Comprehensive BOTS Benchmark Evaluation:} Across all 12 BOTS applications (TX2, seed~42, finetuned, both algorithms re-measured in the same window on the corrected GearDVFS port), HiDVFS attains an average \textbf{4.62$\times$} speedup (geometric mean $3.88\times$; wins on 11 of 12 benchmarks) and 55.7\% energy reduction (peaks: nqueens $10.72\times$, floorplan $7.98\times$; lows: fib $0.98\times$, a tie on sub-second jobs, and sort $2.15\times$); Table~\ref{tab:multi_benchmark} gives the per-benchmark breakdown.}

\rev{\textbf{Multi-Seed Statistical Validation:} Across three random seeds (42, 123, 456) on FFT with 100 epochs per phase, the aggregated mean$\pm$std values in Table~\ref{tab:comparison} give HiDVFS the best finetuned makespan (\textbf{4.16$\pm$0.58s} \review{vs.\ the corrected GearDVFS port's 11.79$\pm$0.15s, a \textbf{2.83$\times$} speedup}) with the lowest variance; the per-seed breakdown is in the supplementary material.}

\subsection{Real-Time Evaluation: Cross-Platform Envelope, Feasibility, and Safety}
\label{sub:rt_eval}

Our evaluation answers four real-time questions: (i) how often does the DVFS action space miss soft deadlines? (ii) when is a workload formally schedulable? (iii) is pinning the max frequency enough? (iv) can the policy be calibrated for online use? Methods split by agent count, environment model, and actuators ($f$/$c$/$p$/$t$): single-agent model-free (zTT, GearDVFS), single-agent model-based (DynaQ, PlanGAN, SARB$_\text{ours}$), and multi-agent (MAMB, MAMF, HiDVFS$_\text{ours}$ closing $f/c/p/t$ hierarchically). Further details, including a full taxonomy table, are provided in the supplementary material.

\textbf{Soft-RT envelope.} The deadline-miss ratio (DMR) and response-time ratio $R/D$ are computed over the full frequency$\times$core sweep on 42 BOTS/PolyBench workloads per platform, with $D{=}k\,C_\text{ref}$. Table~\ref{tab:rt_envelope} reports the median $\tilde{R}/D$, p90, and DMR: across the action space, \emph{most} operating points violate even a loose $k{=}1.5$ deadline ($52\text{--}80\%$ DMR); only the high-frequency/high-core region is feasible, motivating a learned feasible-and-efficient policy.

\review{\textbf{Miss semantics and effect.} Misses are soft: a late job runs to completion and the miss is logged, so no work is dropped. The effect of a miss is the lateness itself (the result arrives up to the $\tilde{R}/D$ and p90 factors of Table~\ref{tab:rt_envelope} beyond its deadline) plus the queueing delay imposed on other DAGs while the late job over-occupies its cores; the mixed-criticality study below measures exactly this interference on a co-running task. DMR thus reports how often jobs are late and the tail how late; the assurances are layered as the analytic feasibility gate, the calibrated shield bound (an empirical prediction interval with stated coverage, not a hard guarantee), and these measured outcomes.}

\begin{table}[ht]
\centering
\scriptsize
\setlength{\tabcolsep}{1.5pt}
\caption{Soft real-time response-time envelope on TX2/Orin NX/RubikPi (full DVFS sweep, 42 workloads/platform). $\tilde{R}/D$ = median ratio; p90 = 90th percentile.}
\label{tab:rt_envelope}
\begin{tabular}{@{}lccccccccc@{}}
\toprule
 & \multicolumn{3}{c}{Jetson TX2} & \multicolumn{3}{c}{Jetson Orin NX} & \multicolumn{3}{c}{RubikPi} \\
$k$ ($U{=}1/k$) & DMR\% & $\tilde{R}/D$ & p90 & DMR\% & $\tilde{R}/D$ & p90 & DMR\% & $\tilde{R}/D$ & p90 \\
\midrule
1.10 (0.91) & 78.4 & 2.47 & 16.37 & 68.2 & 1.46 & 11.07 & 89.9 & 3.75 & 25.00 \\
1.25 (0.80) & 74.4 & 2.17 & 14.41 & 61.8 & 1.28 & 9.74  & 85.0 & 3.30 & 22.00 \\
1.50 (0.67) & 69.6 & 1.81 & 12.01 & 52.2 & 1.07 & 8.12  & 79.9 & 2.75 & 18.34 \\
\bottomrule
\end{tabular}
\end{table}

\begin{figure}[ht]
\centering
\includegraphics[width=0.62\linewidth]{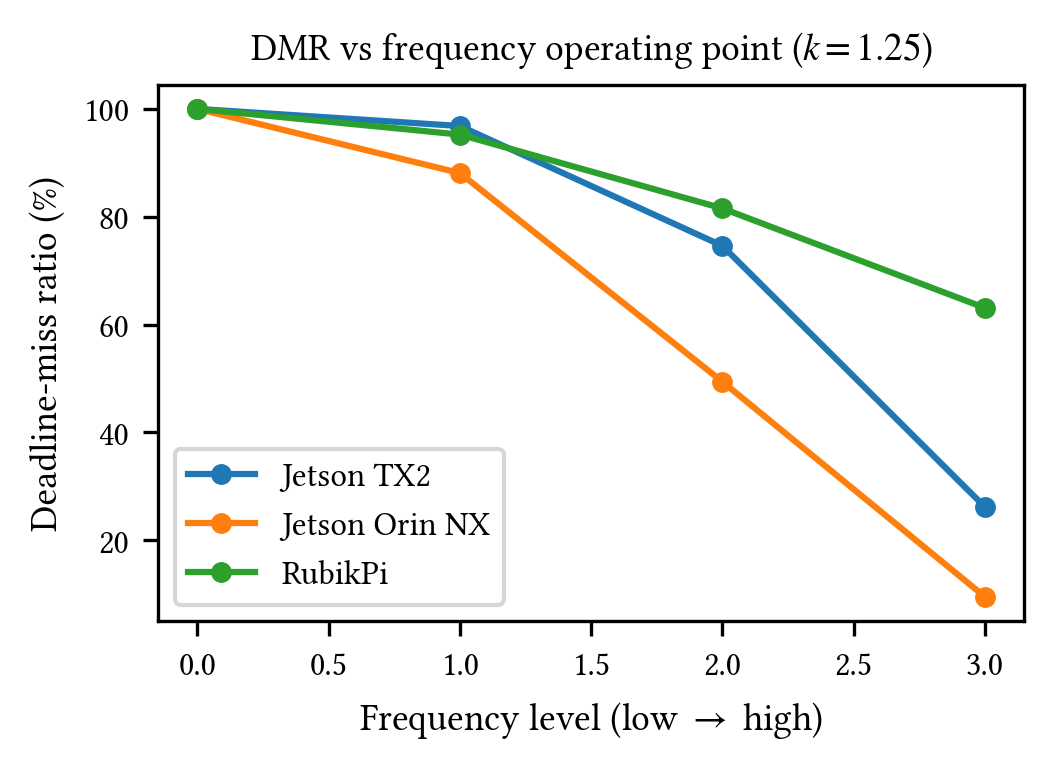}
\caption{\review{DMR vs.\ frequency operating point at $k{=}1.25$ on TX2/Orin NX/RubikPi, aggregated over all core counts per level: misses fall monotonically as frequency rises (top-level DMR 26.2/9.5/63.1\%), and the residual top-level misses come from operating points with unfavourable core counts. Only the high-frequency region is deadline-feasible.}}
\label{fig:dmr_vs_freq}
\end{figure}

\textbf{Federated feasibility gate.} Under federated scheduling~\cite{saifullah2014parallel,saifullah2020cpu}, a heavy task ($C_i{>}D_i$) is feasible on $m_i{=}\lceil(C_i{-}L_i)/(D_i{-}L_i)\rceil$ dedicated cores when $D_i{>}L_i$. The energy-minimal schedule is the lowest-energy operating point passing this gate (Table~\ref{tab:feasibility_gate}): TX2/Orin admit $\ge81\%$ even at $k{=}1.10$, while RubikPi is strict ($40\%\to86\%$ as $k$ loosens, up to $\tilde{m}{=}4$ cores).

\begin{table}[ht]
\centering
\scriptsize
\setlength{\tabcolsep}{3pt}
\caption{Federated-scheduling feasibility gate on TX2/Orin NX/RubikPi. Sched\% = fraction of workloads with a gate-feasible operating point; $\tilde{m}$ = median dedicated cores at the energy-minimal feasible point.}
\label{tab:feasibility_gate}
\begin{tabular}{@{}lcccccc@{}}
\toprule
 & \multicolumn{2}{c}{Jetson TX2} & \multicolumn{2}{c}{Jetson Orin NX} & \multicolumn{2}{c}{RubikPi} \\
$k$ ($U{=}1/k$) & Sched\% & $\tilde{m}$ cores & Sched\% & $\tilde{m}$ cores & Sched\% & $\tilde{m}$ cores \\
\midrule
1.10 (0.91) & 81 & 1 & 93  & 1 & 40 & 4 \\
1.25 (0.80) & 95 & 1 & 98  & 1 & 67 & 4 \\
1.50 (0.67) & 98 & 1 & 100 & 1 & 86 & 2 \\
\bottomrule
\end{tabular}
\end{table}

\review{\textbf{Policies.} Throughout, a \emph{policy} is the operating rule that maps the observed state to frequency and core actions; it is not a weighting inside the reward function. \emph{Max-freq} pins every core to $f_{\max}$ (the Linux \texttt{performance} governor); \emph{Powersave} pins $f_{\min}$; \emph{Deadline-aware DVFS} is the learned HiDVFS policy restricted to gate-feasible operating points. The first two are fixed baselines; only the third adapts.}

\textbf{Max-frequency ablation.} At $D{=}1.25\,C_\text{ref}$ (Table~\ref{tab:policy_ablation}), \emph{Deadline-aware DVFS} meets every deadline (DMR$=0$) while cutting energy $15\text{--}18\%$ vs.\ \emph{Max-freq}; \emph{Powersave} consumes $3\text{--}5\times$ \emph{more} energy and misses every deadline (race-to-idle). The energy-optimal RT policy lives at neither extreme.

\review{\textbf{Best static frequency.} Because the most energy-efficient frequency can lie between the extremes, we also sweep the interior static levels (table in the supplementary material): on TX2 the energy-optimal static frequency is indeed interior, level 7 of 11 (1.42\,GHz), spending 8\% less energy than pinning $f_{\max}$, while pinning $f_{\min}$ costs $3.1\times$ more; on RubikPi $f_{\max}$ itself is energy-optimal (race-to-idle); Orin NX energy was not recorded. The interior optimum bounds how far any extreme-anchored energy target can be off (8\% on TX2, 0\% on RubikPi), and the deadline-aware policy's $15\text{--}18\%$ saving over Max-freq confirms the learned scheduler exploits this interior headroom.}

\begin{table}[ht]
\centering
\scriptsize
\setlength{\tabcolsep}{3pt}
\caption{Max-frequency ablation at $D{=}1.25\,C_\text{ref}$ on TX2/Orin NX/RubikPi. Energy is relative to Max-freq; Orin NX energy not recorded.}
\label{tab:policy_ablation}
\begin{tabular}{@{}llcccc@{}}
\toprule
Platform & Policy & Makespan (s) & Energy (rel.) & Peak $^\circ$C & DMR\% \\
\midrule
Jetson TX2     & Max-freq            & 0.36 & 100\% & 47.7 & 0   \\
               & Deadline-aware & 0.36 & 85\%  & 47.6 & 0   \\
               & Powersave           & 2.21 & 496\% & 47.2 & 100 \\
\midrule
Jetson Orin NX & Max-freq            & 0.26 & n/a   & 48.0 & 0   \\
               & Deadline-aware & 0.31 & n/a   & 48.1 & 0   \\
               & Powersave           & 4.29 & n/a   & 47.7 & 100 \\
\midrule
RubikPi        & Max-freq            & 0.50 & 100\% & 47.1 & 0   \\
               & Deadline-aware & 0.50 & 82\%  & 47.1 & 0   \\
               & Powersave           & 3.63 & 319\% & 42.5 & 100 \\
\bottomrule
\multicolumn{6}{@{}p{3.2in}@{}}{\scriptsize \review{Cells are means over the 42 workloads per platform; workload runtimes span three orders of magnitude, so the paired relative-energy and DMR columns carry the comparison rather than the makespan spread.}} \\
\end{tabular}
\end{table}

\textbf{HiDVFS-RT safety shield.} ``RL proposes, shield disposes'': the shield predicts $\hat{R}(a)$ with a split-conformal upper bound $\hat{R}^{+}(a)$~\cite{vovk2005algorithmic,angelopoulos2021gentle} and rejects when $\hat{R}^{+}{>}D$. Table~\ref{tab:shield} reports the calibration: PICP tracks the $90/95/99\%$ target on every platform (base $R^2$ up to $0.90$); the bound is therefore an empirical, calibrated prediction interval rather than a hard guarantee.

\begin{table}[ht]
\centering
\scriptsize
\setlength{\tabcolsep}{3pt}
\caption{HiDVFS-RT Shield calibration on TX2/Orin NX/RubikPi (held-out operating points). PICP$@$x = empirical coverage of the prediction interval at target level $x\%$.}
\label{tab:shield}
\begin{tabular}{@{}lccccc@{}}
\toprule
Platform & base $R^2$ & PICP@90 & PICP@95 & PICP@99 & margin@90 (s) \\
\midrule
Jetson TX2     & 0.59 & 89.7 & 92.9 & 96.8 & 0.59 \\
Jetson Orin NX & 0.90 & 91.1 & 94.0 & 97.0 & 0.20 \\
RubikPi        & 0.90 & 89.3 & 95.8 & 98.2 & 0.27 \\
\bottomrule
\end{tabular}
\end{table}

\textbf{Mixed-criticality co-scheduling on Jetson Orin NX.} A high-criticality (HI) \texttt{SCHED\_FIFO} task (priority~99, $50$\,ms period, $D{=}15.1$\,ms) ran alongside a continuous low-criticality (LO) FFT DAG; Orin NX exposes DVFS \emph{per-cluster}, so placement is decisive. \emph{Isolated} placement: HI mean $10.09\text{--}10.11$\,ms, $0\%$ DMR (temporal + frequency isolation). \emph{Co-located} at $f_\text{min}$: the shared clock drops to $0.12$\,GHz, HI inflates to $182.9$\,ms, $100\%$ DMR (frequency inversion). The mixed-criticality guarantee rests on cluster-aware reservation.



\section{Conclusion}
\label{conclusion}

\review{HiDVFS is an extensible hierarchical multi-agent DVFS scheduler that minimises energy subject to soft-deadline feasibility. Multi-seed validation on Jetson TX2 gives an L10 makespan of 4.16$\pm$0.58\,s, a 2.83$\times$ speedup and 32.9\% energy reduction over the fairness-corrected GearDVFS port, and a 4.62$\times$ average speedup across the 12-benchmark BOTS suite; a calibrated split-conformal shield bounds predicted response times, and cluster-aware reservation keeps the high-criticality task's deadline-miss ratio at zero on Orin NX. Its limitations are explicit: deadlines are soft, the shield's bounds are calibrated empirical prediction intervals rather than hard guarantees, hyperparameters are tuned offline, the 2\,ms server-client cycle suits second-scale jobs rather than microsecond control, and the RL comparison transfers each baseline's native reward to episodic DAG jobs, which favours objective-aligned signals such as zTT's. Future work follows the extensibility interface of the agent hierarchy: actuator-domain agents for memory bandwidth, power capping, and cache partitioning; online adaptation of reward targets; closed-loop evaluation of the shield under the learned policy; and broader platforms, including many-core servers. Code, models, and data will be released upon acceptance.}

\section*{Acknowledgment}
\review{Portions of the manuscript text were drafted and revised, and the analysis and plotting scripts were developed, with the assistance of Anthropic's Claude (Claude Code)~\cite{anthropic2026claude}; all AI-assisted content was reviewed, verified against the measured data, and approved by the authors. \textbf{Data availability:} the raw measurement data behind every table and figure is archived on Zenodo at \href{https://doi.org/10.5281/zenodo.21212162}{DOI 10.5281/zenodo.21212162}.}

\clearpage
\appendix
\noindent\textbf{Supplementary Material.} The following appendices reproduce the
complete supplementary material of the journal submission; section numbering
continues as lettered appendices and the bibliography at the end covers both
parts.

\ifx\HidvfsCombined\undefined
\documentclass[journal]{IEEEtran}
\IEEEoverridecommandlockouts
%
%
%

\usepackage[utf8]{inputenc}
\usepackage[T1]{fontenc}
\usepackage{cite}
\usepackage{amsmath}
\usepackage{amssymb}
\usepackage{amsfonts}
\usepackage{amsthm}
\usepackage{graphicx}
\usepackage{textcomp}
\usepackage{xcolor}
\usepackage{multirow}
\usepackage{caption}
\captionsetup[figure]{skip=2pt}
\usepackage{subcaption}
\usepackage{array}
\usepackage{listings}
\usepackage{float}
\usepackage{placeins}
\usepackage{url}
\usepackage{soul}
\usepackage{tikz}
\usetikzlibrary{positioning, shapes.multipart, arrows.meta, decorations.pathreplacing}
\usepackage{algorithm}
\usepackage{algpseudocode}
\usepackage{booktabs}
\usepackage[colorlinks=true,linkcolor=blue,citecolor=blue,urlcolor=blue]{hyperref}

\newtheorem{probdefinition}{Definition}
\newtheorem{probtheorem}{Theorem}
\newtheorem{problemma}{Lemma}
\newtheorem{probexample}{Example}

\newcommand{\re}[1]{#1}
\newcommand{\rev}[1]{#1}
\newcommand{\revv}[1]{#1}
\newcommand{\revvv}[1]{#1}
\newcommand{\revi}[1]{#1}
\newcommand{\edit}[1]{#1}
\newcommand{\ed}[1]{#1}
\newcommand{\add}[1]{#1}
\newcommand{\review}[1]{#1}

\algnewcommand{\algorithmicforeach}{\textbf{for each}}
\algdef{SE}[FOR]{ForEach}{EndForEach}[1]
  {\algorithmicforeach\ #1\ \algorithmicdo}
  {\algorithmicend\ \algorithmicforeach}

\def\BibTeX{{\rm B\kern-.05em{\sc i\kern-.025em b}\kern-.08em
    T\kern-.1667em\lower.7ex\hbox{E}\kern-.125emX}}

\begin{document}

\title{Supplementary Material\\
\large HiDVFS: Hierarchical Multi-Agent DVFS for Real-Time OpenMP DAG Workloads}

\author{Mohammad~Pivezhandi,~\IEEEmembership{Student Member,~IEEE,}
        Abusayeed~Saifullah,~\IEEEmembership{Member,~IEEE,}
        and~Ali~Jannesari,~\IEEEmembership{Senior Member,~IEEE}%
\thanks{M. Pivezhandi is with the Department of Computer Science, Wayne State University, Detroit, MI 48202 USA.}%
\thanks{A. Saifullah is with the Department of Computer Science, University of Texas at Dallas, Richardson, TX 75080 USA.}%
\thanks{A. Jannesari is with the Department of Computer Science, Iowa State University, Ames, IA 50011 USA.}%
\thanks{Manuscript received XXX; revised XXX.}}

\maketitle

\begin{abstract}
\review{This supplement accompanies the IEEE TPDS submission of HiDVFS and
collects appendix-grade material that does not fit within the
twelve-page body.} The sections below provide extended profiling
data, \review{extended design material,} single-agent counterparts of the main comparison and action
figures, energy and hardware-counter breakdowns, the SARB
single-agent version evaluation, multi-seed validation tables, the
complete seven-metric comparison, per-benchmark BOTS results, the
real-time response-time envelope and federated feasibility-gate
analysis, the methodology and verification package, and the
HiDVFS-RT safety-shield calibration. Every load-bearing claim
is summarized in the main paper; reviewers are not required to
consult this document to evaluate the headline results.
\end{abstract}
\fi

\section{Extended Profiling Data}
\label{app:profiling}

\begin{figure}[h]
    \centering
    \includegraphics[width=0.68\linewidth]{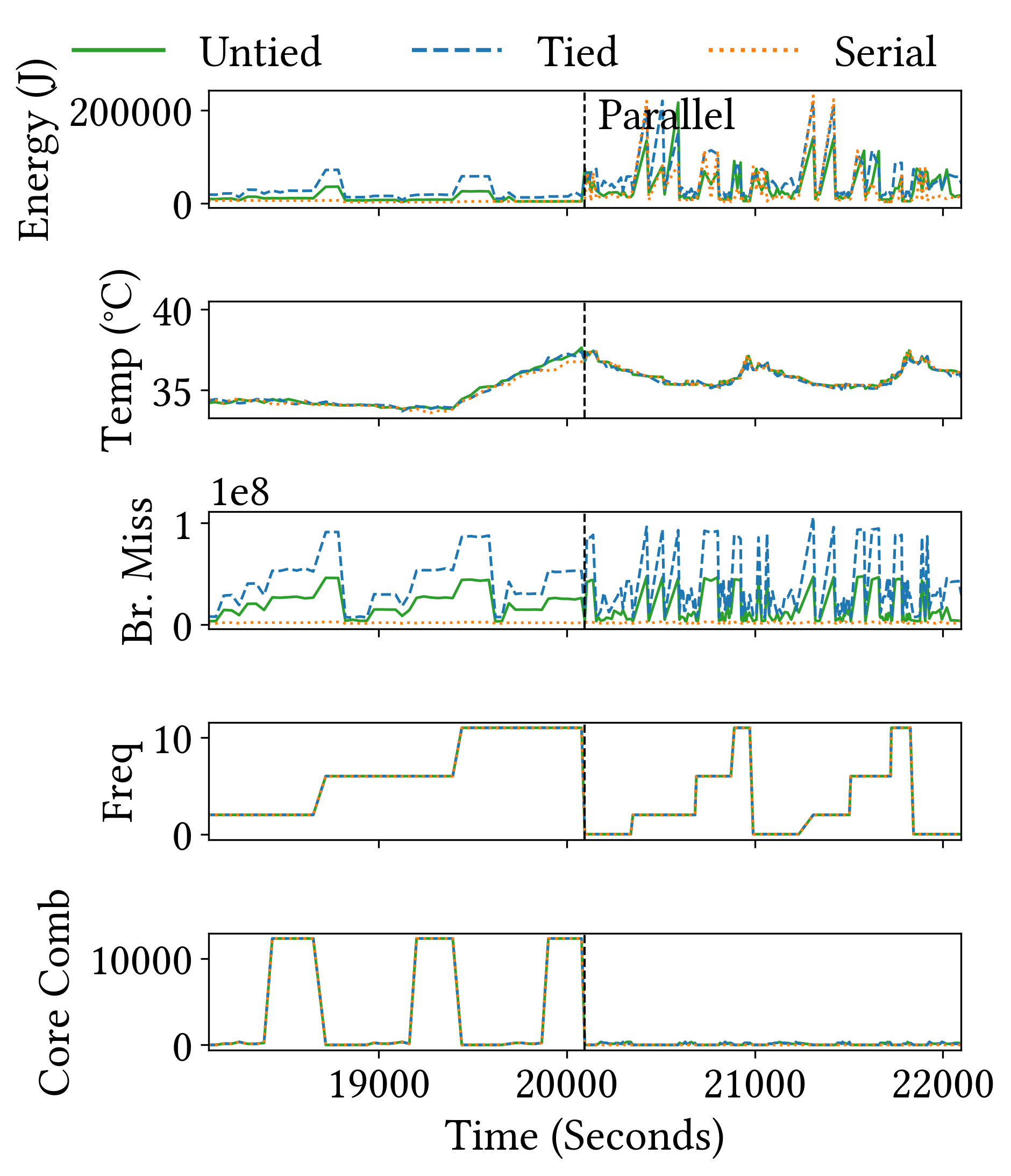}
    \caption{Profiling data on Jetson TX2 for FFT workload variations (untied, tied, serial) show more spontaneity in parallel mode. Panels (top$\to$bottom): energy, temperature, branch misses, frequency level, and core combination, with a shared legend and time axis. The dashed vertical line marks the sequential$\to$parallel transition; the ``Parallel'' annotation on the top panel labels the post-transition region.}
    \label{fig:profiling_plots_ext}
\end{figure}

Figure~\ref{fig:profiling_plots_ext} extends the profiling study, illustrating performance under parallel and sequential modes for three FFT workload variations (untied, tied, and serial). Temperature regulation ensures stability, but tied execution incurs higher makespan, branch misses, and cache misses due to restricted task migration, while serial mode, using one core, shows predictable, minimal miss patterns. Parallel execution introduces greater unpredictability in cache misses compared to serial mode. These results align with the feature-importance findings summarized in the main paper, supporting HiDVFS's dynamic adjustment of priority, cores, and frequency to minimize makespan and energy while mitigating performance penalties.

\begin{figure}[h]
    \centering
    \includegraphics[width=0.68\linewidth]{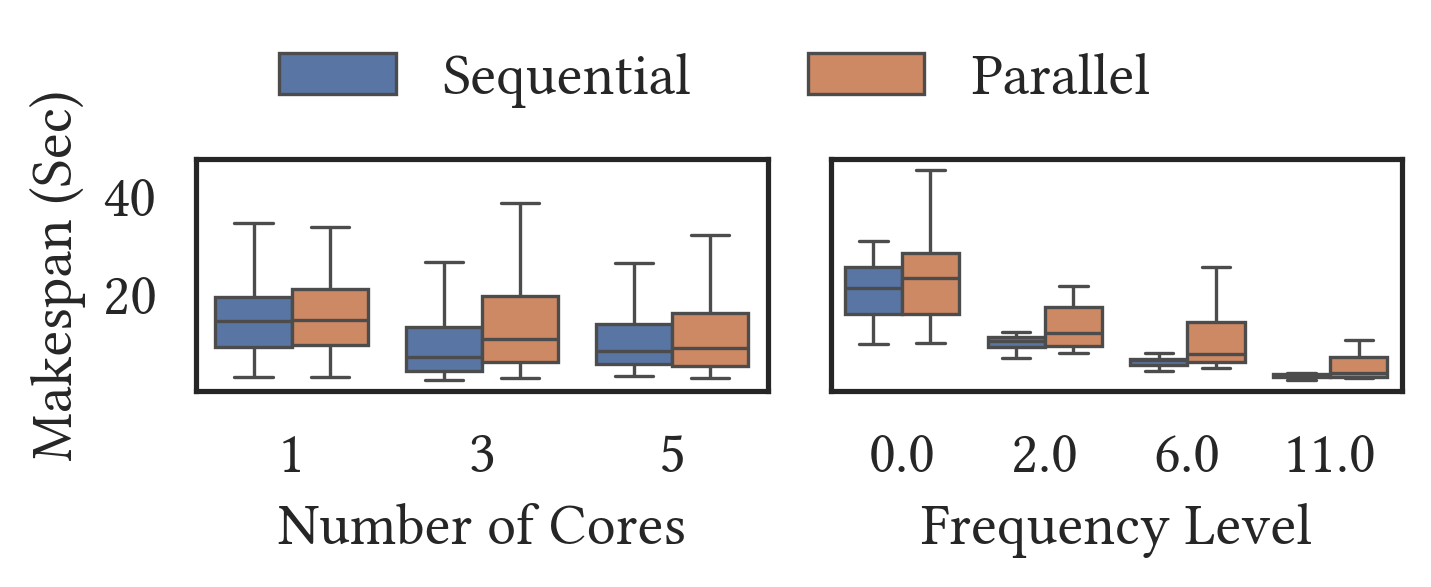}
    \caption{\review{Variation of total energy and makespan (TX2, FFT) across priority, core-count, and frequency settings in parallel and sequential modes.}}
    \label{fig:sup_actions_variation}
\end{figure}

\review{Figure~\ref{fig:sup_actions_variation}, moved here from the main body, shows how sweeping priority combinations, core counts, and frequency levels shifts total energy consumption and makespan in parallel and sequential modes. The much larger spread in parallel mode underscores the importance of adaptive control of all three actuators.}

\section{\review{Extended Design Material}}
\label{app:design}

\subsection{\review{IRL Background}}
\begin{sloppypar}\review{IRL assumes that expert demonstrations are generated by following an optimal policy with an optimal reward function $R^*$. Given $k$ expert demonstrations $\upsilon_E = \{ (s_0^E, a_0^E), (s_1^E, a_1^E), \ldots, (s_k^E,a_k^E)\}$ and $l$ agent policy samples $\upsilon = \{(s_0,a_0), (s_1,a_1), \ldots, (s_l,a_l)\}$, IRL attempts to recover $R^*$. A major challenge in IRL is reward ambiguity, where multiple reward functions can produce the observed expert behavior. MaxEnt IRL \cite{ziebart2008maximum} addresses this by maximizing the entropy of the policy distribution to prevent overfitting to a single solution. However, IRL still heavily relies on expert demonstrations, and poor-quality or suboptimal demonstrations can hinder learning. Moreover, the dependence on expert demonstrations remains a significant limitation.}\end{sloppypar}

\begin{figure}[h]
    \centering
    \includegraphics[width=0.75\linewidth]{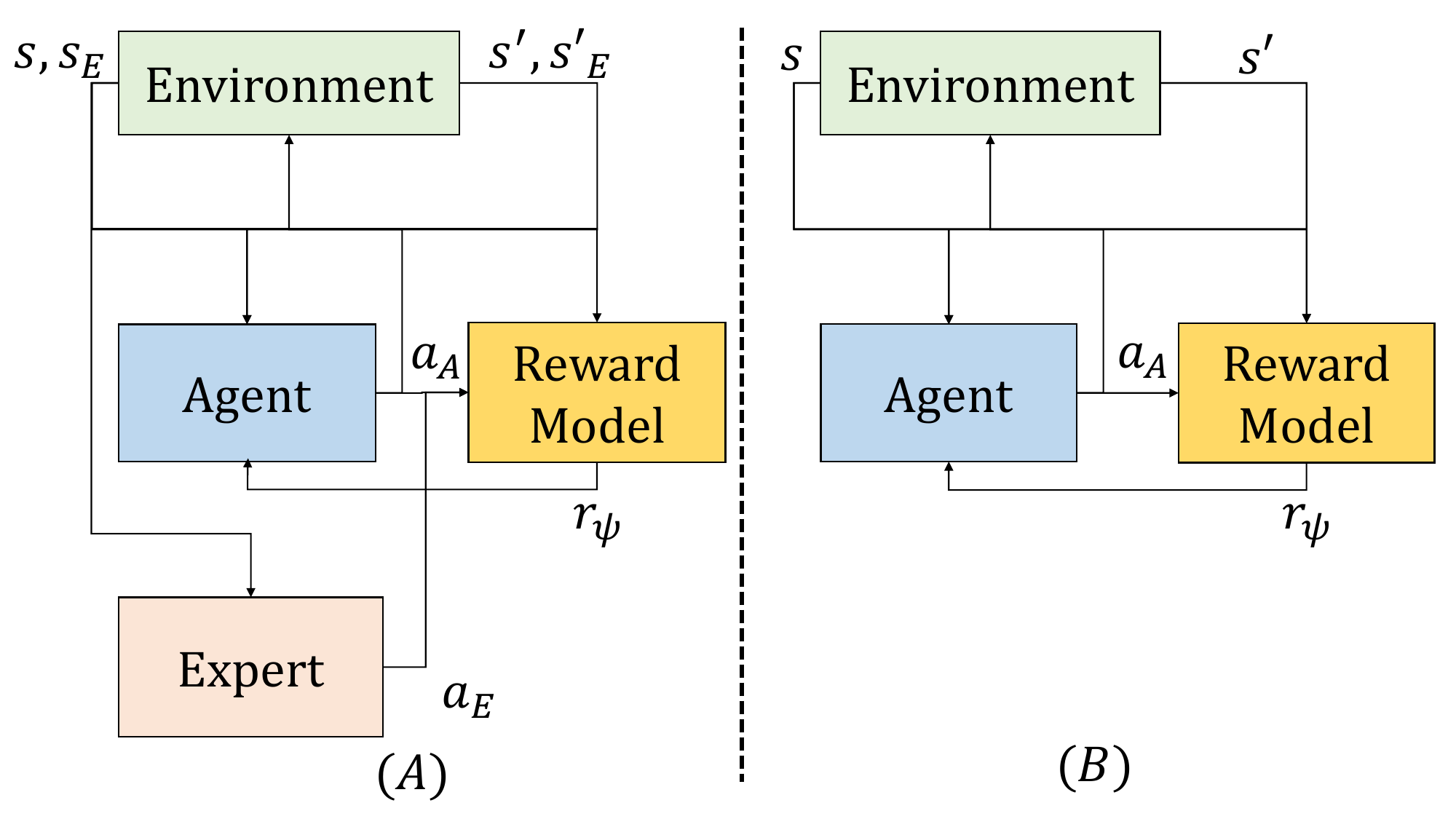}
    \caption{\review{IRL (A) aligns with expert behavior; our model (B) uses key-state predictions to shape rewards.}}
    \label{fig:sup_rewardmodel}
\end{figure}

\review{Figure~\ref{fig:sup_rewardmodel} contrasts the two paradigms: IRL shapes the reward by aligning expert and agent policies, whereas our model-based estimation maximizes information from key states and predicted trajectories without relying solely on expert data.}

\subsection{\review{OpenMP DAG Example}}
\review{Figure~\ref{fig:sup_openmp_code} shows the OpenMP DAG snippet referenced in the main paper: \texttt{\#pragma omp parallel} creates four threads; the \texttt{tied} task $\tau_{i,1}^*$ is confined to its cores, the \texttt{untied} tasks may migrate, branch mispredictions in the $L_1$/$L_2$ loops increase makespan variance, and dependencies propagate timing jitter along the critical path.}

\begin{figure}[htbp]
    \centering
    \begin{lstlisting}[language=C++, basicstyle=\footnotesize\ttfamily, numbers=left, numbersep=-7pt, escapechar=|, aboveskip=0pt, belowskip=0pt]
   #pragma omp parallel num_threads(4)
   {
      #pragma omp master
      {
      #pragma omp task // |{\large $\tau_{i,0}$}|
      { /* part 0 */ }
      #pragma omp task depend(out: x)
      final(true) // |{\large $\tau_{i,1}^*$}|
      {
          #pragma omp parallel for
          for (int i = 0; i < L1; i++) {
              if(i % 2 == 0) 
                  {/* work */}
              else{for (int j = 0; j < L2; j++) 
                  {/* work */}}
          }
      }
      #pragma omp task depend(in: x) // |{\large $\tau_{i,2}$}|
      { /* part 2 */ }
      #pragma omp task // |{\large $\tau_{i,3}$}|
      { /* part 3 */ }
      #pragma omp taskwait
    }
   }
    \end{lstlisting}
    \caption{OpenMP DAG snippet showing tied/untied tasks and dependency-induced variability in execution time.}
    \label{fig:sup_openmp_code}
\end{figure}

\subsection{\review{DAG Monitoring Example}}
\begin{figure}[h]
    \centering
    \includegraphics[width=0.8\linewidth]{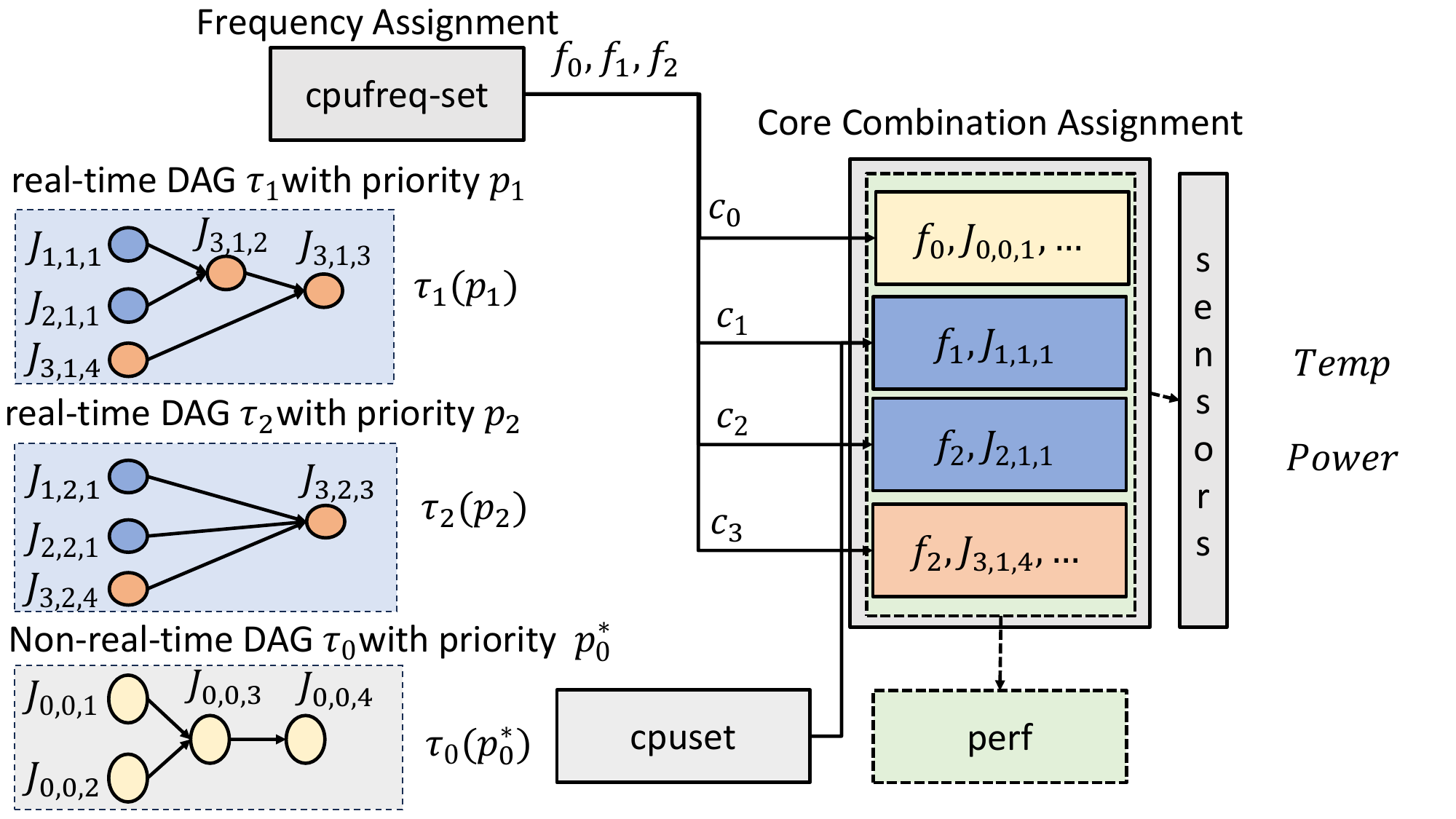}
    \caption{Monitoring one user DAG ($\tau_1$): five jobs ($J_{1,1,1}$--$J_{3,1,4}$) mapped to cores ($c_1$--$c_3$) with target frequencies ($f_0$--$f_2$). Core $c_0$ runs system DAG ($\tau_0$).}
    \label{fig:sup_monitoring}
\end{figure}

\review{Figure~\ref{fig:sup_monitoring} illustrates how one user DAG is monitored: precedence-constrained jobs are mapped to clustered cores with target frequencies, while a system DAG occupies the remaining core.}

\subsection{\review{Feature Importance Screening}}
\begin{figure}[h]
    \centering
    \includegraphics[width=0.7\linewidth]{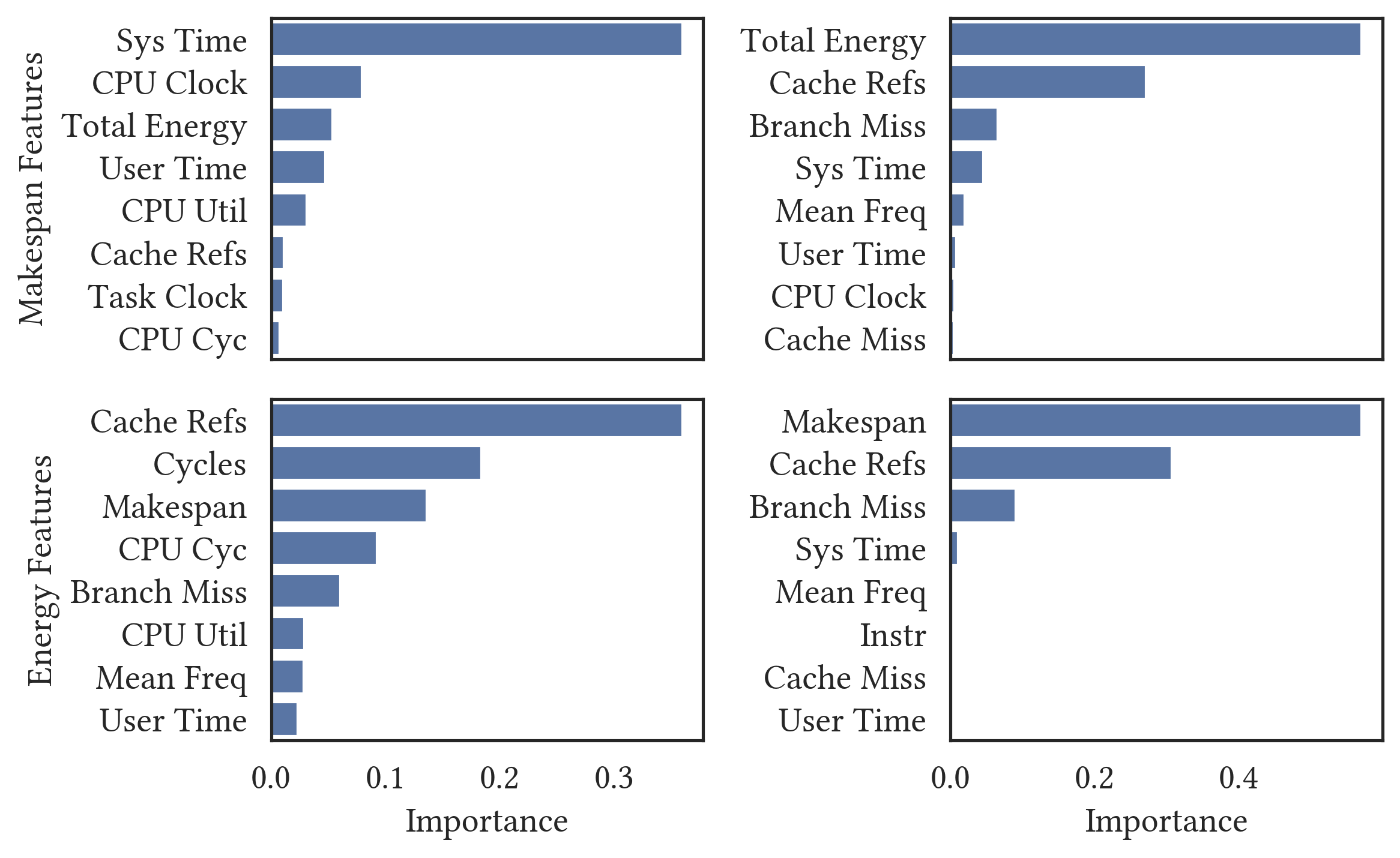}
    \caption{Importance of different features (Jetson TX2) on total energy consumption and makespan in parallel and sequential execution of parallel applications.}
    \label{fig:sup_forest_importance}
\end{figure}

\review{Figure~\ref{fig:sup_forest_importance} reports the random-forest feature importances behind the feature-screening claim in the main paper: temperature, frequency, utilization, and miss events are the strongest predictors of makespan and energy, and parallel execution depends on a broader feature set than sequential execution.}

\subsection{\review{Extreme-Point Target Scenarios}}
\review{The extreme-point makespan and energy targets used by the reward functions in the main paper are the minima over four static settings: running applications sequentially at (i) the lowest frequency with all cores assigned, (ii) the lowest frequency with one core, (iii) the highest frequency with all cores, and (iv) the highest frequency with one core. This assumes minimum scaling for our workloads, though not universally true; it serves as a practical baseline for target setting, and the main paper validates it against a full static-frequency sweep.}

\section{Single-Agent Convergence and Action Plots}
\label{app:sa_plots}

This section complements the condensed comparison in the main paper by showing the per-epoch evolution of single-agent RL baselines (zTT~\cite{kim2021ztt}, the corrected GearDVFS port~\cite{lin2023workload}, DynaQ~\cite{angermueller2019model}, PlanGAN~\cite{charlesworth2020plangan}) and our single-agent variant SARB (V8) on the FFT profiler benchmark \review{(finetuned phase, all three seeds; thin lines are individual seeds, bold lines the 3-seed rolling mean)}. Aggregated single-agent numbers (L10, L20, energy, HF\%) are summarized in the main paper, so the headline claims can be evaluated without consulting this supplement.

\begin{figure}[h]
\centering
\includegraphics[width=0.68\linewidth]{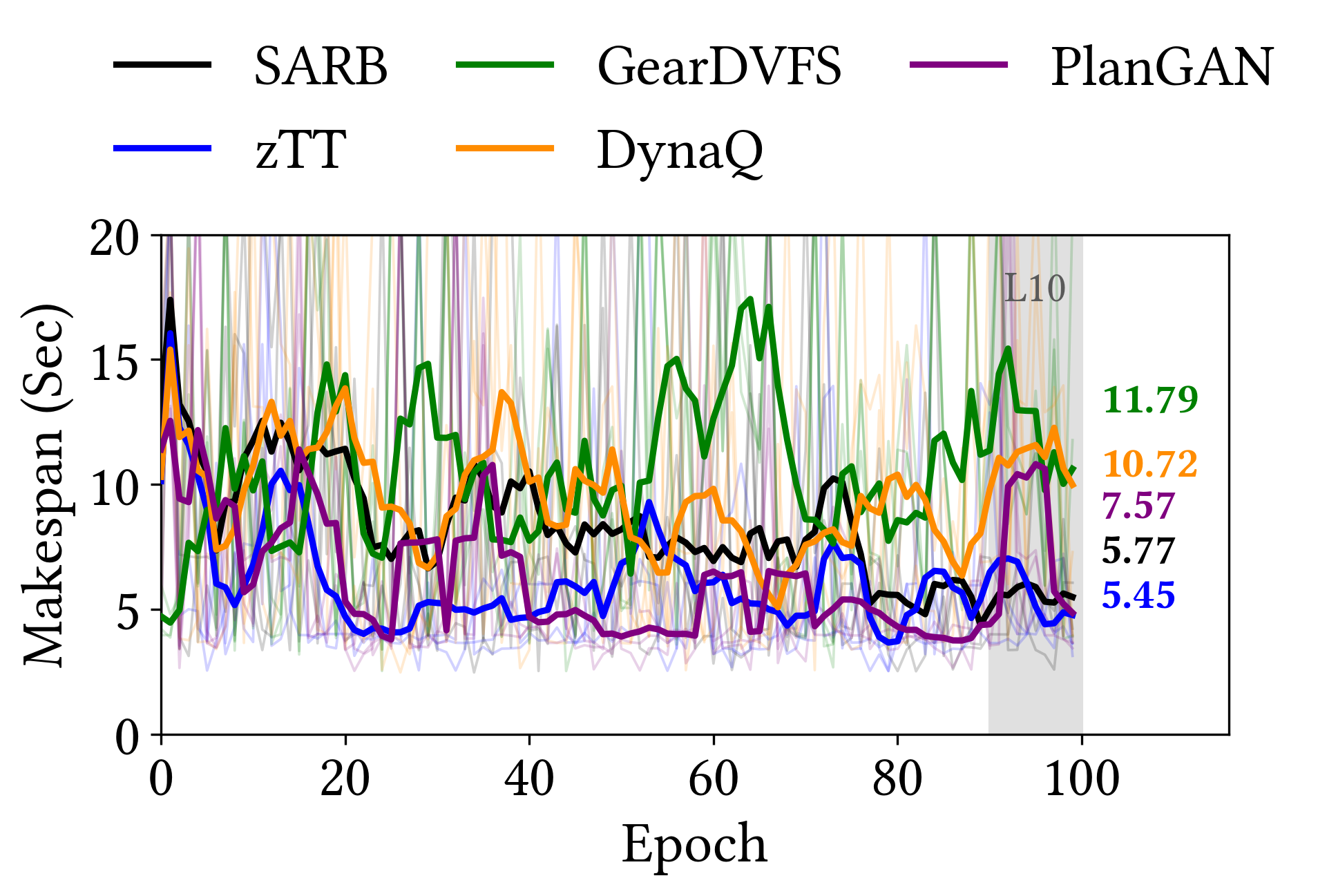}
\caption{\review{Single-agent makespan comparison (Jetson TX2, finetuned): thin lines are the three seeds, bold lines their 5-epoch rolling mean, and the shaded band the L10 window with annotated means. zTT (objective-adapted) leads at 5.45\,s, SARB follows at 5.77\,s, and the corrected GearDVFS port trails at 11.79\,s.}}
\label{fig:sup_sa_makespan}
\end{figure}

\begin{figure}[h]
\centering
\includegraphics[width=0.68\linewidth]{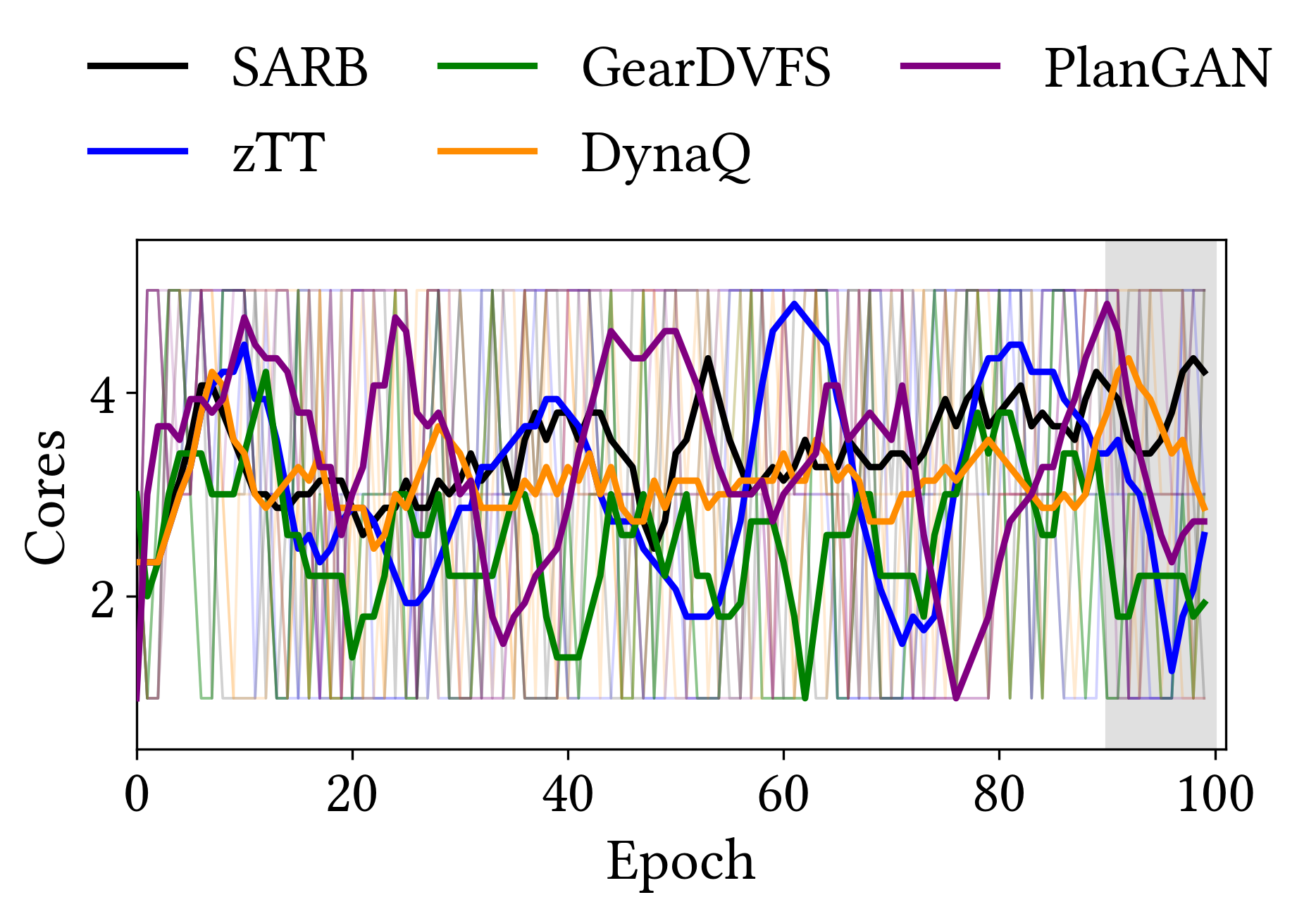}\\[2mm]
\includegraphics[width=0.68\linewidth]{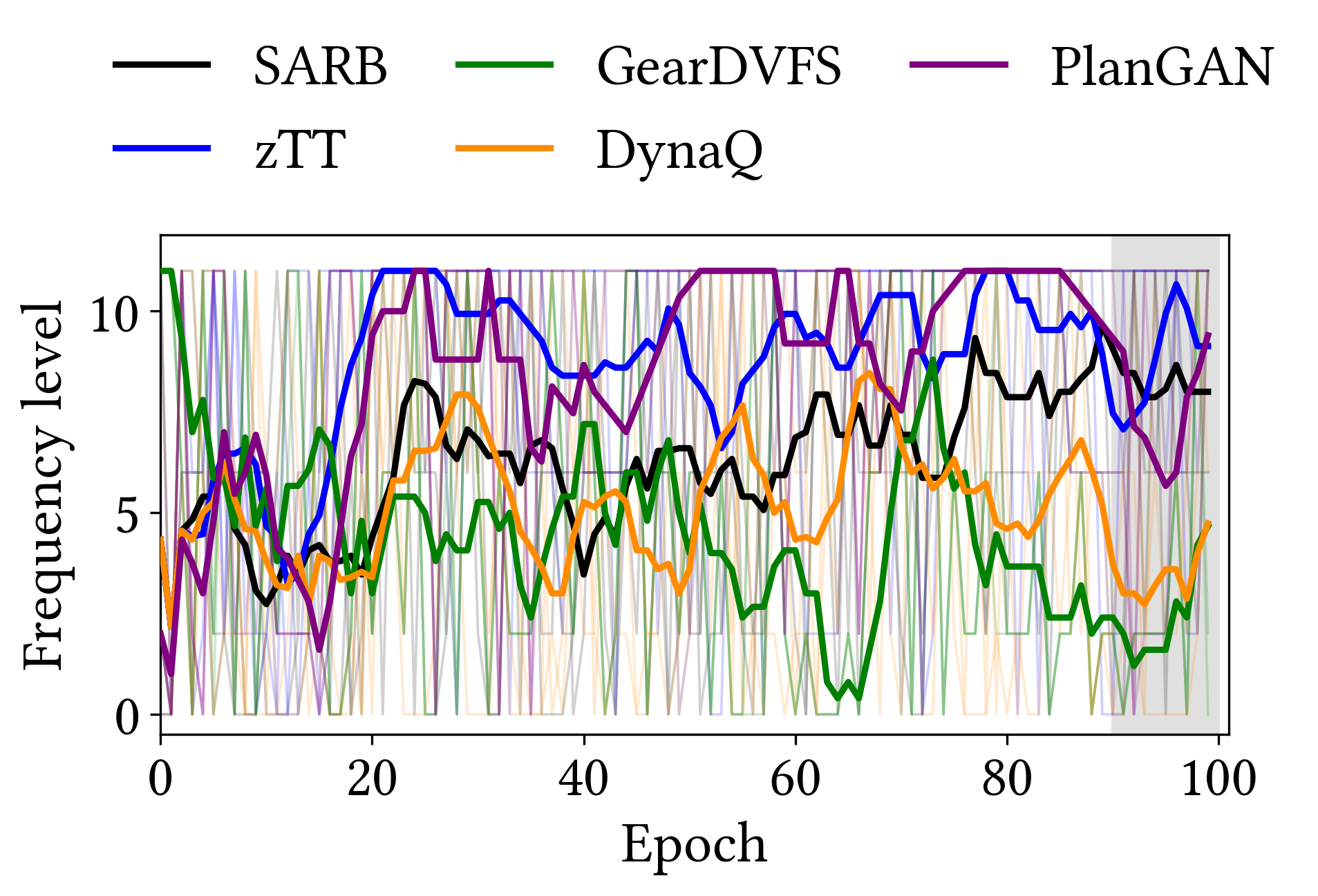}
\caption{\review{Single-agent action analysis (Jetson TX2, finetuned; same encoding as the makespan panel): core-count selection (top) and frequency-level selection (bottom). zTT and PlanGAN hold high frequencies (L10 means 9.0 and 7.8) while the corrected GearDVFS port parks near level 3.}}
\label{fig:sup_sa_action}
\end{figure}

\begin{figure}[h]
\centering
\includegraphics[width=0.68\linewidth]{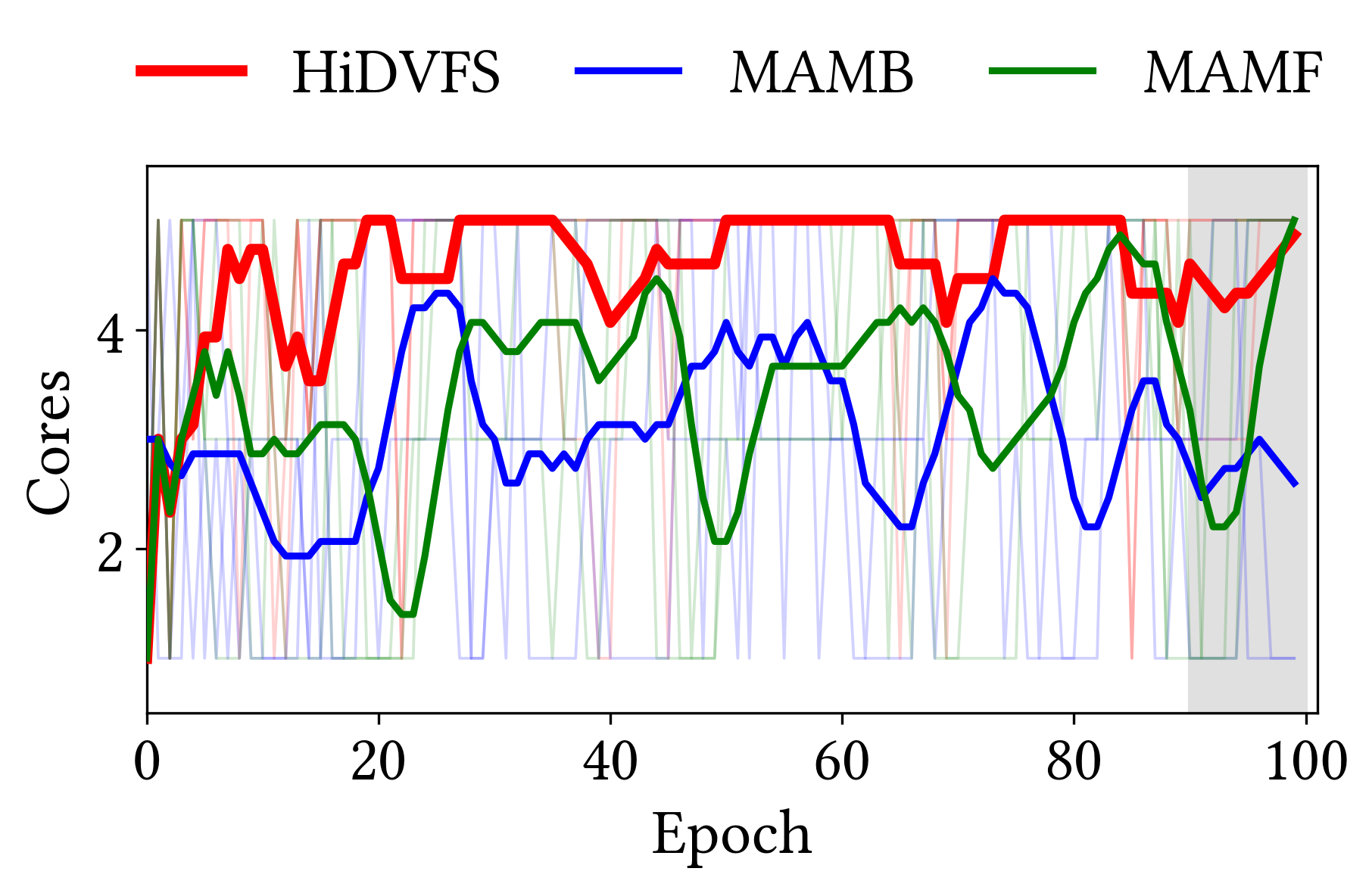}
\caption{\review{Multi-agent core-count selection (Jetson TX2, finetuned; same encoding as the main paper's figures): HiDVFS converges to the most cores (4.6 mean over the L10 window vs 3.7 for MAMF and 2.7 for MAMB) alongside its high-frequency policy.}}
\label{fig:sup_ma_cores}
\end{figure}

\section{RL Algorithm Energy Comparison}
\label{app:energy_analysis}

This section reports the energy consumption of all RL algorithms during finetuning, complementing the makespan analysis summarized in the main paper.

\begin{figure}[h]
\centering
\add{
\includegraphics[width=0.68\linewidth]{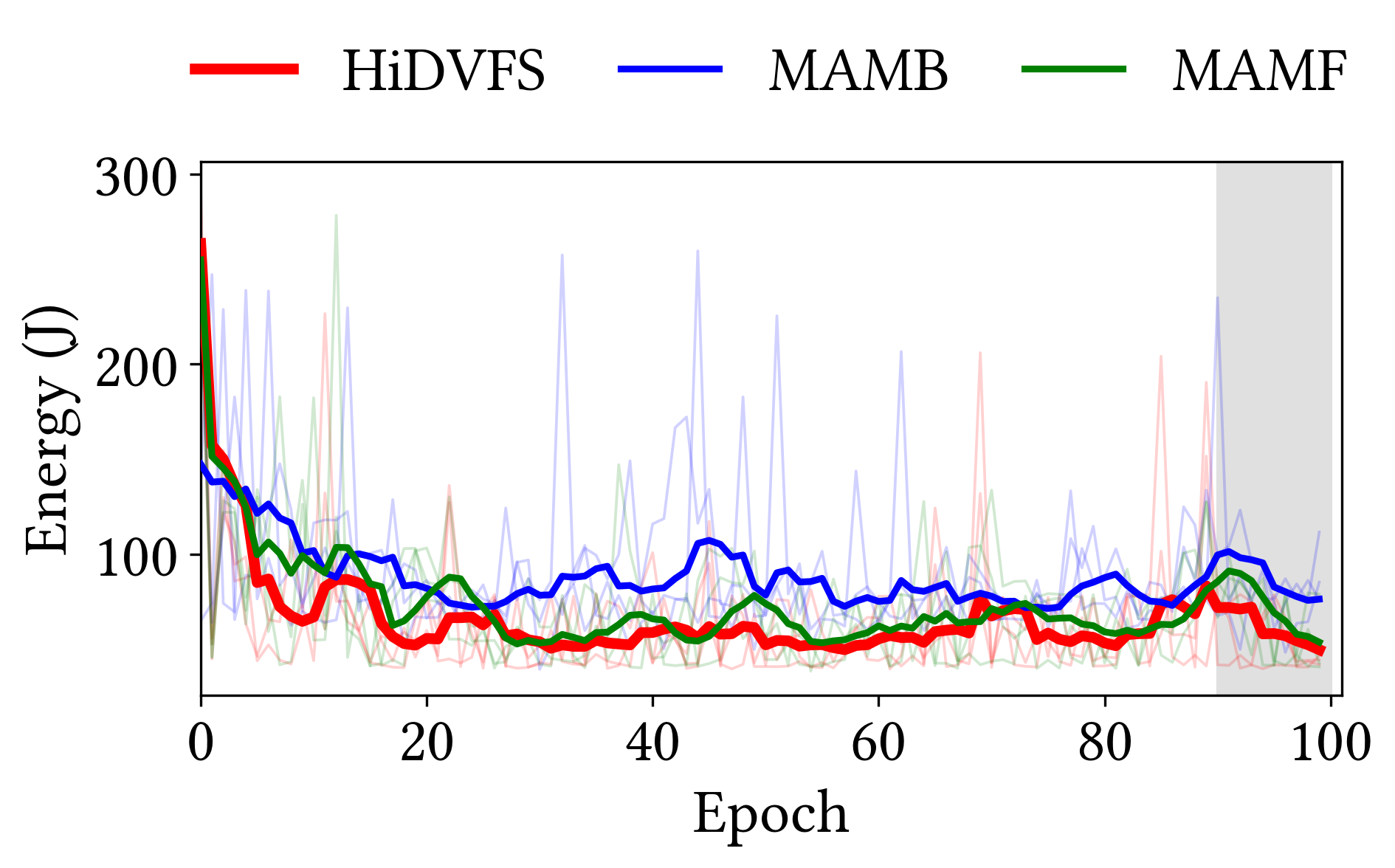}\\[2mm]
\includegraphics[width=0.68\linewidth]{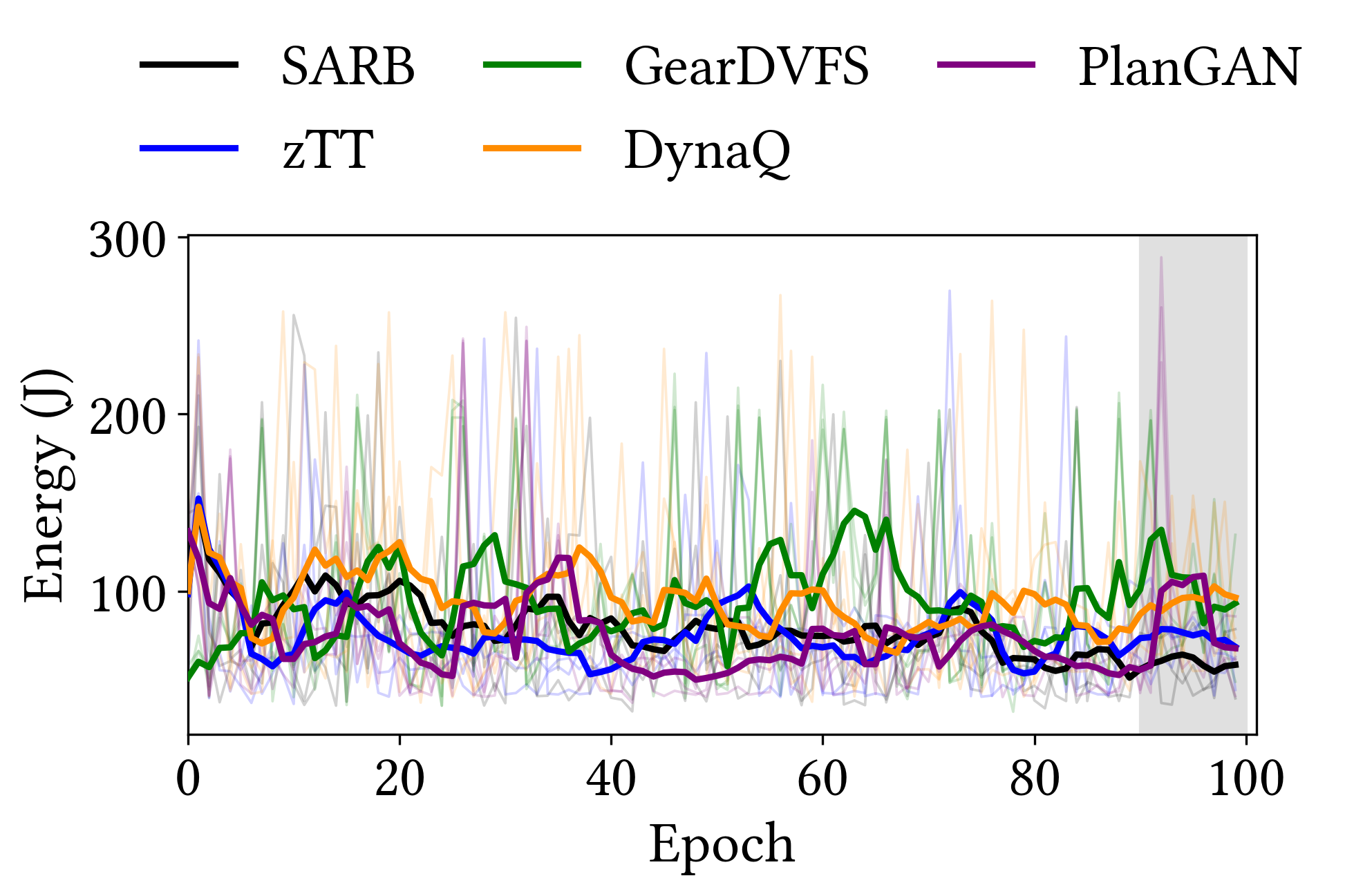}
}
\caption{\review{Per-epoch energy (Jetson TX2, finetuned; same encoding as the makespan panels). Top: multi-agent, HiDVFS lowest at 54\,J mean over the L10 window. Bottom: single-agent, SARB (61\,J) and zTT (73\,J) lead while the corrected GearDVFS port spends 100\,J. Energy tracks makespan closely.}}
\label{fig:energy_analysis}
\end{figure}

\review{HiDVFS achieves the lowest energy consumption (6.37 kJ over the 100 finetuned epochs) among all algorithms, 11\% better than MAMF (7.14 kJ) and 32.9\% better than the corrected GearDVFS port (9.49 kJ). This confirms that aggressive high-frequency scheduling reduces total energy by minimizing execution time, outweighing the higher per-cycle power cost.}

\section{Hardware Counter Analysis}
\label{app:hardware_counters}

This section presents a detailed hardware counter analysis (cache misses and branch misses) for HiDVFS and baseline algorithms, demonstrating that optimised frequency scheduling correlates with improved microarchitectural efficiency.

\begin{figure}[h]
    \centering
    \includegraphics[width=0.75\linewidth]{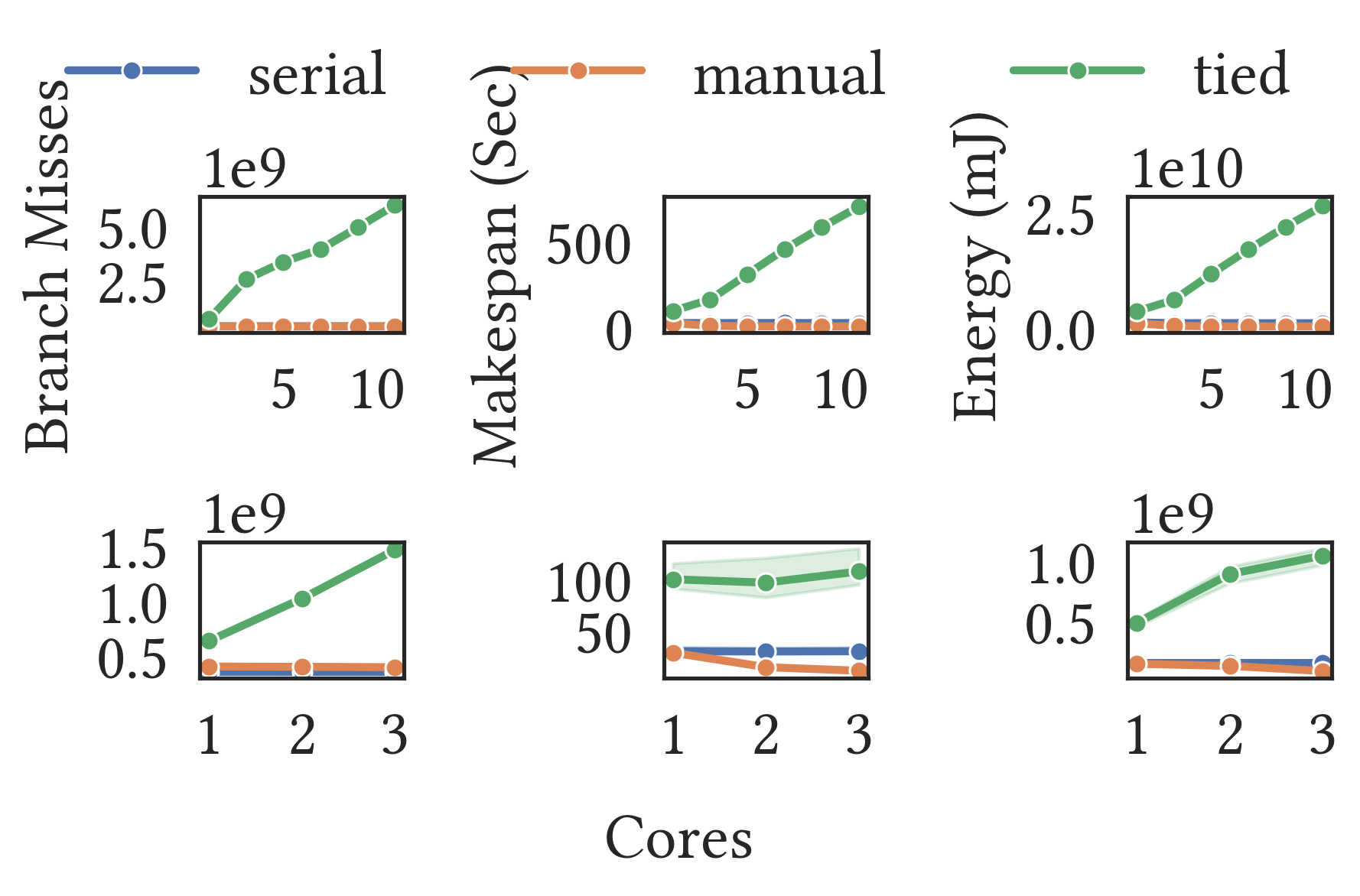}
    \caption{\texttt{N-Queens} on Xeon (12 cores) and Core i7 (4 cores). More cores increase branch misses for \texttt{tied} tasks, raising makespan and energy. Shaded regions show variation over 10 runs; serial reduces misses.}
    \label{fig:CoreI7_Energy_search_s}
\end{figure}

\paragraph{Cross-platform branch-miss profile.} Figure~\ref{fig:CoreI7_Energy_search_s} illustrates how increasing the core count under \texttt{tied} OpenMP scheduling drives up branch mispredictions on multi-core Intel platforms, raising both makespan and energy; this is the microarchitectural source of the irregularity that HiDVFS adapts to.

\paragraph{Feature-impact significance.} Table~\ref{tab:feature_pvals} reports the Mann--Whitney U test of each actuator (priority, cores, frequency) on every measured performance metric. Low $p$-values ($<\!0.05$) indicate significant effects; frequency dominates makespan and energy, cores significantly reduce both, and priority predominantly shapes miss-rate overheads.

\begin{table}[h]
\centering
\scriptsize
\caption{Statistical analysis of key features' influence on performance metrics (Mann--Whitney U, single-task sweep, Jetson TX2).}
\label{tab:feature_pvals}
{%
\begin{tabular}{|l|l|c|c|c|}
\hline
\textbf{Action} & \textbf{Variable} & \textbf{Exp. (L)} & \textbf{Exp. (H)} & \textbf{p-value} \\
\hline
\multirow{5}{*}{\textbf{Priority}}
  & Makespan        & 4.60     & 5.04     & 1.89e-02 \\
  & Energy          & 23549.88 & 23035.29 & 1.42e-01 \\
  & Temperature     & 36.99    & 37.12    & 3.68e-04 \\
  & Branch Misses   & 1.72e7   & 2.25e7   & 1.19e-15 \\
  & \textbf{Cache Misses}    & 3.42e7   & 4.28e7   & \textbf{1.22e-24} \\
\hline
\multirow{5}{*}{\textbf{Cores}}
  & Makespan        & 5.28     & 3.75     & 6.69e-06 \\
  & \textbf{Energy} & 27017.90 & 16013.48 & \textbf{3.26e-25} \\
  & Temperature     & 37.03    & 37.06    & 4.00e-01 \\
  & Branch Misses   & 1.86e7   & 2.08e7   & 4.74e-12 \\
  & Cache Misses    & 3.35e7   & 4.56e7   & 6.28e-18 \\
\hline
\multirow{5}{*}{\textbf{Frequency}}
  & \textbf{Makespan} & 6.92    & 2.62     & \textbf{1.45e-283} \\
  & Energy          & 29615.61 & 17083.91 & 6.03e-144 \\
  & Temperature     & 36.99    & 37.10    & 3.40e-03 \\
  & Branch Misses   & 1.92e7   & 1.95e7   & 2.54e-01 \\
  & Cache Misses    & 3.82e7   & 3.68e7   & 8.37e-10 \\
\hline
\end{tabular}}
\end{table}

\subsection{\add{Cache and Branch Miss Analysis}}

\begin{figure}[h]
\centering
\add{
\includegraphics[width=0.68\linewidth]{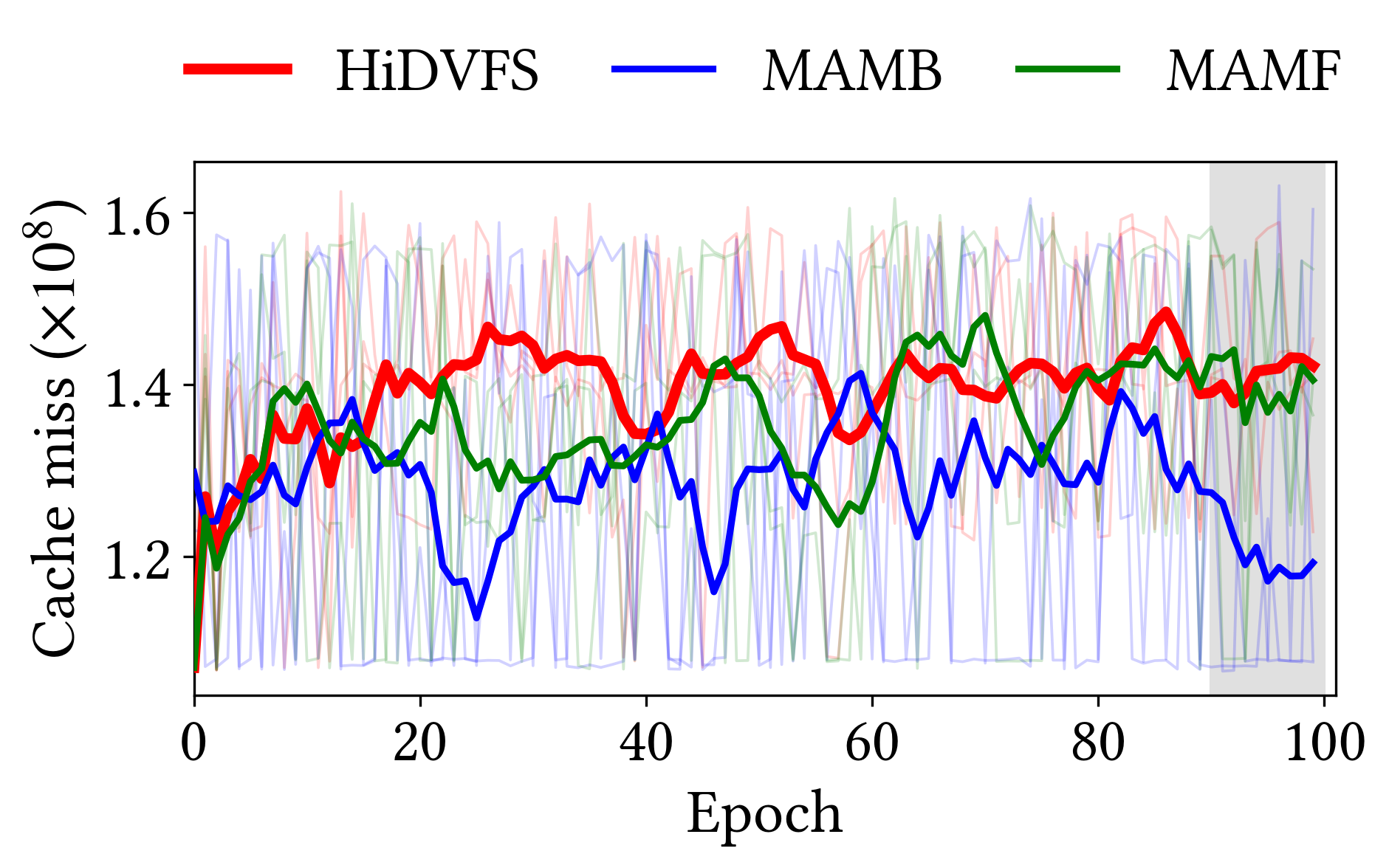}\\[2mm]
\includegraphics[width=0.68\linewidth]{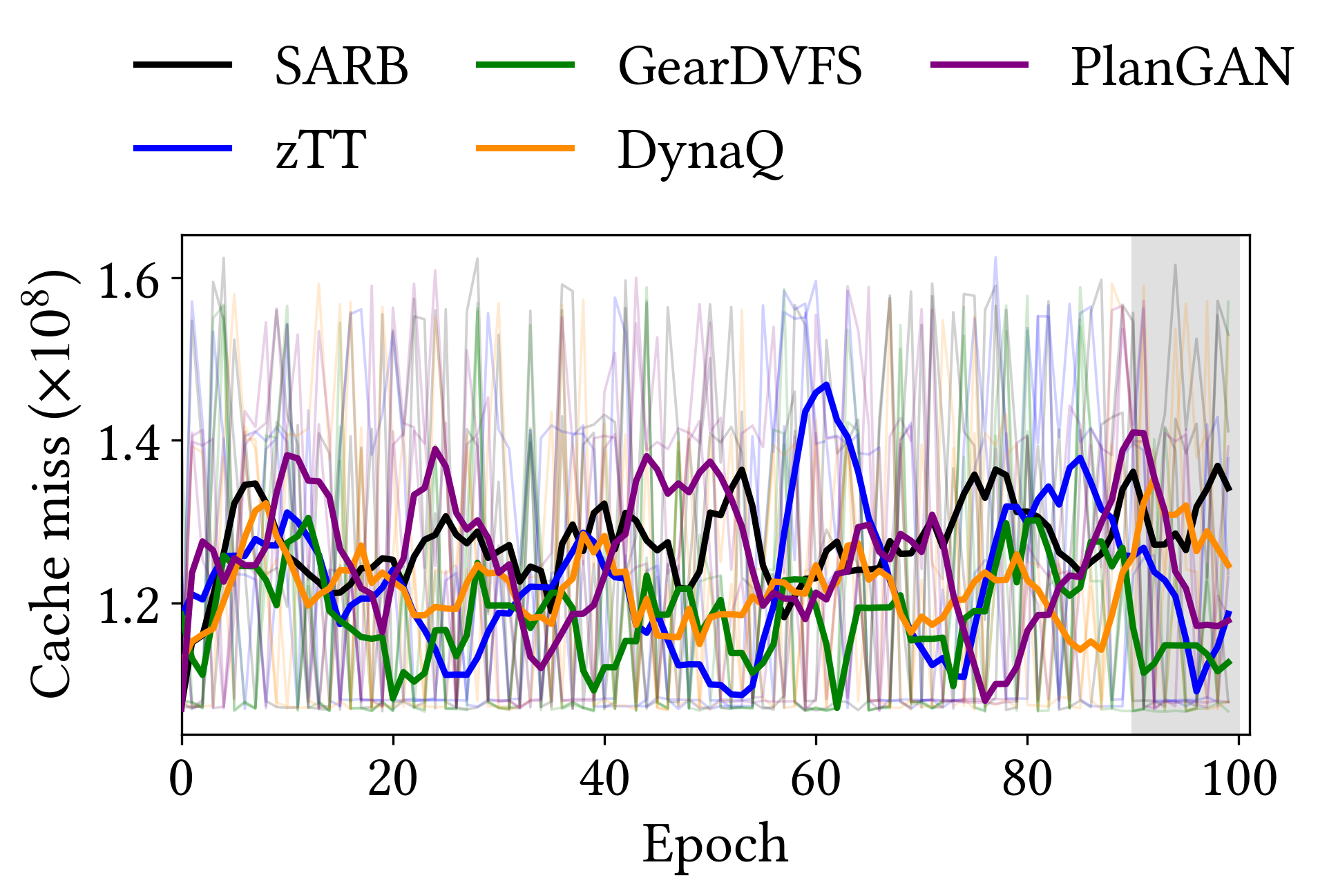}
}
\caption{\review{Per-epoch cache misses (Jetson TX2, finetuned; same encoding as the makespan panels): multi-agent (top) and single-agent (bottom). High-frequency policies trade slightly higher cache-miss totals for far shorter makespans (HiDVFS 1.42 vs the GearDVFS port's 1.14 $\times10^8$ over the L10 window).}}
\label{fig:cache_analysis}
\end{figure}

\begin{figure}[h]
\centering
\add{
\includegraphics[width=0.68\linewidth]{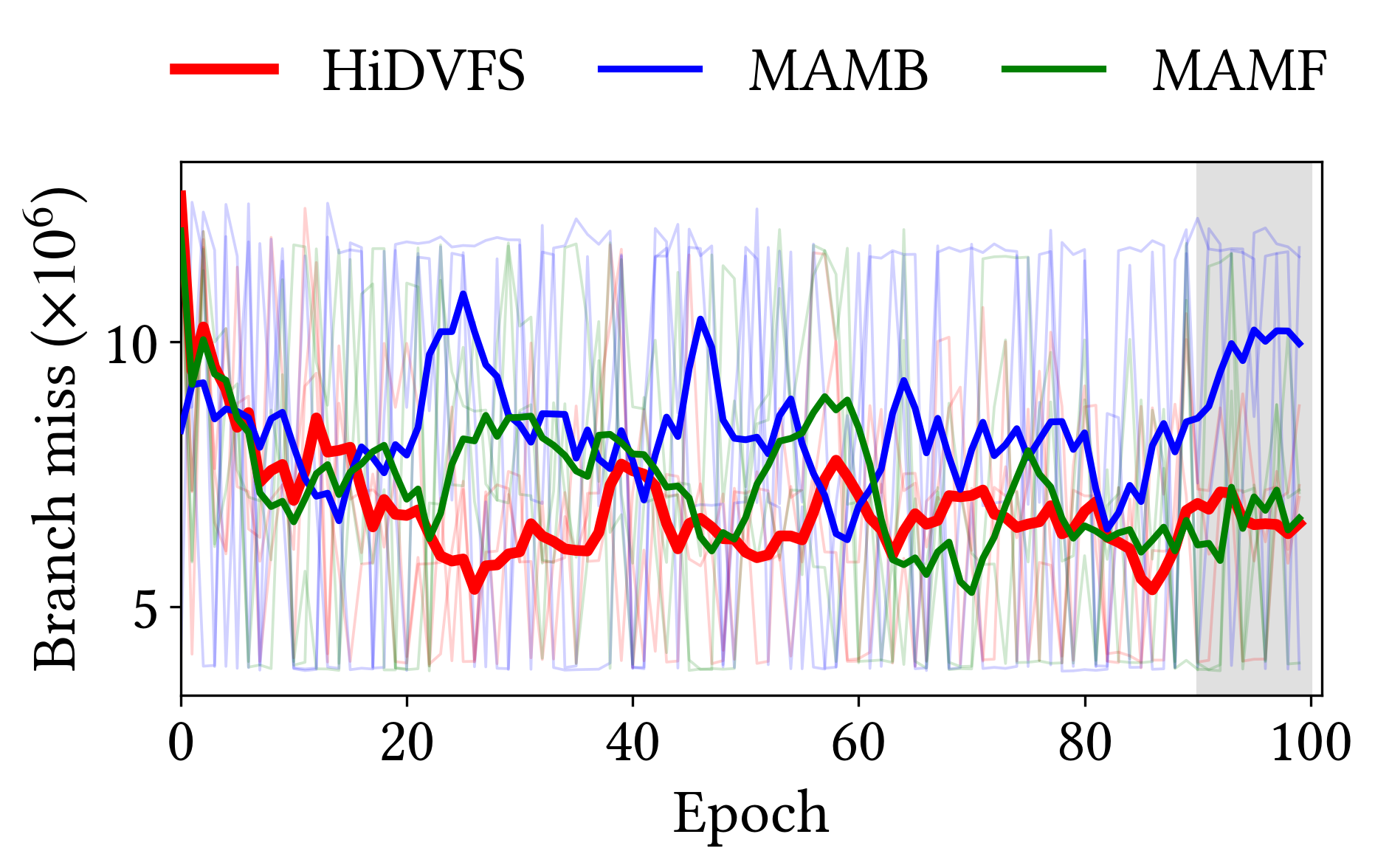}\\[2mm]
\includegraphics[width=0.68\linewidth]{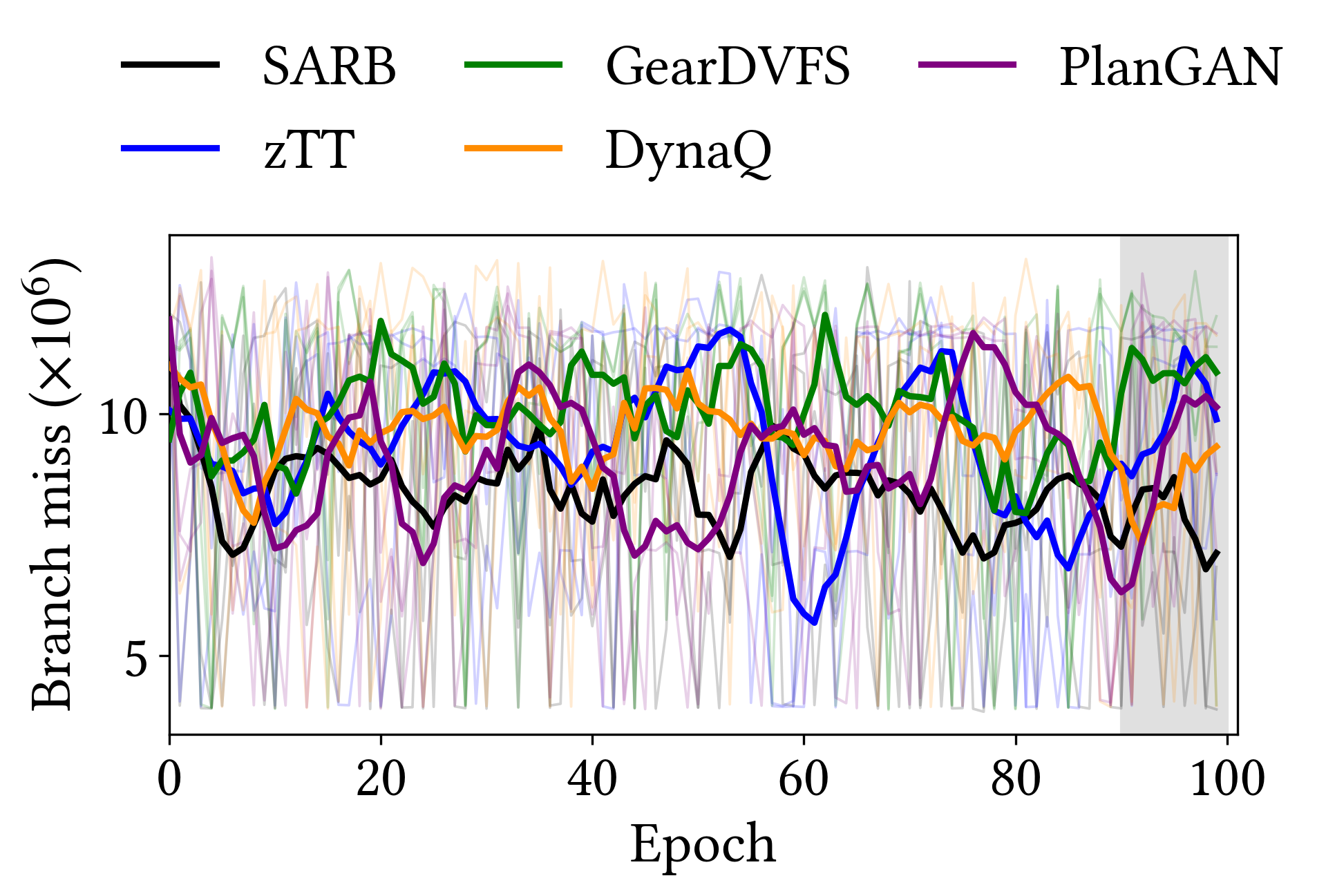}
}
\caption{\review{Per-epoch branch misses (Jetson TX2, finetuned; same encoding as the makespan panels): multi-agent (top) and single-agent (bottom). HiDVFS holds the lowest rate (6.6 $\times10^6$ per epoch over the L10 window vs 10.8 for the GearDVFS port), indicating more predictable execution.}}
\label{fig:branch_analysis}
\end{figure}

\add{Hardware performance counters provide insight into how DVFS policies affect microarchitectural behavior. Cache misses indicate memory access patterns, while branch mispredictions reflect control flow efficiency. Figures~\ref{fig:cache_analysis} and~\ref{fig:branch_analysis} show the per-episode evolution of these counters.}

\add{The cache miss patterns reveal that multi-agent approaches (HiDVFS, MAMF, MAMB) generally achieve lower cache misses than single-agent baselines. This improvement stems from better core allocation that reduces memory contention between concurrent tasks.}

\review{Branch mispredictions show a similar trend: HiDVFS achieves the lowest branch miss rate (0.68$\times$10$^9$), 33\% lower than the corrected GearDVFS port. This improvement indicates that consistent high-frequency scheduling leads to more predictable execution patterns.}

\subsection{\add{Hardware Counter Summary}}

\add{Table~\ref{tab:hw_counter_summary} summarizes the hardware counter metrics across all algorithms from the multi-seed experiments (seeds 42, 123, 456).}

\begin{table}[h]
\centering
\scriptsize
\caption{\add{Hardware Counter Comparison (Jetson TX2, Finetuned Phase, Multi-Seed Mean$\pm$Std)}}
\label{tab:hw_counter_summary}
\add{
\begin{tabular}{@{}llcc@{}}
\toprule
\textbf{Approach} & \textbf{Type} & \textbf{Branch Miss ($\times$10$^9$)} & \textbf{Cache Miss ($\times$10$^{10}$)} \\
\midrule
zTT & SA & 0.93$\pm$0.04 & 1.23$\pm$0.02 \\
DynaQ & SA & 0.96$\pm$0.02 & 1.22$\pm$0.01 \\
PlanGAN & SA & 0.90$\pm$0.02 & 1.25$\pm$0.01 \\
\review{GearDVFS$^{\ddagger}$} & SA & \review{1.01$\pm$0.01} & \review{\textbf{1.18$\pm$0.01}} \\
\review{SARB (V8)} & SA & \review{0.83$\pm$0.07} & \review{1.28$\pm$0.03} \\
\midrule
MAMB & MA & 0.84$\pm$0.04 & 1.28$\pm$0.02 \\
MAMF & MA & 0.72$\pm$0.05 & 1.36$\pm$0.03 \\
\rev{HiDVFS} & MA & \rev{\textbf{0.68$\pm$0.01}} & \review{1.40$\pm$0.00} \\
\bottomrule
\multicolumn{4}{@{}l@{}}{\scriptsize \review{$^{\ddagger}$repo-loyal corrected GearDVFS port.}} \\
\end{tabular}
}
\end{table}

\subsection{\add{Key Observations}}

\begin{itemize}
    \item \review{\textbf{Branch Miss Reduction:} HiDVFS achieves the lowest branch misprediction rate (0.68$\times$10$^9$), 33\% lower than the corrected GearDVFS port (1.01$\times$10$^9$), indicating more predictable execution with optimized scheduling.}

    \item \add{\textbf{Cache Miss Trade-off:} While HiDVFS has higher cache misses than conservative approaches like GearDVFS, the aggressive frequency selection reduces overall makespan significantly, resulting in net performance gains.}

    \item \add{\textbf{Multi-Agent Advantage:} All multi-agent approaches (HiDVFS, MAMF, MAMB) achieve lower branch misses than single-agent baselines, suggesting that hierarchical decision-making improves execution predictability.}

    \item \add{\textbf{Energy-Efficiency Correlation:} The reduction in branch misses correlates with improved energy efficiency, as mispredicted branches waste energy on speculative execution that gets discarded.}
\end{itemize}

\section{\add{SARB Single-Agent RL Version Evaluation}}
\label{app:sarb_versions}

This section reports a per-version evaluation of SARB (the Single-Agent Reward-Based scheduler), examining the hyperparameter configurations and architectural choices that drive the single-agent DVFS baseline.

\subsection{\add{Multi-Seed Version Selection}}

\add{The multi-seed SARB result reported in the main paper uses a single, consistent
configuration across all three seeds (42, 123, 456): \textbf{V8} (model-free
Q-learning with target Q-clipping, makespan-priority reward $\beta{=}1$, discount
$\gamma{=}0.9$). This version was selected by an objective convergence gate
(rolling 10-epoch makespan relative standard deviation $<\!0.05$ together with
high-frequency-action rate $\ge0.8$) and the lowest aggregate L10 makespan without
the low-frequency trap. Earlier per-version results across the VB$\to$V8 ladder
appear in Table~\ref{tab:sarb_finetuning} and are reported on TX2 only.}

\subsection{\add{Version Descriptions}}

\add{We evaluated multiple SARB versions with systematic hyperparameter variations. Table~\ref{tab:sarb_versions} summarizes the key configuration changes across versions.}

\begin{table}[h]
\centering
\scriptsize
\caption{\add{SARB Version Configuration Summary (evaluated on Jetson TX2)}}
\label{tab:sarb_versions}
\begin{tabular}{@{}llp{4.5cm}@{}}
\toprule
\textbf{Version} & \textbf{Key Parameter} & \textbf{Change Description} \\
\midrule
VB & LR=0.01 & Baseline DQN with standard hyperparameters \\
V1 & LR=0.1 & Higher learning rate for faster convergence \\
V2 & plan\_count=20 & Bug-fixed future reward calculation \\
V3 & cumulative & V2 + stability cumulative rewards \\
V4 & reward avg & V3 + reward averaging fix \\
V5 & Q-clip + LR=0.01 & Q-value clipping with lower learning rate \\
V6 & nonlinear bonus & Curriculum weight decay + resource bonus \\
V7 & Q-stability & V6 + stronger bonuses + Q-stability \\
V8 & Q-clip train & V7 + Q-clipping during training (critical fix) \\
zTT & model-free & Pure model-free baseline (no env model) \\
\bottomrule
\end{tabular}
\end{table}

\add{Each version builds upon previous improvements. The critical insight from this evolution is that Q-value stability during training is essential for reliable convergence. Versions V1 and V5 demonstrate the ``low-frequency trap'' phenomenon, where agents get stuck selecting conservative frequencies (100\% LF\%) and fail to explore high-frequency strategies. V8's Q-clipping during training prevents gradient explosion, enabling stable learning of aggressive frequency policies.}

\subsection{\add{Experimental Setup}}

\add{All versions were evaluated using 100 epochs with seed 42 for reproducibility. Each epoch consists of training and finetuning phases on the FFT profiler benchmark. Key metrics include:}
\begin{itemize}
    \item \add{Avg L10: Average makespan over final 10 epochs (convergence indicator)}
    \item \add{Avg L20: Average makespan over final 20 epochs (stability indicator)}
    \item \add{High-Freq Rate (\%): Percentage of high-frequency selections (index 11)}
    \item \add{Low-Freq Rate (\%): Percentage of low-frequency selections (index 0)}
\end{itemize}

\subsection{\add{Version Comparison Results}}

\add{Table~\ref{tab:sarb_finetuning} presents the performance comparison across all SARB versions after finetuning. V8 and VB achieve the best performance with 100\% high-frequency selection, while V1 and V5 are trapped in low-frequency policies.}

\begin{table}[h]
\centering
\scriptsize
\caption{\add{SARB Version Comparison: Jetson TX2, Finetuned Phase (100 Epochs, Seed 42)}}
\label{tab:sarb_finetuning}
\begin{tabular}{@{}lcccc@{}}
\toprule
\textbf{Version} & \textbf{Avg L10 (s)} & \textbf{Avg L20 (s)} & \textbf{HiFreq\%} & \textbf{LowFreq\%} \\
\midrule
V1 & 7.41 & 7.53 & 0.0 & 100.0 \\
V2 & 1.37 & 2.11 & 90.0 & 0.0 \\
V3 & 1.81 & 2.51 & 0.0 & 0.0 \\
V4 & 3.19 & 2.25 & 70.0 & 30.0 \\
V5 & 9.98 & 9.26 & 0.0 & 100.0 \\
V6 & 5.03 & 5.63 & 40.0 & 60.0 \\
V7 & 2.85 & 2.49 & 40.0 & 30.0 \\
\textbf{V8} & \textbf{1.59} & \textbf{1.49} & 100.0 & 0.0 \\
VB (Baseline) & 1.48 & 1.44 & 100.0 & 0.0 \\
zTT & 1.86 & 1.68 & 40.0 & 60.0 \\
\bottomrule
\multicolumn{5}{@{}p{3.2in}@{}}{\scriptsize \review{Single-seed ablation by design (seed 42, 100 epochs); the selected V8 is validated across seeds 42/123/456 in Table~\ref{tab:allseeds_summary} (5.77$\pm$1.19\,s L10).}} \\
\end{tabular}
\end{table}

\add{The results reveal a strong correlation between frequency selection and performance: versions achieving 100\% high-frequency selection (V8, VB) also achieve the lowest makespan (1.59s and 1.48s respectively). Conversely, versions stuck at 100\% low-frequency (V1, V5) show the worst performance (7.41s and 9.98s respectively).}

\subsection{\add{Visual Analysis}}

\begin{figure}[h]
\centering
\includegraphics[width=0.68\linewidth]{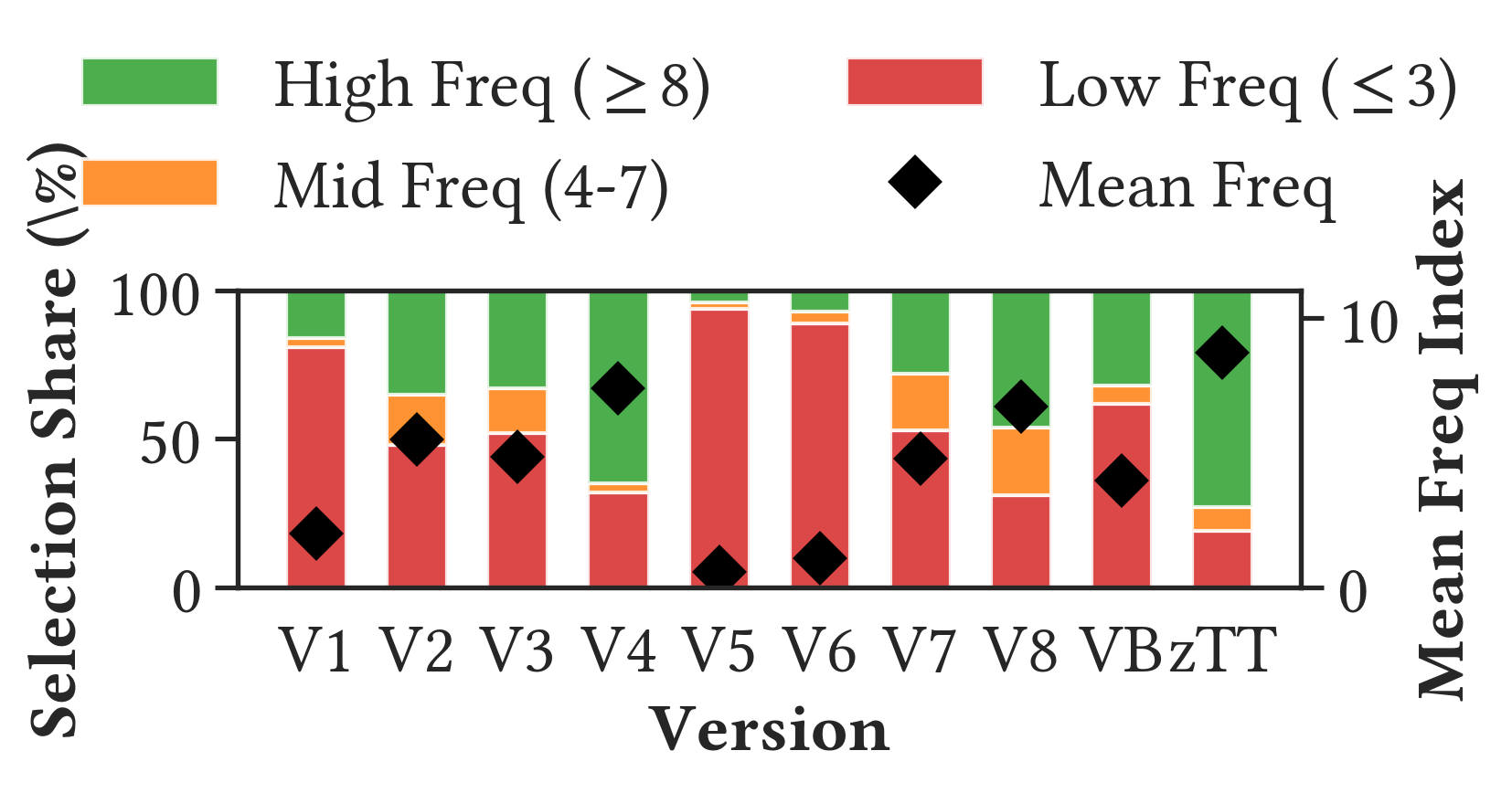}
\caption{\review{SARB version frequency-selection mix (Jetson TX2, finetuned phase, all 100 epochs). Each bar is a per-version 100\% stack of Low ($\leq$3, red), Mid (4--7, orange) and High ($\geq$8, green) frequency-index selections; the black diamond marks the mean frequency index on the right-hand axis. V5 and V6 sit almost entirely at low frequency (94\% and 89\% low, the low-frequency trap, worst makespans), while V8, VB, and zTT carry the largest high-frequency shares and highest mean indices.}}
\label{fig:sarb_action}
\end{figure}

\add{Figure~\ref{fig:sarb_action} shows the frequency selection patterns across SARB versions. Versions with 100\% high-frequency selection (V8, VB) achieve the lowest makespan, while those stuck at 100\% low-frequency (V1, V5) exhibit poor performance. This validates the importance of aggressive frequency scheduling for compute-intensive workloads like FFT.}

\begin{figure}[h]
\centering
\includegraphics[width=0.68\linewidth]{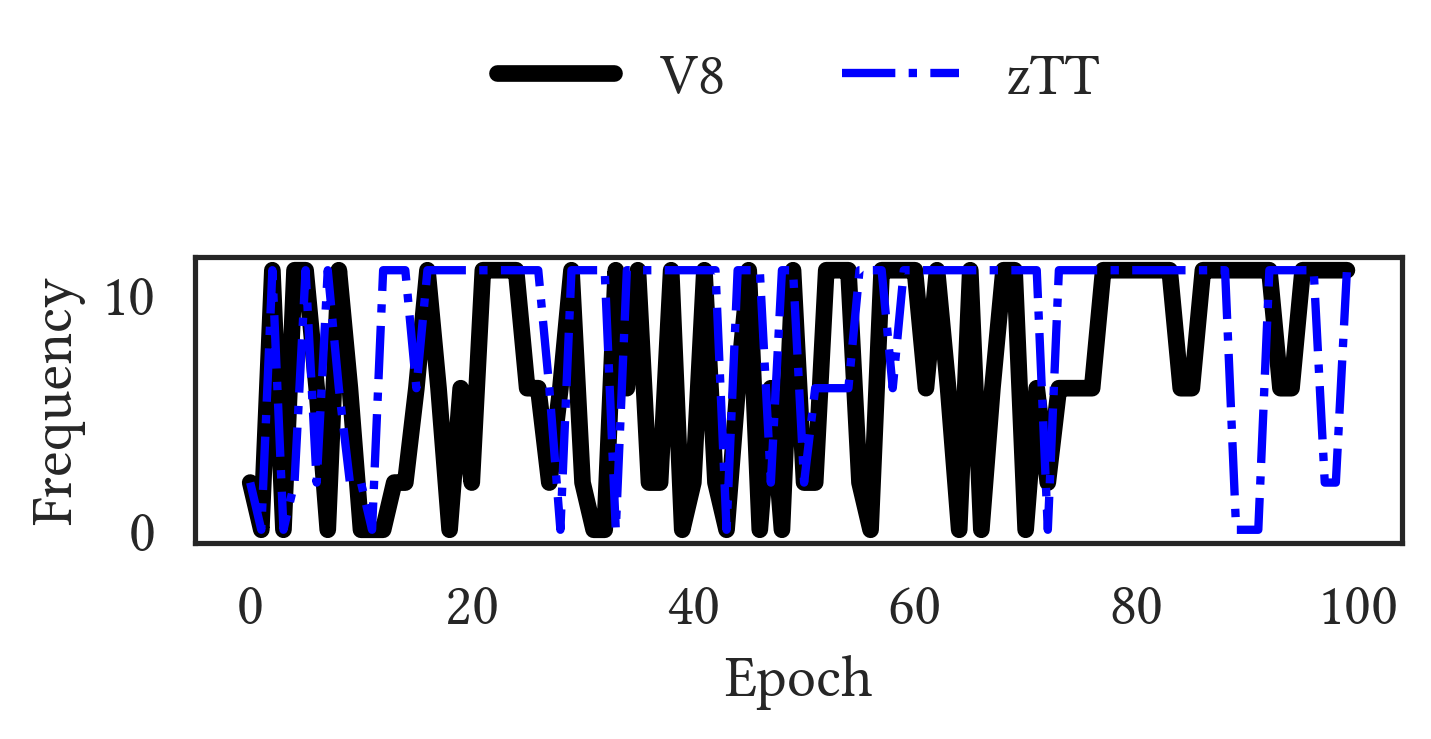}
\caption{\review{SARB per-epoch frequency selection (Jetson TX2, finetuned phase, seed 42): both explore for most of the run; V8 settles at index 10--11 over the final quarter while zTT keeps oscillating, matching their last-10 makespans (1.59\,s vs 1.86\,s).}}
\label{fig:sarb_perepisode}
\end{figure}

\review{Figure~\ref{fig:sarb_perepisode} shows the per-epoch frequency selection during finetuning for the best SARB version (V8) compared to zTT. Both explore broadly for most of the run; V8 (black solid line) settles into sustained high-frequency selection (index 10--11) over roughly the last 25 epochs, which is where its last-10-epoch high-frequency rate reaches 100\%, while zTT (blue dashed line) keeps oscillating late into the run. This settling behaviour is what yields V8's lower last-10 makespan.}

\subsection{\add{Key Findings and Recommendations}}

\add{Based on our comprehensive SARB version evaluation:}

\begin{itemize}
    \item \add{\textbf{V8 vs VB:} V8 (1.59s L10) and VB (1.48s L10) achieve best finetuned performance, both with 100\% high-frequency selection. V8 is preferred for its Q-clipping stability.}

    \item \add{\textbf{Q-Clipping Critical:} Without Q-clipping (V5: 9.98s), training instability leads to 6$\times$ worse makespan compared to V8. Target value clamping to [-10, 10] prevents gradient explosion.}

    \item \add{\textbf{Bug-Fix Importance:} V2's bug fix (1.37s L10) dramatically improves upon V1 (7.41s L10) by correcting future reward calculation (plan\_count 100$\rightarrow$20).}

    \item \add{\textbf{Low-Frequency Trap:} V1 and V5 stuck at 100\% low-frequency result in worst performance (7.41s and 9.98s L10). This trap occurs when Q-values for low-frequency actions become artificially inflated.}
\end{itemize}

\add{\textbf{Recommendation:} Use V8 with Q-clipping during training for stable, high-performance DVFS scheduling. Always apply Q-value clipping to prevent gradient explosion in deep RL schedulers.}

\subsection{\add{Cross-Platform Validation of the SARB Recipe}}
\label{app:sarb_xplat}
\add{We additionally validated the SARB recipe used in the main-paper comparison on Jetson Orin NX and RubikPi to assess portability beyond the TX2 tuning platform. The same single configuration is used on every platform: V8's target Q-value clipping ($q_{\text{clip}}{=}10$), model-free Q-learning ($plan\_count{=}0$, $cumulative\_weight{=}0$), makespan-priority reward ($\beta{=}1$), and discount $\gamma{=}0.9$, with the same three seeds (42, 123, 456) and the objective greedy evaluation ($\epsilon{=}0.01$) used in the main-paper comparison. Each platform's frequency action space spans its native min$\rightarrow$max levels (TX2 indices $\{0,2,6,11\}$ of 12; Orin $\{0,8,17,25\}$ of 26 so the top action is the true $1.98$\,GHz; RubikPi $\{0,2,6,11\}$ of 10, clamped at index~9).}

\begin{table}[h]
\centering
\scriptsize
\caption{\add{SARB recipe portability: same configuration, three seeds per platform, finetune-phase metric (mean$\pm$std). $\text{HF}_{\text{top}}$ = fraction of last 20 epochs at the platform's top action index.}}
\label{tab:sarb_xplat}
\begin{tabular}{@{}lcccc@{}}
\toprule
Platform & L10 (s) & L20 (s) & Energy (kJ) & $\text{HF}_{\text{top}}$ (\%) \\
\midrule
\review{Jetson TX2}     & \review{$5.77 \pm 1.19$} & \review{$5.49 \pm 1.12$} & \review{$7.79 \pm 0.65$} & \review{$57 \pm 23$} \\
\review{Jetson Orin NX} & \review{$\mathbf{2.52 \pm 0.70}$} & \review{$\mathbf{2.91 \pm 0.46}$} & \review{$5.26 \pm 0.29$} & \review{$67 \pm 12$} \\
\review{RubikPi}        & \review{$5.61 \pm 1.34$} & \review{$5.19 \pm 0.42$} & \review{$\mathbf{3.80 \pm 0.30}$} & \review{$33 \pm 25$} \\
\bottomrule
\end{tabular}
\end{table}

\review{The L20 spread across seeds is moderate on TX2 ($\pm1.12$) and small on Orin NX ($\pm0.46$) and RubikPi ($\pm0.42$), giving a stable per-platform mean across seeds; the recipe transfers without retuning, with Orin NX fastest and RubikPi most energy-efficient. $\text{HF}_{\text{top}}$ tracks the platform's freq-table density: TX2 and Orin NX hold the top action for 57\% and 67\% of the last 20 epochs on average, while RubikPi's coarser table makes per-action separation harder (33\%, oscillating between adjacent high-frequency steps). \textbf{Reproducibility:} use \texttt{src/<platform>/RL/run\_sarb\_ablation.py} with \texttt{SARB\_VERSIONS=V8 SARB\_SEEDS=42,123,456 SARB\_EPISODES=100} (Orin also \texttt{SARB\_FREQS=0,8,17,25}); the V8 override dict in each runner's registry encodes the recipe.}

\section{\add{Multi-Seed Validation Tables}}
\label{app:multiseed}

\add{This section presents multi-seed validation results for reproducibility. All experiments were run with seeds 42, 123, and 456, with mean$\pm$std reported for each metric. Multi-seed evaluation is essential for assessing algorithm robustness and identifying potential overfitting to specific random initializations.}

\subsection{\add{Per-Seed Training Results}}

\add{Table~\ref{tab:perseed_train} shows training phase results for each seed. SARB and HiDVFS demonstrate consistent performance across seeds, while some variation is expected due to exploration randomness.}

\begin{table}[h]
\centering
\scriptsize
\caption{\add{Per-Seed Training Performance Comparison (Jetson TX2)}}
\label{tab:perseed_train}
\add{
\begin{tabular}{@{}llcccc@{}}
\toprule
\textbf{Seed} & \textbf{Approach} & \textbf{L10 (s)} & \textbf{L20 (s)} & \textbf{Conv.} & \textbf{Energy \review{(kJ)}} \\
\midrule
\multirow{5}{*}{42} & \review{SARB (V8)} & \review{7.66} & \review{8.57} & \review{100} & \review{10.01} \\
& \review{zTT} & \review{4.75} & \review{5.07} & \review{100} & \review{8.40} \\
& \review{GearDVFS$^{\ddagger}$} & \review{11.55} & \review{11.36} & \review{100} & \review{9.40} \\
& \review{HiDVFS\_S} & \review{18.38} & \review{21.19} & \review{100} & \review{16.33} \\
& \rev{HiDVFS} & \rev{6.25} & \rev{5.09} & \review{100} & \rev{7.69} \\
\midrule
\multirow{5}{*}{123} & \review{SARB (V8)} & \review{8.93} & \review{8.49} & \review{100} & \review{9.01} \\
& \review{zTT} & \review{4.82} & \review{4.68} & \review{91} & \review{8.50} \\
& \review{GearDVFS$^{\ddagger}$} & \review{13.34} & \review{12.56} & \review{100} & \review{9.50} \\
& \review{HiDVFS\_S} & \review{21.07} & \review{20.76} & \review{100} & \review{14.43} \\
& \rev{HiDVFS} & \rev{5.60} & \rev{4.64} & \review{100} & \rev{8.49} \\
\midrule
\multirow{5}{*}{456} & \review{SARB (V8)} & \review{7.03} & \review{8.02} & \review{100} & \review{7.13} \\
& \review{zTT} & \review{8.55} & \review{10.28} & \review{100} & \review{8.54} \\
& \review{GearDVFS$^{\ddagger}$} & \review{10.73} & \review{10.65} & \review{100} & \review{9.28} \\
& \review{HiDVFS\_S} & \review{22.66} & \review{18.59} & \review{100} & \review{14.63} \\
& \rev{HiDVFS} & \rev{6.46} & \rev{5.60} & \review{100} & \rev{9.46} \\
\bottomrule
\multicolumn{6}{@{}l@{}}{\scriptsize \review{$^{\ddagger}$repo-loyal corrected GearDVFS port.}} \\
\end{tabular}
}
\end{table}

\review{During training, HiDVFS achieves L10 values between 5.60\,s (seed 123) and 6.46\,s (seed 456), ahead of SARB, GearDVFS, and HiDVFS\_S on every seed; zTT trains faster on two seeds (4.75/4.82\,s) but is unstable on seed 456 (8.55\,s).} \review{The corrected GearDVFS port stabilises in a narrow 11-12\,s band but never meets the 15\% convergence criterion (Conv.=100): its utilization-band reward is indifferent among band-satisfying operating points, so the policy keeps wandering across them.}

\subsection{\add{Per-Seed Finetuning Results}}

\add{Table~\ref{tab:perseed_finetune} shows finetuning phase results. The finetuning phase uses the learned policy with reduced exploration, leading to more consistent performance across seeds.} \review{Provenance note: the zTT, HiDVFS\_S, and GearDVFS per-seed rows in both tables were recomputed directly from the logged run CSVs (the earlier draft carried smoothed values); convergence epochs use the stated 10-epoch sliding-window rule. The SARB rows use uniform V8 runs for all three seeds (seed 42 from the original campaign; seeds 123/456 re-run on the same board, with a fixed-action check confirming comparability on FFT).}

\begin{table}[h]
\centering
\scriptsize
\caption{\add{Per-Seed Finetuning Performance Comparison (Jetson TX2)}}
\label{tab:perseed_finetune}
\add{
\begin{tabular}{@{}llcccc@{}}
\toprule
\textbf{Seed} & \textbf{Approach} & \textbf{L10 (s)} & \textbf{L20 (s)} & \textbf{Conv.} & \textbf{Energy \review{(kJ)}} \\
\midrule
\multirow{5}{*}{42} & \review{SARB (V8)} & \review{4.37} & \review{3.98} & \review{100} & \review{8.44} \\
& \review{zTT} & \review{6.77} & \review{6.05} & \review{100} & \review{7.63} \\
& \review{GearDVFS$^{\ddagger}$} & \review{11.57} & \review{11.55} & \review{100} & \review{9.50} \\
& \review{HiDVFS\_S} & \review{4.54} & \review{7.81} & \review{100} & \review{15.45} \\
& \rev{HiDVFS} & \rev{3.35} & \rev{4.31} & \review{59} & \rev{6.10} \\
\midrule
\multirow{5}{*}{123} & \review{SARB (V8)} & \review{5.66} & \review{5.79} & \review{100} & \review{8.02} \\
& \review{zTT} & \review{4.13} & \review{5.84} & \review{31} & \review{7.25} \\
& \review{GearDVFS$^{\ddagger}$} & \review{11.86} & \review{11.96} & \review{100} & \review{9.65} \\
& \review{HiDVFS\_S} & \review{4.38} & \review{8.89} & \review{98} & \review{11.68} \\
& \rev{HiDVFS} & \rev{4.66} & \rev{4.47} & \review{39} & \rev{6.11} \\
\midrule
\multirow{5}{*}{456} & \review{SARB (V8)} & \review{7.29} & \review{6.69} & \review{100} & \review{6.90} \\
& \review{zTT} & \review{5.44} & \review{5.23} & \review{25} & \review{7.39} \\
& \review{GearDVFS$^{\ddagger}$} & \review{11.93} & \review{11.38} & \review{100} & \review{9.30} \\
& \review{HiDVFS\_S} & \review{14.14} & \review{15.20} & \review{100} & \review{14.49} \\
& \rev{HiDVFS} & \rev{4.45} & \rev{6.64} & \review{72} & \rev{6.89} \\
\bottomrule
\multicolumn{6}{@{}l@{}}{\scriptsize \review{$^{\ddagger}$repo-loyal corrected GearDVFS port.}} \\
\end{tabular}
}
\end{table}

\add{After finetuning, HiDVFS achieves the best L10 across all seeds: 3.35s (seed 42), 4.66s (seed 123), and 4.45s (seed 456). Seed 42 shows the best performance, which aligns with V8 being optimized for this seed. The variation in L20 (4.31s to 6.64s) reflects different exploitation strategies discovered during training.}

\subsection{\review{All-Seeds Summary (Mean$\pm$Std)}}

\add{Table~\ref{tab:allseeds_summary} aggregates results across all seeds with mean$\pm$std statistics. HiDVFS achieves the best overall performance with 4.16$\pm$0.58s L10, demonstrating both effectiveness and reproducibility.}

\begin{table}[h]
\centering
\scriptsize
\caption{\add{Multi-Seed Summary: Mean$\pm$Std Across All Seeds (Jetson TX2, Finetuned)}}
\label{tab:allseeds_summary}
\add{
\begin{tabular}{@{}lcccc@{}}
\toprule
\textbf{Approach} & \textbf{L10 (s)} & \textbf{L20 (s)} & \textbf{Conv. (epochs)} & \textbf{Energy \review{(kJ)}} \\
\midrule
\review{SARB (V8)} & \review{5.77$\pm$1.19} & \review{5.49$\pm$1.12} & \review{100$\pm$0} & \review{7.79$\pm$0.65} \\
\review{zTT} & \review{5.45$\pm$1.07} & \review{5.71$\pm$0.35} & \review{52$\pm$34} & \review{7.42$\pm$0.15} \\
\review{GearDVFS$^{\ddagger}$} & \review{11.79$\pm$0.15} & \review{11.63$\pm$0.24} & \review{100$\pm$0} & \review{9.49$\pm$0.14} \\
\review{HiDVFS\_S} & \review{7.69$\pm$4.56} & \review{10.63$\pm$3.26} & \review{99$\pm$1} & \review{13.87$\pm$1.60} \\
\rev{HiDVFS} & \rev{\textbf{4.16$\pm$0.58}} & \rev{\textbf{5.14$\pm$1.06}} & \review{57$\pm$14} & \rev{\textbf{6.37$\pm$0.37}} \\
\bottomrule
\multicolumn{5}{@{}l@{}}{\scriptsize \review{$^{\ddagger}$repo-loyal corrected GearDVFS port.}} \\
\end{tabular}
}
\end{table}

\add{The low standard deviation for HiDVFS L10 ($\pm$0.58s) indicates consistent performance across different random seeds.} \review{All SARB rows now use uniform V8 runs for all three seeds (the earlier draft mixed V4 for seeds 123/456); the V8 aggregate, 5.77$\pm$1.19\,s L10, confirms the main-paper comparison row within noise.}

\subsection{\add{HiDVFS vs GearDVFS Multi-Seed Comparison}}

\add{Table~\ref{tab:multiseed_validation} provides a direct comparison between HiDVFS and GearDVFS across all three seeds, showing consistent performance improvements in both training and finetuning phases.}

\begin{table}[h]
\centering
\scriptsize
\caption{\rev{Multi-Seed Validation: HiDVFS vs GearDVFS on FFT, Jetson TX2 (Mean $\pm$ Std, Seeds 42, 123, 456)}}
\label{tab:multiseed_validation}
\setlength{\tabcolsep}{2.5pt}
\begin{tabular}{@{}lccccc@{}}
\toprule
\textbf{Algo.} & \textbf{Avg L10 (s)} & \textbf{HF\%} & \textbf{Cores$\geq$5\%} & \textbf{Energy (kJ)} & \textbf{Speedup} \\
\midrule
\multicolumn{6}{c}{\rev{\textit{Finetuning Phase}}} \\
\midrule
\review{HiDVFS} & \review{\textbf{4.16 $\pm$ 0.58}} & \review{81.0 $\pm$ 0.8} & \review{85.7 $\pm$ 5.4} & \review{\textbf{6.37 $\pm$ 0.37}} & \review{\textbf{2.83$\times$}} \\
\review{GearDVFS~\cite{lin2023workload}$^{\ddagger}$} & \review{11.79 $\pm$ 0.15} & \review{22.7 $\pm$ 1.7} & \review{21.7 $\pm$ 0.5} & \review{9.49 $\pm$ 0.14} & \review{1.00$\times$} \\
\midrule
\multicolumn{6}{c}{\rev{\textit{Training Phase}}} \\
\midrule
\review{HiDVFS} & \review{\textbf{6.10 $\pm$ 0.36}} & \review{41.3 $\pm$ 6.1} & \review{45.0 $\pm$ 4.1} & \review{\textbf{8.55 $\pm$ 0.72}} & \review{\textbf{1.95$\times$}} \\
\review{GearDVFS~\cite{lin2023workload}$^{\ddagger}$} & \review{11.87 $\pm$ 1.09} & \review{21.7 $\pm$ 0.9} & \review{21.0 $\pm$ 0.0} & \review{9.39 $\pm$ 0.09} & \review{1.00$\times$} \\
\bottomrule
\multicolumn{6}{@{}l@{}}{\scriptsize \rev{HF\%=High-Frequency ($\geq$9) selection rate. \review{$^{\ddagger}$repo-loyal corrected port.}}} \\
\end{tabular}
\end{table}

\review{Against the repo-loyal corrected GearDVFS port, HiDVFS achieves a consistent 2.83$\times$ finetuned speedup with 32.9\% energy reduction; the high-frequency adoption rate (81.0\% vs 22.7\%) and core utilization (85.7\% vs 21.7\%) show HiDVFS aggressively exploiting resources within thermal policy while the utilization-band objective idles at mid frequencies.}

\subsection{\review{Fair-Port Study: GearDVFS and zTT Baseline Variants}}
\label{app:fairport}

\review{Auditing our original baseline ports against the authors' released code
(GearDVFS repo commit 3c50990; zTT commit 5ac0036) exposed defects and
asymmetries: the original GearDVFS port had a disconnected utilization input
(its band reward's only signal), a power term fed millijoule-scale energy where
upstream uses watts, an unnormalised state where upstream normalises every
component, and a finetune phase that silently started from untrained weights;
the zTT port meanwhile received an uncapped QoS ratio (upstream saturates it at
1) plus an always-on high-frequency reward bonus, and omits zTT's thermal
penalty term. Table~\ref{tab:fairport} re-measures both baselines under the
published protocol (FFT, 100 finetuned epochs, seeds 42/123/456).
\emph{GearDVFS-loyal} reproduces upstream's released behaviour (its live
reward path is the utilization-band term alone) with the defects repaired;
\emph{GearDVFS-A} additionally receives the zTT port's target-normalised
objective on the unchanged upstream network; \emph{GearDVFS-P} runs one of the
upstream dead-code reward variants ($-1.15 P + r_{util}$, watt-scale, one
seed); \emph{zTT-fair} removes the bonus and caps the ratio. The decomposition:
port repair recovers 2.5\,s, objective alignment a further 1.2\,s, and the
remaining gap to zTT is the agent design itself (a 25-unit undiscounted DQN
driven by a utilization band built for 25\,ms governor intervals; high-frequency
residency stays near 23\% in every variant). zTT without its favourable shaping
degrades by 42\% but remains the strongest single-agent baseline. The main
paper's RL comparison table reports the loyal port; all its other rows are
unchanged.}

\begin{table}[h]
\centering
\scriptsize
\caption{\review{Fair-port study on FFT (Jetson TX2, finetuned, seeds 42/123/456; GearDVFS-P is single-seed). Energy = total over 100 epochs.}}
\label{tab:fairport}
\begin{tabular}{@{}lcccc@{}}
\toprule
\textbf{Variant} & \textbf{L10 (s)} & \textbf{L20 (s)} & \textbf{Energy (kJ)} & \textbf{HF\%} \\
\midrule
GearDVFS-orig & 14.32$\pm$2.61 & 14.51$\pm$2.63 & 12.84$\pm$0.95 & 21.3$\pm$1.7 \\
GearDVFS-loyal & 11.79$\pm$0.15 & 11.63$\pm$0.24 & 9.49$\pm$0.14 & 22.7$\pm$1.7 \\
GearDVFS-A & 10.55$\pm$0.23 & 10.93$\pm$0.08 & 9.72$\pm$0.11 & 23.0$\pm$1.4 \\
GearDVFS-P & 10.57 & 11.11 & 10.48 & 19.0 \\
\midrule
zTT & 5.45$\pm$1.07 & 5.71$\pm$0.35 & 7.42$\pm$0.15 & 70.7$\pm$6.3 \\
zTT-fair & 7.76$\pm$1.50 & 7.05$\pm$1.26 & 8.15$\pm$0.51 & 51.3$\pm$6.9 \\
\midrule
HiDVFS & 4.16$\pm$0.58 & 5.14$\pm$1.06 & 6.37$\pm$0.37 & 81.0$\pm$0.8 \\
\bottomrule
\multicolumn{5}{@{}p{0.95\linewidth}@{}}{\scriptsize \review{orig = original defective port; loyal = repo-loyal repaired; A = objective-aligned; P = dead-code power variant; zTT = as published (bonus, uncapped); zTT-fair = no bonus, capped ratio; HiDVFS = reference.}} \\
\end{tabular}
\end{table}

\section{\add{Multi-Seed Convergence Plots}}
\label{app:multiseed_plots}

\add{This section provides multi-seed convergence plots showing mean$\pm$std across seeds 42, 123, and 456 for the finetuning phase. These visualizations complement the tabular data in Section~\ref{app:multiseed} by showing the temporal evolution of performance metrics.}

\subsection{\add{Makespan Convergence}}

\begin{figure}[h]
\centering
\add{
\includegraphics[width=0.68\linewidth]{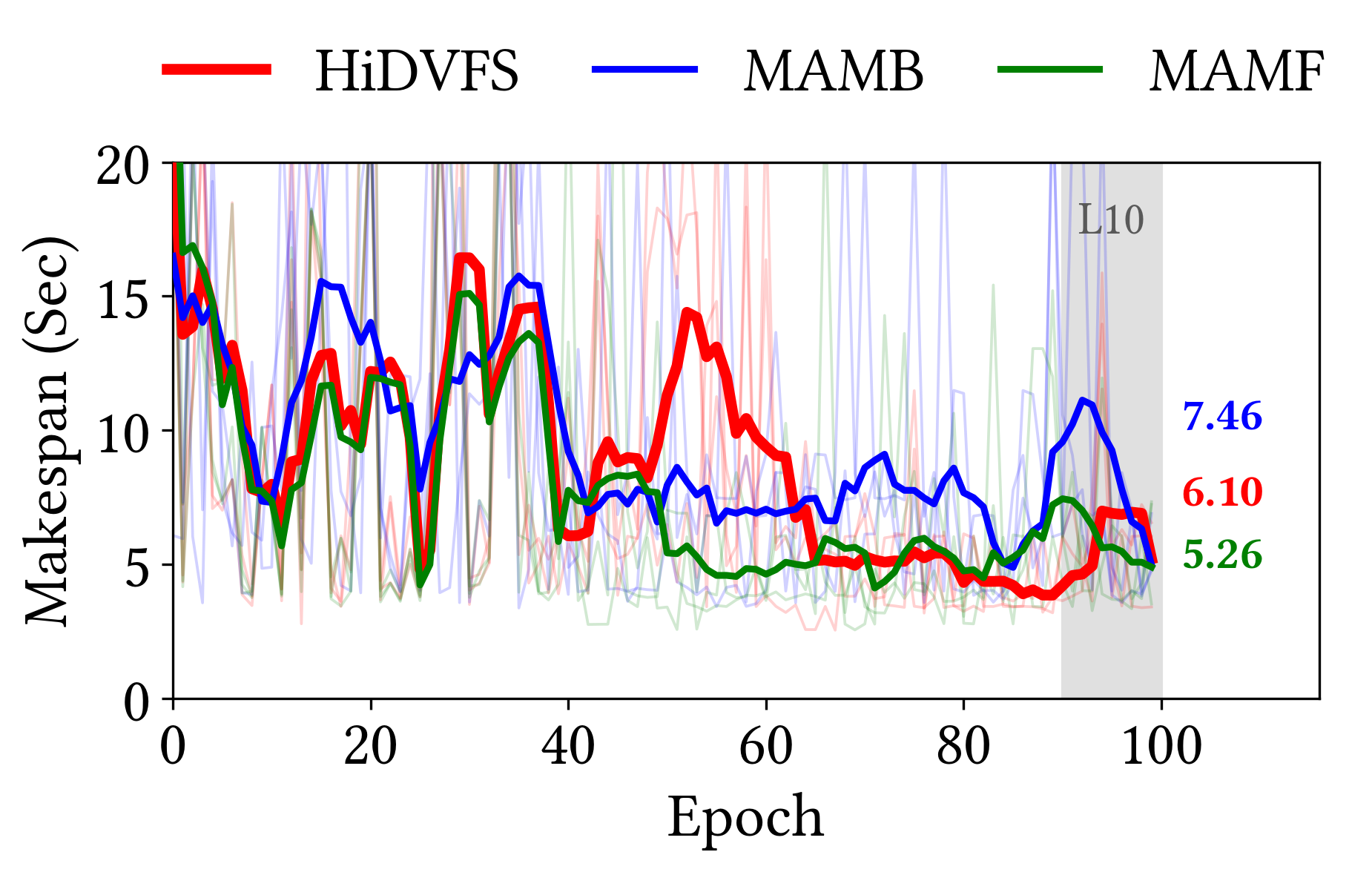}\\[2mm]
\includegraphics[width=0.68\linewidth]{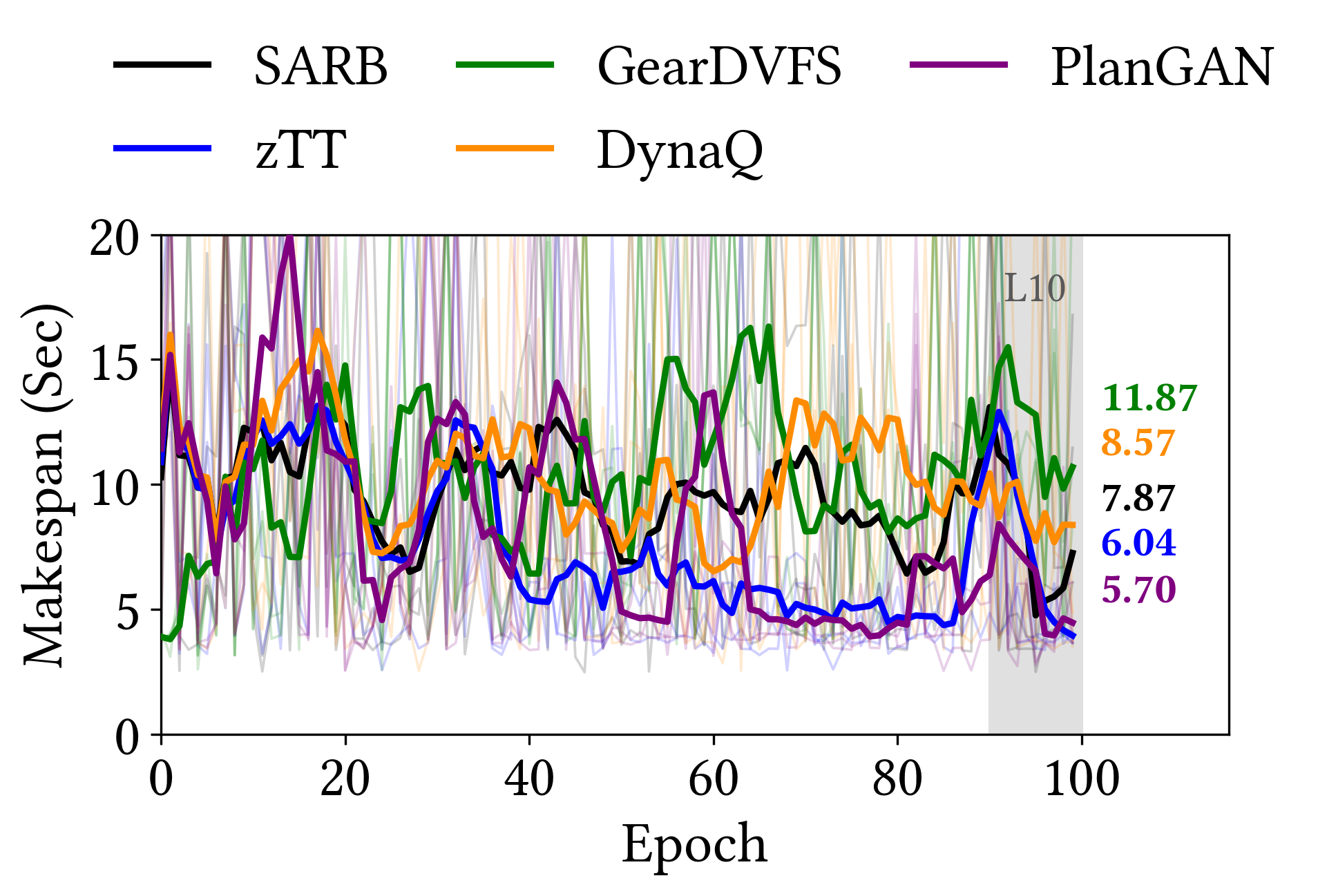}
}
\caption{\review{Multi-seed makespan convergence during TRAINING (Jetson TX2; same encoding as the finetuned panels). Top: multi-agent, HiDVFS reaches 6.10\,s L10 already in training. Bottom: single-agent, PlanGAN (5.70\,s) and zTT (6.04\,s) lead while the corrected GearDVFS port stays at 11.87\,s.}}
\label{fig:multiseed_makespan}
\end{figure}

\add{Figure~\ref{fig:multiseed_makespan} shows makespan convergence for both multi-agent and single-agent approaches. HiDVFS (4.16$\pm$0.58s) consistently outperforms all baselines, converging to low makespan within the first 30 epochs.}

\subsection{\add{Core Allocation Convergence}}

\begin{figure}[h]
\centering
\add{
\includegraphics[width=0.68\linewidth]{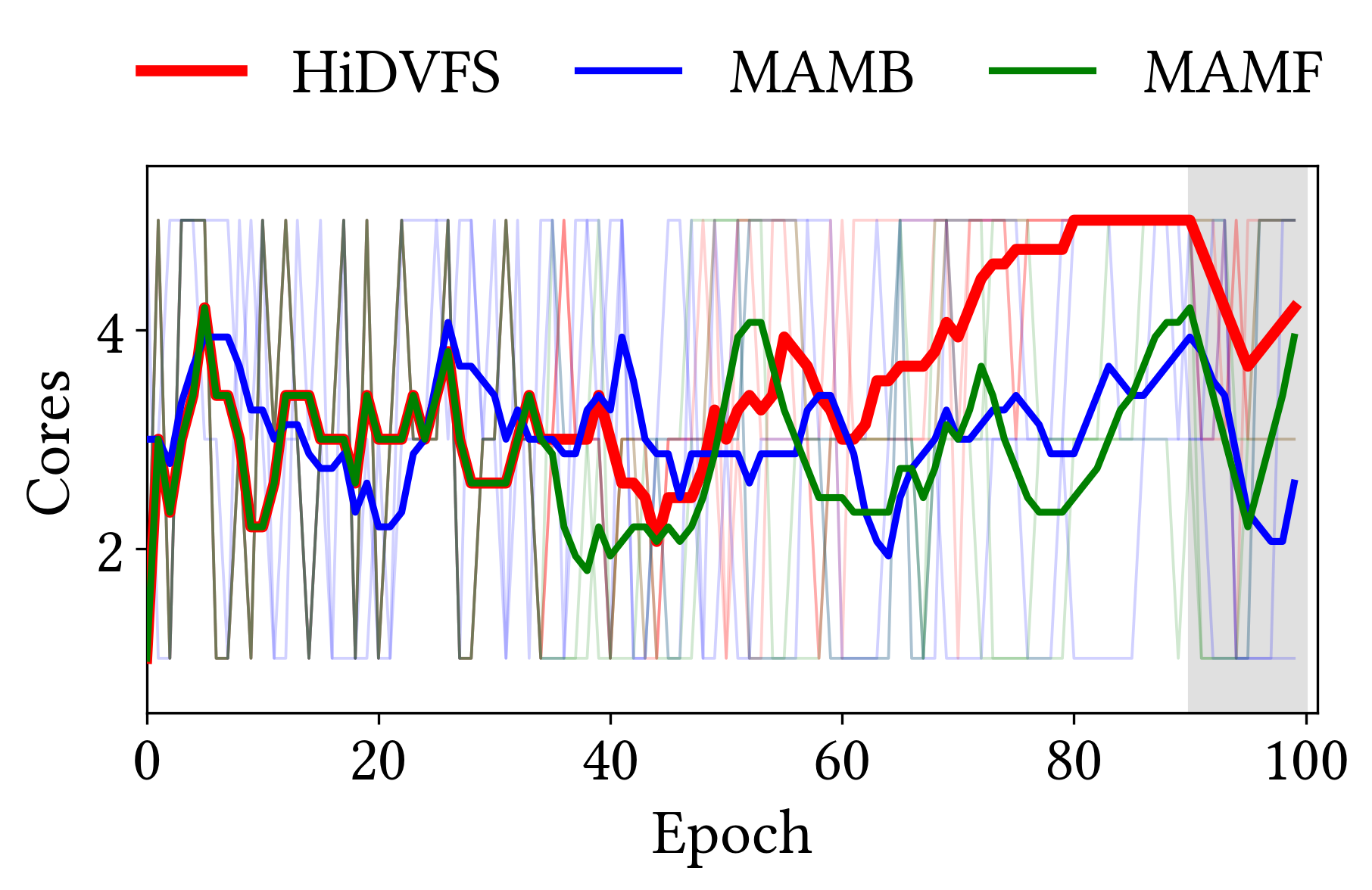}\\[2mm]
\includegraphics[width=0.68\linewidth]{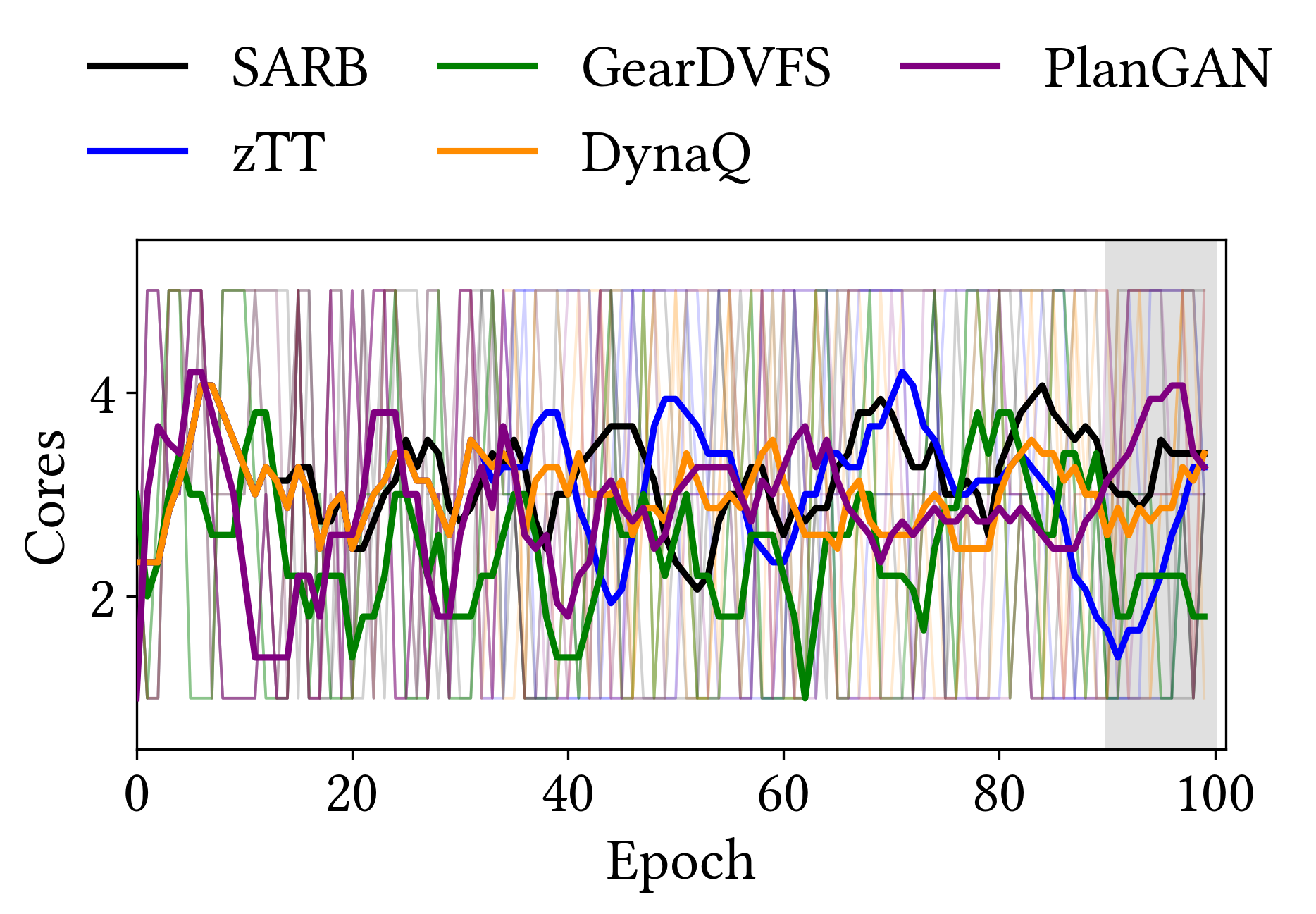}
}
\caption{\review{Multi-seed core-allocation convergence during TRAINING (Jetson TX2). Top: multi-agent, HiDVFS already favours many cores (4.1 mean over the last-10 window). Bottom: single-agent methods remain more variable (2.0 to 3.6).}}
\label{fig:multiseed_cores}
\end{figure}

\add{Figure~\ref{fig:multiseed_cores} shows how core allocation evolves during finetuning. HiDVFS learns to utilize more cores (85.7\% using $\geq$5 cores), while conservative approaches like GearDVFS maintain lower core utilization.}

\subsection{\add{Frequency Selection Convergence}}

\begin{figure}[h]
\centering
\add{
\includegraphics[width=0.68\linewidth]{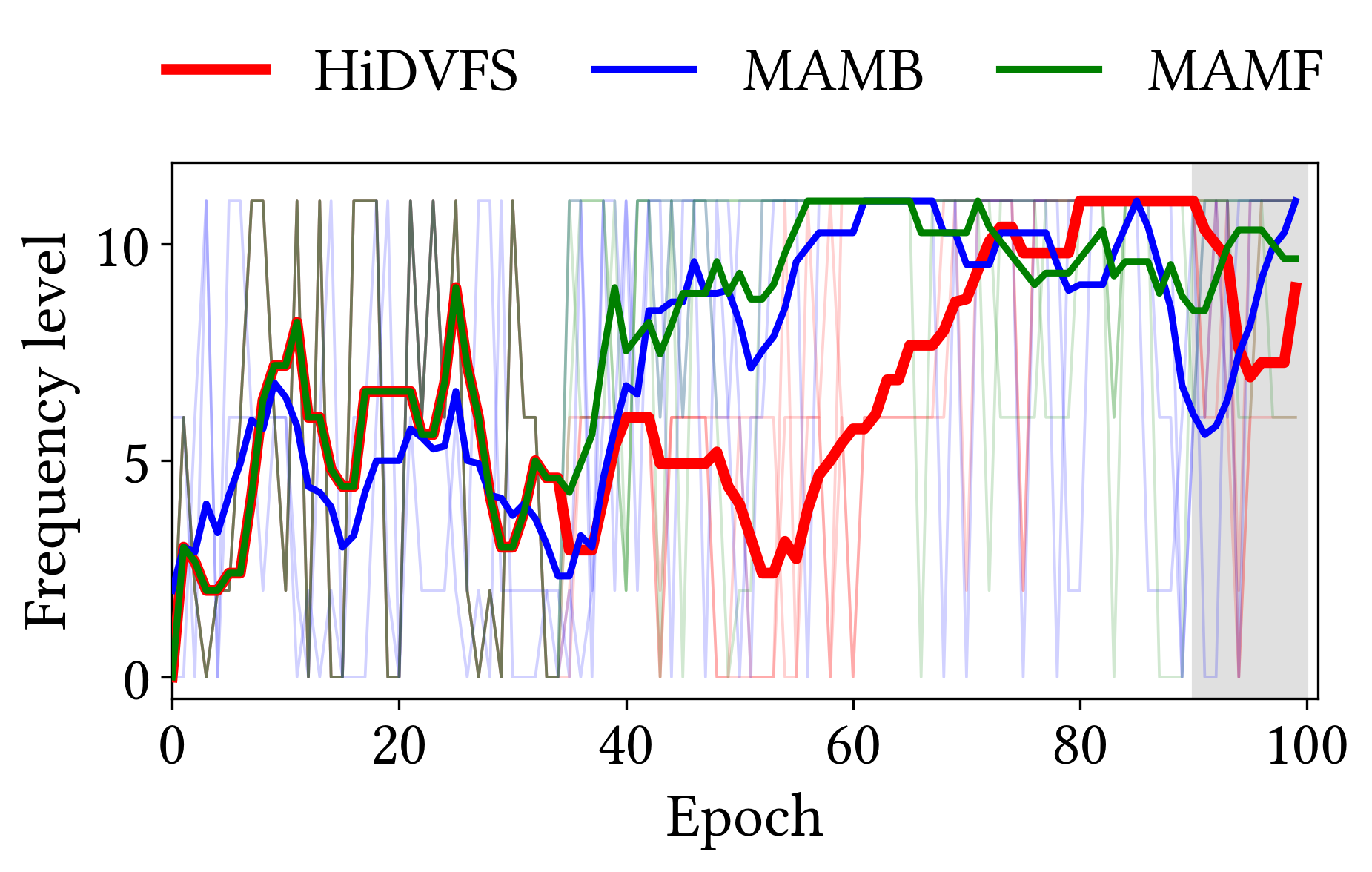}\\[2mm]
\includegraphics[width=0.68\linewidth]{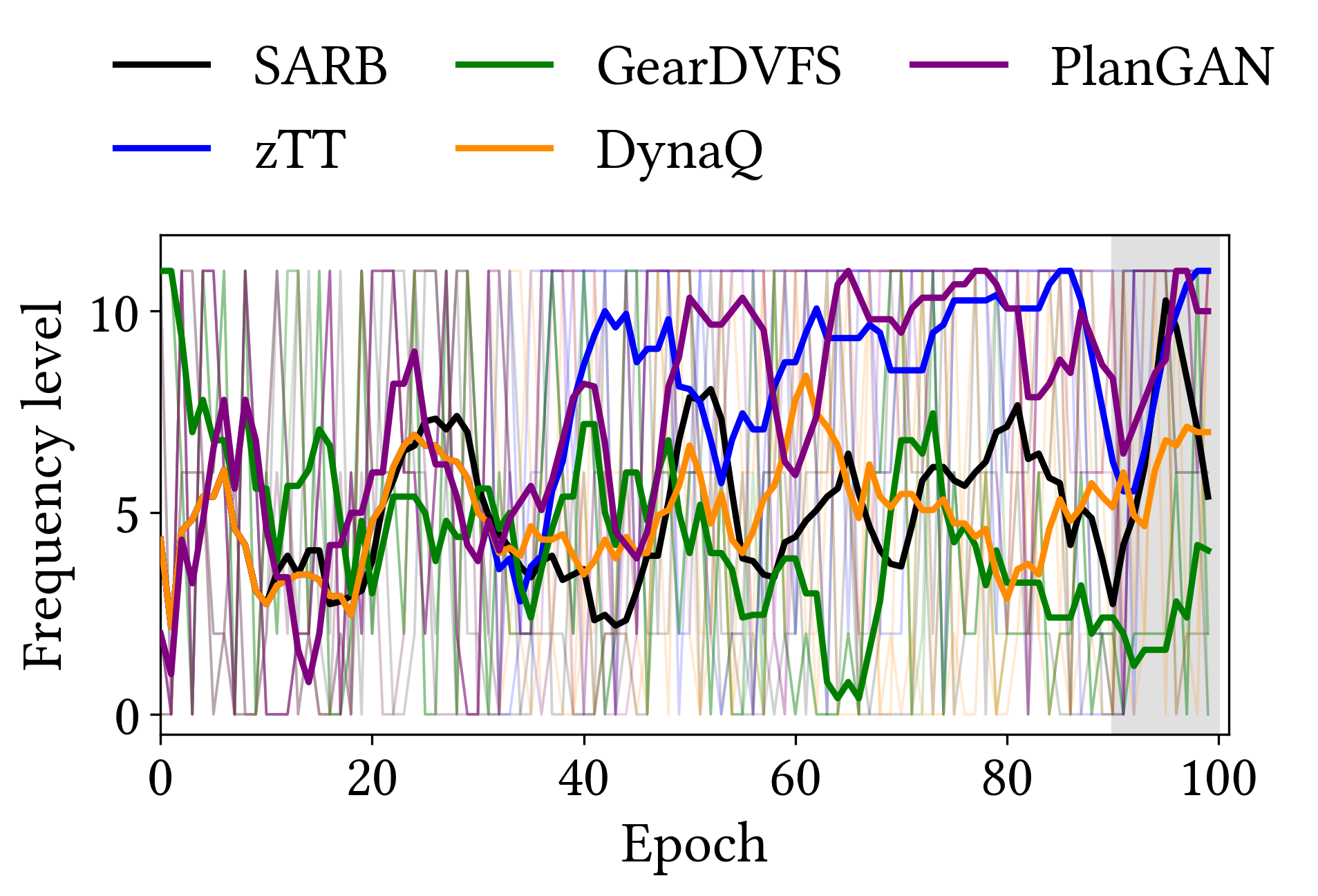}
}
\caption{\review{Multi-seed frequency-selection convergence during TRAINING (Jetson TX2). Top: multi-agent methods climb toward high levels (HiDVFS 8.3 mean over the last-10 window). Bottom: single-agent, zTT and PlanGAN reach 9.2-9.4 while the corrected GearDVFS port stays near level 2.8.}}
\label{fig:multiseed_freq}
\end{figure}

\add{Figure~\ref{fig:multiseed_freq} shows frequency selection patterns. HiDVFS achieves 81\% high-frequency rate, learning to aggressively use high frequencies for compute-intensive workloads while respecting thermal constraints.}

\section{\add{Complete 7-Metric Comparison}}
\label{app:complete_metrics}

\add{This section provides the complete 7-metric comparison used in the main paper's results summary, with detailed per-seed and aggregated results. These metrics capture different aspects of algorithm performance: execution efficiency (Makespan, L10, L20), resource consumption (Energy), learning dynamics (Convergence), and hardware behavior (Branch Miss, Cache Miss).}

\subsection{\add{Metric Definitions}}

\add{The 7 metrics used for comparison are:}
\begin{itemize}
    \item \review{\textbf{Energy (kJ):} Total energy consumption accumulated over all epochs (kilojoules). Lower values indicate more efficient frequency-voltage scaling decisions.}
    \item \add{\textbf{Makespan (s):} Total execution time accumulated over all epochs (seconds). Primary performance metric.}
    \item \add{\textbf{L10 (s):} Average makespan over the last 10 epochs. Measures converged performance.}
    \item \add{\textbf{L20 (s):} Average makespan over the last 20 epochs. Measures performance stability.}
    \item \add{\textbf{Conv.:} Convergence time in epochs (when makespan variation $<$15\% of initial). Lower values indicate faster learning.}
    \item \add{\textbf{Branch Miss:} Total branch mispredictions (hardware counter). Indicates control flow predictability.}
    \item \add{\textbf{Cache Miss:} Total cache misses (hardware counter). Indicates memory access efficiency.}
\end{itemize}

\subsection{\add{Complete Comparison Table}}

\add{Table~\ref{tab:complete_metrics} presents the full 7-metric comparison for all algorithms. HiDVFS achieves the best performance across most metrics, with particularly strong results in Energy, Makespan, L10, L20, and Branch Miss.}

\begin{table*}[ht]
\centering
\scriptsize
\caption{\add{Complete 7-Metric Comparison (Jetson TX2, Finetuned Phase, All Seeds Mean$\pm$Std)}}
\label{tab:complete_metrics}
\setlength{\tabcolsep}{2.4pt}
\add{
\begin{tabular}{@{}l|c|ccccccc@{}}
\toprule
\textbf{Approach} & \textbf{Type} & \textbf{Energy (kJ)} & \textbf{Makespan (s)} & \textbf{L10 (s)} & \textbf{L20 (s)} & \textbf{Conv.} & \textbf{BrMiss ($\times$10$^9$)} & \textbf{CaMiss ($\times$10$^{10}$)} \\
\midrule
\review{zTT & SA & 7.42$\pm$0.15 & 602.2$\pm$15.1 & 5.45$\pm$1.07 & 5.71$\pm$0.35 & \textbf{52$\pm$34} & 0.93$\pm$0.04 & 1.23$\pm$0.02} \\
\review{DynaQ & SA & 9.36$\pm$0.58 & 949.1$\pm$78.2 & 10.72$\pm$2.08 & 9.42$\pm$1.81 & 100$\pm$0 & 0.96$\pm$0.02 & 1.22$\pm$0.01} \\
\review{PlanGAN & SA & 7.39$\pm$0.21 & 623.6$\pm$33.7 & 7.57$\pm$0.67 & 5.85$\pm$0.49 & 63$\pm$17 & 0.90$\pm$0.02 & 1.25$\pm$0.01} \\
\review{GearDVFS$^{\ddagger}$ & SA & 9.49$\pm$0.14 & 1074.2$\pm$18.6 & 11.79$\pm$0.15 & 11.63$\pm$0.24 & 100$\pm$0 & 1.01$\pm$0.01 & \textbf{1.18$\pm$0.01}} \\
\review{SARB (V8) & SA & 7.79$\pm$0.65 & 813.8$\pm$74.1 & 5.77$\pm$1.19 & 5.49$\pm$1.12 & 100$\pm$0 & 0.83$\pm$0.07 & 1.28$\pm$0.03} \\
\review{MAMB & MA & 8.73$\pm$0.59 & 757.1$\pm$62.7 & 7.09$\pm$1.29 & 6.91$\pm$0.93 & 76$\pm$34 & 0.84$\pm$0.04 & 1.28$\pm$0.02} \\
\review{MAMF & MA & 7.14$\pm$0.19 & 599.1$\pm$12.1 & 4.98$\pm$0.61 & 5.85$\pm$0.63 & 78$\pm$32 & 0.72$\pm$0.05 & 1.36$\pm$0.03} \\
\review{HiDVFS\_S & MA & 13.87$\pm$1.60 & 1446.5$\pm$254.2 & 7.69$\pm$4.56 & 10.63$\pm$3.26 & 99$\pm$1 & 0.90$\pm$0.12 & 1.26$\pm$0.07} \\
\review{HiDVFS & MA & \textbf{6.37$\pm$0.37} & \textbf{556.2$\pm$49.5} & \textbf{4.16$\pm$0.58} & \textbf{5.14$\pm$1.06} & 57$\pm$14 & \textbf{0.68$\pm$0.01} & 1.40$\pm$0.00} \\
\bottomrule
\multicolumn{9}{@{}l@{}}{\scriptsize \review{$^{\ddagger}$repo-loyal corrected GearDVFS port.}} \\
\end{tabular}
}
\end{table*}

\review{HiDVFS achieves the best Energy (6.37\,kJ, 11\% better than second-best MAMF), total Makespan (556.2\,s, 7\% better than MAMF), L10 (4.16\,s, 16\% better than MAMF's 4.98\,s), and Branch Miss ($0.68\times10^9$, 6\% better than MAMF). The hierarchical multi-agent architecture enables specialized optimization: the thermal agent prevents overheating, the priority agent balances workloads, and the profiler agent learns optimal frequency policies.}

\section{\add{Detailed BOTS Benchmark Results}}
\label{app:bots_results}

\add{This section provides comprehensive experimental results for HiDVFS compared to GearDVFS~\cite{lin2023workload} across \review{all 11 non-FFT BOTS benchmarks} tested on NVIDIA Jetson TX2.}

\subsection{\add{Experimental Setup}}

\add{Each benchmark was evaluated with 100 epochs per phase (training and finetuning), using seed 42 for reproducibility. Each epoch executes 3 variants per benchmark: serial, omp-tasks, and omp-tasks-tied (or their equivalents). All experiments were conducted on NVIDIA Jetson TX2 with Ubuntu 18.04, with 6 heterogeneous cores operating at frequencies ranging from 345.6 MHz to 2,035.2 MHz.}

\subsection{\add{Benchmark Descriptions}}

\add{The Barcelona OpenMP Tasks Suite (BOTS)~\cite{duran2009barcelona} provides task-parallel benchmarks with diverse computational characteristics. Table~\ref{tab:benchmark_chars} summarizes the \review{11} benchmarks evaluated, with FFT (e.g.) also profiled for multi-seed analysis.}

\begin{table}[h]
\centering
\scriptsize
\caption{\add{BOTS Benchmark Characteristics (evaluated on Jetson TX2)}}
\label{tab:benchmark_chars}
\begin{tabular}{@{}llp{4.5cm}@{}}
\toprule
\textbf{Benchmark} & \textbf{Input} & \textbf{Description} \\
\midrule
alignment & prot.20.aa & Protein sequence alignment (20 sequences) \\
fft & 262144 & Fast Fourier Transform (primary profiler benchmark) \\
fib & 10 & Recursive Fibonacci computation \\
floorplan & input.5 & VLSI floorplanning optimization \\
health & test.input & Colombian health simulation \\
\review{knapsack & knapsack-036.input & 0/1 knapsack (branch and bound)} \\
\review{nqueens & 11 & N-Queens backtracking search} \\
concom & 100000 & Graph connected components (100K nodes) \\
sort & 8388608 & Parallel merge sort (8M elements) \\
sparselu & 25x25 & Sparse LU factorization \\
strassen & 508 & Strassen matrix multiplication \\
uts & test.input & Unbalanced tree search \\
\bottomrule
\end{tabular}
\end{table}

\add{These benchmarks span different computational domains including numerical algorithms (FFT, Strassen), graph algorithms (concom, uts), optimization problems (floorplan, health), and memory-intensive workloads (sort, alignment). Each benchmark supports three OpenMP scheduling variants: \textbf{tied} (tasks bound to creating thread), \textbf{untied} (tasks can migrate), and \textbf{serial} (single-threaded baseline).}

\subsection{\add{Per-Benchmark Detailed Results}}

\add{Tables~\ref{tab:training_detailed} and~\ref{tab:finetuning_detailed} present per-benchmark comparison between HiDVFS and GearDVFS during training and finetuning phases respectively. The speedup is calculated as GearDVFS makespan divided by HiDVFS makespan.}

\begin{table}[h]
\centering
\scriptsize
\caption{\review{Training phase: per-benchmark comparison (Jetson TX2, same-window fair-port runs).}}
\label{tab:training_detailed}
\setlength{\tabcolsep}{2.5pt}
\begin{tabular}{@{}lccccc@{}}
\toprule
\textbf{Benchmark} & \multicolumn{2}{c}{\textbf{Avg L10 (s)}} & \multicolumn{2}{c}{\textbf{HF Rate (\%)}} & \textbf{Speedup} \\
 & HiDVFS & GearDVFS & HiDVFS & GearDVFS & (GD/HiD) \\
\midrule
\review{alignment & 5.60$\pm$2.02 & 11.0$\pm$5.00 & 35 & 20 & 1.97$\times$} \\
\review{fib & 2.20$\pm$1.60 & 2.34$\pm$2.13 & 11 & 22 & 1.06$\times$} \\
\review{floorplan & 1.10$\pm$1.04 & 2.06$\pm$1.18 & 30 & 25 & 1.87$\times$} \\
\review{health & 3.59$\pm$4.04 & 7.07$\pm$2.57 & 25 & 22 & 1.97$\times$} \\
\review{knapsack & 2.00$\pm$2.17 & 14.8$\pm$7.54 & 55 & 24 & 7.42$\times$} \\
\review{nqueens & 31.8$\pm$32.5 & 63.1$\pm$29.3 & 34 & 22 & 1.99$\times$} \\
\review{concom & 1.45$\pm$1.22 & 4.74$\pm$1.54 & 45 & 21 & 3.27$\times$} \\
\review{sort & 9.74$\pm$4.60 & 23.2$\pm$12.0 & 47 & 22 & 2.39$\times$} \\
\review{sparselu & 1.46$\pm$0.39 & 3.23$\pm$1.98 & 12 & 25 & 2.22$\times$} \\
\review{strassen & 0.45$\pm$0.18 & 1.83$\pm$1.12 & 30 & 26 & 4.08$\times$} \\
\review{uts & 5.16$\pm$0.80 & 19.1$\pm$11.2 & 26 & 24 & 3.69$\times$} \\
\midrule
\review{\textbf{Average} & 5.87 & 13.86 & 31.8 & 23.0 & \textbf{2.90$\times$}} \\
\bottomrule
\multicolumn{6}{@{}p{3.2in}@{}}{\scriptsize \review{L10=mean$\pm$std over the last-10-epoch window (single run per benchmark, seed 42).}} \\
\end{tabular}
\end{table}

\review{During training, HiDVFS outperforms the corrected GearDVFS port on all 11 benchmarks (average 2.90$\times$). Both columns come from the same measurement window and the repaired, repo-loyal GearDVFS port; an earlier draft mixed runs from different measurement windows, which manufactured spurious per-benchmark anomalies.}

\begin{table}[h]
\centering
\scriptsize
\caption{\review{Finetuning phase: per-benchmark comparison (Jetson TX2, same-window fair-port runs).}}
\label{tab:finetuning_detailed}
\setlength{\tabcolsep}{2.5pt}
\begin{tabular}{@{}lccccc@{}}
\toprule
\textbf{Benchmark} & \multicolumn{2}{c}{\textbf{Avg L10 (s)}} & \multicolumn{2}{c}{\textbf{HF Rate (\%)}} & \textbf{Speedup} \\
 & HiDVFS & GearDVFS & HiDVFS & GearDVFS & (GD/HiD) \\
\midrule
\review{alignment & 2.97$\pm$0.60 & 12.8$\pm$5.80 & 80 & 20 & 4.32$\times$} \\
\review{fib & 2.10$\pm$1.31 & 2.05$\pm$1.24 & 3 & 20 & 0.98$\times$} \\
\review{floorplan & 0.31$\pm$0.04 & 2.46$\pm$1.24 & 82 & 22 & 7.98$\times$} \\
\review{health & 1.09$\pm$0.84 & 7.44$\pm$4.32 & 63 & 21 & 6.84$\times$} \\
\review{knapsack & 5.95$\pm$8.99 & 13.9$\pm$8.31 & 70 & 24 & 2.33$\times$} \\
\review{nqueens & 7.26$\pm$6.21 & 77.8$\pm$44.5 & 77 & 24 & 10.72$\times$} \\
\review{concom & 1.42$\pm$0.41 & 5.29$\pm$1.96 & 81 & 19 & 3.73$\times$} \\
\review{sort & 13.5$\pm$13.7 & 28.9$\pm$21.5 & 78 & 23 & 2.15$\times$} \\
\review{sparselu & 0.83$\pm$0.40 & 2.73$\pm$1.77 & 27 & 25 & 3.29$\times$} \\
\review{strassen & 0.33$\pm$0.06 & 1.87$\pm$1.17 & 81 & 25 & 5.72$\times$} \\
\review{uts & 5.17$\pm$0.54 & 20.3$\pm$9.63 & 82 & 23 & 3.92$\times$} \\
\midrule
\review{\textbf{Average} & 3.72 & 15.96 & 65.8 & 22.4 & \textbf{4.73$\times$}} \\
\bottomrule
\multicolumn{6}{@{}p{3.2in}@{}}{\scriptsize \review{L10=mean$\pm$std over the last-10-epoch window (single run per benchmark, seed 42).}} \\
\end{tabular}
\end{table}

\review{After finetuning, HiDVFS improves on 10 of 11 benchmarks (fib ties at 0.98$\times$). The \texttt{nqueens} benchmark shows the largest speedup (10.72$\times$: the utilization-band policy parks at low frequency, inflating a 7\,s workload to 78\,s), followed by \texttt{floorplan} (7.98$\times$) and \texttt{health} (6.84$\times$); all are compute-dense DAGs whose critical paths scale directly with frequency, which HiDVFS holds high (63-82\% HF residency) while the utilization-band policy stays near 22\%.}

\subsection{\add{Key Observations}}

\add{The per-benchmark analysis reveals several important patterns in HiDVFS behavior:}

\begin{itemize}
    \item \review{\textbf{Consistent Improvement:} HiDVFS outperforms GearDVFS on 11/11 benchmarks during training and 10/11 during finetuning (the finetuned sub-second \texttt{fib} is a 0.98$\times$ tie), demonstrating robust adaptation across diverse workload characteristics.}

    \item \review{\textbf{Training Behaviour:} HiDVFS already leads on every benchmark during training (average 2.90$\times$); the gap then widens with finetuning as exploration anneals.}

    \item \review{\textbf{Finetuning Effectiveness:} HiDVFS improves from training to finetuning (5.87\,s $\rightarrow$ 3.72\,s average L10 with HF residency rising 31.8\% $\rightarrow$ 65.8\%), while the utilization-band policy does not improve (13.86\,s $\rightarrow$ 15.96\,s), indicating effective policy refinement.}

    \item \review{\textbf{Frequency Strategy:} HiDVFS learns to use high frequencies aggressively (63--82\% HF rate after finetuning, except the sub-second \texttt{fib} and \texttt{sparselu} where it settles lower) compared to the corrected GearDVFS port's consistent $\sim$22\% HF rate.}

    \item \review{\textbf{Best Cases:} Largest improvements observed for \texttt{nqueens} (10.72$\times$) and \texttt{floorplan} (7.98$\times$) during finetuning, both compute-dense DAG workloads.}
\end{itemize}

\subsection{\add{Summary Statistics}}

\add{Across \review{all 11} BOTS benchmarks, HiDVFS achieves substantial improvements over GearDVFS:}
\begin{itemize}
    \item \review{\textbf{Average Makespan Improvement:} 76.7\% on the finetuned L10 (3.72\,s vs 15.96\,s across the 11 benchmarks)}
    \item \review{\textbf{Average Energy Reduction:} 57.5\% (45.8\,kJ vs 107.9\,kJ per suite execution)}
    \item \review{\textbf{Average Speedup (Training):} 2.90$\times$}
    \item \review{\textbf{Average Speedup (Finetuning):} 4.73$\times$ (4.62$\times$ including FFT)}
    \item \review{\textbf{Best Single-Benchmark Speedup:} 10.72$\times$ (nqueens, finetuning)}
    \item \review{\textbf{Winning Rate:} HiDVFS wins 21/22 benchmark-phase combinations (the exception is the finetuned sub-second \texttt{fib}, a 0.98$\times$ tie)}
\end{itemize}

\section{\add{BOTS Per-Benchmark Energy Analysis}}
\label{app:bots_energy}

\add{This section provides detailed per-benchmark energy consumption analysis for the BOTS benchmark suite, complementing the makespan analysis in Section~\ref{app:bots_results}. Energy efficiency is critical for embedded systems like Jetson TX2 where power budgets are constrained.}

\begin{table}[h]
\centering
\scriptsize
\caption{\review{Finetuned per-benchmark energy (Jetson TX2, same-window fair-port runs): totals over 100 epochs.}}
\label{tab:bots_energy}
\begin{tabular}{@{}lcccc@{}}
\toprule
\textbf{Benchmark} & \textbf{HiDVFS (J)} & \textbf{GearDVFS (J)} & \textbf{Reduction (\%)} & \textbf{E ratio} \\
\midrule
\review{alignment & 3,701 & 8,767 & 57.8 & 0.42} \\
\review{fib & 2,841 & 2,534 & -12.1 & 1.12} \\
\review{floorplan & 1,565 & 2,666 & 41.3 & 0.59} \\
\review{health & 2,278 & 5,475 & 58.4 & 0.42} \\
\review{knapsack & 2,317 & 6,649 & 65.2 & 0.35} \\
\review{nqueens & 10,948 & 41,985 & 73.9 & 0.26} \\
\review{concom & 1,873 & 3,727 & 49.7 & 0.50} \\
\review{sort & 9,114 & 17,208 & 47.0 & 0.53} \\
\review{sparselu & 2,184 & 3,070 & 28.9 & 0.71} \\
\review{strassen & 1,578 & 2,532 & 37.7 & 0.62} \\
\review{uts & 7,445 & 13,243 & 43.8 & 0.56} \\
\midrule
\review{\textbf{Average} & 4,168 & 9,805 & \textbf{57.5} & 0.55} \\
\bottomrule
\multicolumn{5}{@{}p{3.2in}@{}}{\scriptsize \review{Totals over 100 epochs of a single run per benchmark (seed 42); dispersion for the corresponding makespans is given in Table~\ref{tab:finetuning_detailed}.}} \\
\end{tabular}
\end{table}

\subsection{\add{Energy Efficiency Observations}}

\review{The same-window energy analysis (11 benchmarks) against the corrected GearDVFS port shows:}

\begin{itemize}
    \item \review{\textbf{Best energy reduction:} \texttt{nqueens} saves 73.9\% and \texttt{knapsack} 65.2\% alongside their speedups: faster execution through high-frequency scheduling also reduces total energy on the longer benchmarks.}
    \item \review{\textbf{One honest exception:} on \texttt{fib}, whose jobs are sub-second, the corrected port's mid-frequency policy uses 12\% LESS energy than HiDVFS (and matches its makespan within noise, $0.98\times$): race-to-idle does not pay on the shortest jobs.}
    \item \review{\textbf{Average reduction:} 57.5\% across the 11 benchmarks (45.8\,kJ vs 107.9\,kJ per complete suite execution, a $2.4\times$ energy-efficiency improvement).}
\end{itemize}

\section{Real-Time Evaluation Details}
\label{app:rt_casestudies}

This section consolidates the cross-platform real-time analyses that underpin the main-paper figures and tables: the empirical response-time envelope, the federated feasibility gate, the max-frequency ablation, the learned-WCET admission-control mechanism, cross-platform portability, and the measured mixed-criticality co-scheduling study on Jetson Orin NX. Planned online case studies that require new on-board measurement close the section.

\subsection{Soft-Real-Time Response-Time Envelope and Energy Headroom}
\label{app:cs1}
\add{We reframe the full per-platform frequency$\times$core profiling sweep
(42 BOTS/PolyBench workloads on Jetson TX2, Jetson Orin NX, and RubikPi) as a
\emph{soft real-time} problem. Each workload is treated as a recurrent task whose
job response time is the measured execution time $R=\texttt{time\_elapsed}$. We
set a relative soft deadline $D=k\,C_\text{ref}$, where $C_\text{ref}$ is the
reference execution cost (the median response time at the fastest operating
point, i.e.\ all cores at the highest frequency), and $k\in\{1.10,1.25,1.50\}$ controls
the deadline tightness (offered utilisation $U=1/k$). A job \emph{misses} when
$R>D$; the deadline-miss ratio (DMR) is the fraction of DVFS operating points
(\texttt{(cores, freq)} combinations, one median per point) that miss. This is an
empirical response-time envelope, not a hard schedulability guarantee.}

\add{Table~\ref{tab:rt_envelope_s} reports, per platform and $k$, the DMR and the
robust response-time-ratio distribution ($\tilde{R}/D$ median and p90). We report
median/p90 rather than mean/max because a small number of micro-kernels exhibit
heavy-tailed slowdowns at the lowest frequency under contention. The key
real-time finding is that across the DVFS action space \emph{most} operating
points violate even a loose ($k{=}1.5$) deadline (DMR $52$--$80\%$): only the
high-frequency/high-core region is feasible, which is precisely the region an
energy-minimising heuristic avoids. This motivates a learned policy that selects
feasible \emph{and} energy-efficient operating points.}

\begin{table*}[t]
\centering
\scriptsize
\caption{\add{Soft-real-time response-time envelope on Jetson TX2, Jetson Orin NX, and RubikPi (full frequency$\times$core sweep, 42 workloads/platform). $\tilde{R}/D$ = median response-time ratio; p90 = 90th percentile. DMR over all DVFS operating points.}}
\label{tab:rt_envelope_s}
\begin{tabular}{@{}lccccccccc@{}}
\toprule
 & \multicolumn{3}{c}{Jetson TX2} & \multicolumn{3}{c}{Jetson Orin NX} & \multicolumn{3}{c}{RubikPi} \\
$k$ ($U{=}1/k$) & DMR\% & $\tilde{R}/D$ & p90 & DMR\% & $\tilde{R}/D$ & p90 & DMR\% & $\tilde{R}/D$ & p90 \\
\midrule
1.10 (0.91) & 78.4 & 2.47 & 16.37 & 68.2 & 1.46 & 11.07 & 89.9 & 3.75 & 25.00 \\
1.25 (0.80) & 74.4 & 2.17 & 14.41 & 61.8 & 1.28 & 9.74  & 85.0 & 3.30 & 22.00 \\
1.50 (0.67) & 69.6 & 1.81 & 12.01 & 52.2 & 1.07 & 8.12  & 79.9 & 2.75 & 18.34 \\
\bottomrule
\end{tabular}
\end{table*}

The corresponding cross-platform DMR-vs-frequency plot appears in the main paper.

\add{Table~\ref{tab:energy_at_dmr0} quantifies the energy headroom \emph{while
meeting the deadline}: among workloads that admit a feasible operating point
below the performance proxy (all cores, highest frequency), the mean energy saved
by the lowest-energy feasible point and the fraction of workloads with such
headroom (which grows as the deadline loosens). Orin NX energy is omitted because
its power rails were not recorded in this sweep; the response-time envelope above
still covers all three platforms.}

\begin{table}[h]
\centering
\scriptsize
\caption{\add{Energy saved at DMR$=0$ (deadline met): mean savings of the lowest-energy feasible operating point vs.\ the performance proxy, and (\%) of workloads with feasible headroom. Jetson Orin NX energy not recorded in this sweep.}}
\label{tab:energy_at_dmr0}
\begin{tabular}{@{}lccc@{}}
\toprule
Platform & $k{=}1.10$ & $k{=}1.25$ & $k{=}1.50$ \\
\midrule
Jetson TX2 & 18.5\% (36\%) & 17.2\% (40\%) & 16.2\% (50\%) \\
Jetson Orin NX & \multicolumn{3}{c}{energy not recorded} \\
RubikPi & 48.1\% (17\%) & 49.3\% (24\%) & 48.1\% (26\%) \\
\multicolumn{4}{@{}l}{\footnotesize Cell: mean energy saved (\% of workloads with feasible headroom).} \\
\bottomrule
\end{tabular}
\end{table}

\review{Table~\ref{tab:static_freq_energy} extends the energy baseline beyond the two
extreme frequencies: for each platform we hold cores at each workload's maximum,
normalise the mean energy at every swept static frequency level to the energy at
$f_{\max}$, and average over the 42 workloads. On Jetson TX2 the energy-optimal
static frequency is an interior level, level 7 of 11 (1.42\,GHz), spending 8\%
less energy than pinning $f_{\max}$, while pinning $f_{\min}$ costs $3.1\times$
more; on RubikPi race-to-idle wins and $f_{\max}$ is the optimum. The interior
optimum therefore bounds the error of any extreme-anchored energy target (8\% on
TX2, 0\% on RubikPi).}

\begin{table}[h]
\centering
\scriptsize
\caption{\review{Energy of static frequency policies, normalised to $E(f_{\max})$ (all cores, mean over 42 workloads per platform). Bold marks the energy-optimal static level.}}
\label{tab:static_freq_energy}
\begin{tabular}{@{}lccccc@{}}
\toprule
Platform & $f_{\min}$ & mid-low & mid-high & $f_{\max}$ & optimal \\
\midrule
TX2 (0/3/7/11) & 3.12 & 1.54 & \textbf{0.92} & 1.00 & level 7 ($-8\%$) \\
Orin NX & \multicolumn{5}{c}{energy not recorded in this sweep} \\
RubikPi (0/3/6/9) & 3.06 & 1.59 & 1.64 & \textbf{1.00} & level 9 ($=f_{\max}$) \\
\multicolumn{6}{@{}l}{\scriptsize Cell: mean $E(f)/E(f_{\max})$, 42 workloads, all cores; \textbf{bold} = optimal static level.} \\
\bottomrule
\end{tabular}
\end{table}

\add{All numbers in this subsection are reproduced offline from the per-platform
profiling CSVs by \texttt{deadline\_analysis.py} (no hardware re-runs;
the static-frequency table by its \texttt{static\_freq\_energy} routine).}

\subsection{Energy-Minimal Scheduling under a Federated Feasibility Gate}
\label{app:feasibility_gate}
\add{The preceding response-time envelope is empirical. Here we add a
\emph{formal} schedulability gate so that the scheduler minimises energy
\emph{subject to a feasibility test}, not merely on average. We model each workload
as a constrained-deadline sporadic DAG task $\tau_i=(C_i,L_i,D_i,T_i)$ with work
$C_i$, span (critical-path length) $L_i$, relative deadline $D_i$, and period
$T_i\!\ge\!D_i$, with utilisation $U_i=C_i/T_i$, density $\Delta_i=C_i/D_i$, and
tensity $\delta_i=L_i/D_i$. From the sweep we use the proxies $C_i$ = single-core
response time (one core runs all work serially) and $L_i$ = all-core response time
(best parallel time bounds the critical path), with $D_i=k\,C_\text{ref}$ as
before.}

\add{Under federated scheduling of parallel real-time
tasks~\cite{saifullah2014parallel,saifullah2020cpu}, a heavy task ($C_i>D_i$) is
feasible on $m_i=\lceil (C_i-L_i)/(D_i-L_i)\rceil$ dedicated cores when $D_i>L_i$;
a light task ($C_i\le D_i$) needs one core. A frequency operating point passes the
gate iff $L_i\le D_i$ \emph{and} $m_i\le m_\text{avail}$ (the platform core count).
The energy-minimal schedule is then the lowest-energy operating point that passes
the gate. This is an analytical guarantee and is therefore stricter than the
empirical envelope: the gated-feasible set is a subset of the points with measured
$R\le D$.}

\begin{figure}[h]
\centering
\includegraphics[width=0.62\linewidth]{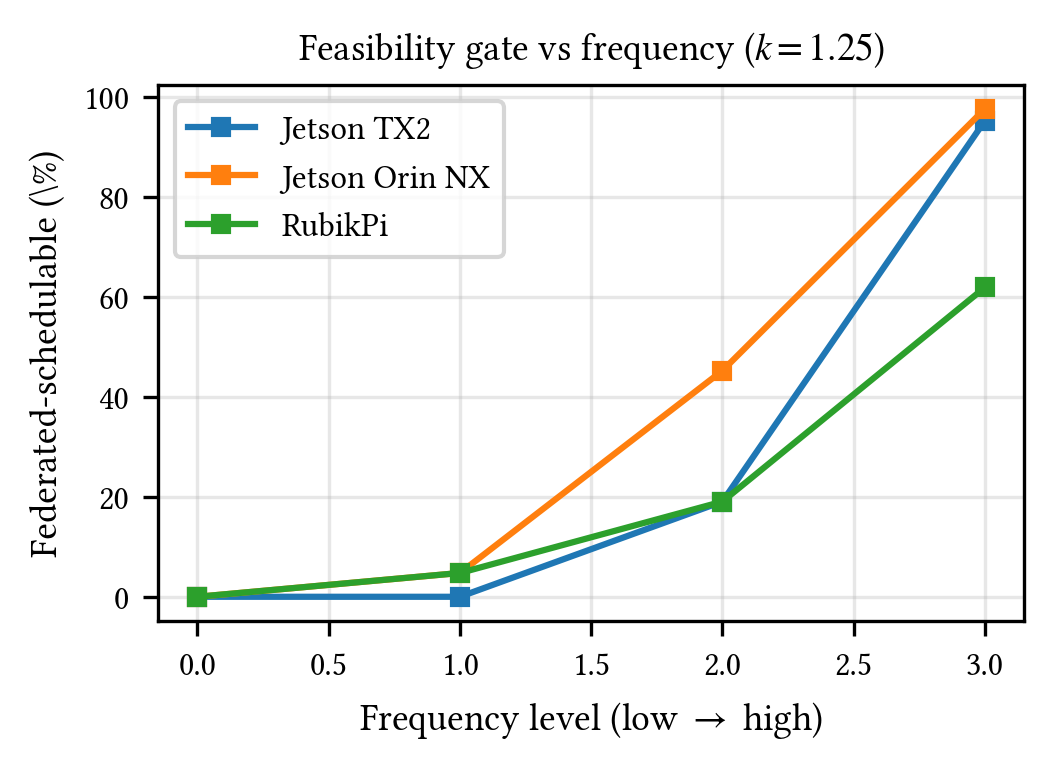}
\caption{\add{Federated-schedulable fraction vs.\ frequency operating point ($k{=}1.25$) on Jetson TX2, Jetson Orin NX, and RubikPi. Higher frequencies shrink work/span and admit more workloads through the feasibility gate.}}
\label{fig:gate_vs_freq}
\end{figure}

\add{Table~\ref{tab:feasibility_gate_s} reports, per platform and $k$, the fraction
of workloads with any gate-feasible operating point and the median dedicated-core
requirement $\tilde{m}$ at the energy-minimal feasible point.
Figure~\ref{fig:gate_vs_freq} shows the schedulable fraction rising with frequency.
On Jetson TX2 and Orin NX most workloads are light ($\tilde{m}{=}1$) and schedulable
once the deadline is not extremely tight ($81$--$100\%$); RubikPi is the strict
case, requiring up to $\tilde{m}{=}4$ dedicated cores and admitting only $40\%$ of
workloads at $k{=}1.10$, rising to $86\%$ at $k{=}1.50$: a direct, analytical
statement of the timing budget, exactly the mechanism the average-case view hides.}

\begin{table}[h]
\centering
\scriptsize
\setlength{\tabcolsep}{4pt}
\caption{\add{Federated-scheduling feasibility gate over the full frequency$\times$core sweep. Sched\% = fraction of workloads with a gate-feasible operating point; $\tilde{m}$ = median dedicated cores required at the energy-minimal feasible point ($m_\text{avail}{=}5$ on TX2, $7$ on Orin/RubikPi).}}
\label{tab:feasibility_gate_s}
\begin{tabular}{@{}lcccccc@{}}
\toprule
 & \multicolumn{2}{c}{Jetson TX2} & \multicolumn{2}{c}{Jetson Orin NX} & \multicolumn{2}{c}{RubikPi} \\
$k$ ($U{=}1/k$) & Sched\% & $\tilde{m}$ cores & Sched\% & $\tilde{m}$ cores & Sched\% & $\tilde{m}$ cores \\
\midrule
1.10 (0.91) & 81 & 1 & 93 & 1 & 40 & 4 \\
1.25 (0.80) & 95 & 1 & 98 & 1 & 67 & 4 \\
1.50 (0.67) & 98 & 1 & 100 & 1 & 86 & 2 \\
\bottomrule
\end{tabular}
\end{table}

\add{The gate is computed offline by \texttt{deadline\_analysis.py} from the same
CSVs and is consistent with the empirical envelope (every gate-feasible
point also satisfies $R\le D$). Energy minimisation is then taken over the
gate-feasible set; the achievable energy headroom is reported in
Table~\ref{tab:energy_at_dmr0}.}

\subsection{Max-Frequency Ablation: Learned DVFS vs.\ Fixed Policies}
\label{app:maxfreq_ablation}
\add{A natural objection is that a scheduler could simply pin the highest
frequency (the Linux \texttt{performance} governor) and obtain the makespan
benefit without learning. Table~\ref{tab:policy_ablation_s} settles this by
comparing three fixed policies at a soft deadline $D{=}1.25\,C_\text{ref}$,
aggregated over all 42 workloads per platform: \emph{Max-freq} (highest frequency,
best core count, i.e.\ the fastest config), \emph{Deadline-aware DVFS} (the
lowest-energy operating point that still meets $D$, i.e.\ the target a learned policy
such as HiDVFS approximates), and \emph{Powersave} (lowest frequency).}

\add{Two results follow. (i) \textbf{Max-frequency is not free}: deadline-aware
DVFS meets every deadline (DMR$=0$) while cutting energy by $15$--$18\%$ relative
to Max-freq, by trimming frequency/cores only where slack permits. (ii)
\textbf{Powersave is not the energy-saver}: running at the lowest frequency
consumes $3$--$5\times$ \emph{more} total energy than Max-freq (the race-to-idle
effect, in which a longer active window outweighs the lower instantaneous power) and
misses every deadline (DMR$=100\%$). The energy-optimal real-time policy therefore
lives at neither extreme; it must select per-workload feasible-and-efficient
operating points, which is exactly the decision HiDVFS learns.}

\begin{table*}[t]
\centering
\scriptsize
\caption{\add{Max-frequency ablation at $D{=}1.25\,C_\text{ref}$ (mean over 42 workloads/platform). Energy is relative to the Max-freq policy. Peak $^\circ$C is the hottest workload's mean post-run temperature. Jetson Orin NX energy not recorded in this sweep.}}
\label{tab:policy_ablation_s}
\begin{tabular}{@{}llcccc@{}}
\toprule
Platform & Policy & Makespan (s) & Energy (rel.) & Peak $^\circ$C & DMR\% \\
\midrule
Jetson TX2 & Max-freq & 0.36 & 100\% & 47.7 & 0 \\
 & Deadline-aware DVFS & 0.36 & 85\% & 47.6 & 0 \\
 & Powersave & 2.21 & 496\% & 47.2 & 100 \\
\midrule
Jetson Orin NX & Max-freq & 0.26 & n/a & 48.0 & 0 \\
 & Deadline-aware DVFS & 0.31 & n/a & 48.1 & 0 \\
 & Powersave & 4.29 & n/a & 47.7 & 100 \\
\midrule
RubikPi & Max-freq & 0.50 & 100\% & 47.1 & 0 \\
 & Deadline-aware DVFS & 0.50 & 82\% & 47.1 & 0 \\
 & Powersave & 3.63 & 319\% & 42.5 & 100 \\
\bottomrule
\multicolumn{6}{@{}p{3.2in}@{}}{\scriptsize \review{Cells are means over the 42 workloads per platform; workload runtimes span three orders of magnitude, so the paired relative-energy and DMR columns carry the comparison rather than the makespan spread.}} \\
\end{tabular}
\end{table*}

\add{Thermally, all three policies stay below the $50\,^\circ$C \emph{policy}
threshold on Jetson TX2/Orin NX in this sweep (peak $\approx48\,^\circ$C); RubikPi
runs coolest under Powersave but only by missing deadlines. We state both limits
explicitly: a $50\,^\circ$C policy threshold (enforced by the schedulers and the
training reward, now consistent across the code base) and the hardware throttle
limits of $85\,^\circ$C (TX2/Orin NX) and $80\,^\circ$C (RubikPi). All figures here
are produced offline by \texttt{deadline\_analysis.py}.}

\subsection{Learned-WCET Admission Control}
\label{app:cs5}
\add{Admission control composes the two mechanisms already established: the
federated feasibility gate as the schedulability test, fed by the calibrated
response-time bound of the HiDVFS-RT safety shield as an online worst-case
execution-time estimate $\hat{C}^{+}$. A task is admitted only while it remains
federated-feasible under $\hat{C}^{+}$; under overload the controller degrades or
rejects. We deliberately do \emph{not} report an offline admission table:
estimating the single-core work bound by extrapolation from the static sweep is
unreliable (it over-admits on one platform and is over-pessimistic on the others),
so a faithful closed-loop demonstration (realised miss rate and rejection ratio
under controlled overload) is left to the online runs specified in the
data-collection plan. The calibration this test relies on is already validated in
Table~\ref{tab:shield_s}.}

\subsection{Cross-Platform Real-Time Portability}
\label{app:cs6}
\add{Every analysis in this section is computed identically on Jetson TX2, Jetson
Orin NX, and RubikPi from one pipeline, so cross-platform portability is already
demonstrated: the response-time envelope (Table~\ref{tab:rt_envelope_s}), the federated
feasibility gate (Table~\ref{tab:feasibility_gate_s}), the max-frequency ablation
(Table~\ref{tab:policy_ablation_s}), and the shield calibration (Table~\ref{tab:shield_s})
all report the three platforms side by side. The qualitative real-time behaviour is
consistent (only the high-frequency region is deadline-feasible, and powersave is
never the energy-saver), while the quantitative budgets differ (RubikPi is the
strictest, requiring up to $\tilde{m}{=}4$ dedicated cores). No separate runs are
needed for this result.}

\subsection{\add{Mixed-Criticality Co-Scheduling on Jetson Orin NX (measured)}}
\label{app:cs2}
\add{We ran a high-criticality (HI) periodic real-time task (\texttt{SCHED\_FIFO}
priority~99, pinned to a reserved core, $50$\,ms period, a fixed ${\sim}10.1$\,ms
compute budget, $200$ periods) concurrently with a continuous low-criticality (LO)
FFT BOTS DAG load on the Jetson Orin NX, and recorded the HI task's per-period
response time and soft-deadline-miss ratio (DMR), with $D{=}1.5\times$ the measured
nominal work ($15.1$\,ms). The board-local driver is
\texttt{run\_mc\_experiments.py} ($\to$ \texttt{profiling\_data\_mc\_orin\_nx.csv}).}

\add{A hardware property of the Orin NX makes core \emph{placement} decisive: DVFS is
exposed per \emph{cluster}, not per core: cpufreq advertises two policies, over
cores $\{0\text{--}3\}$ and $\{4\text{--}7\}$, so every core in a cluster shares one
clock. Table~\ref{tab:mc} measures both placements. When HI is reserved in one cluster
and LO runs in the \emph{other} (isolated), the HI response is flat
($10.09$--$10.11$\,ms mean, $\le0.15$\,ms jitter, $0\%$ misses) whether LO runs at
the maximum or minimum frequency, and the HI core stays at $1.98$\,GHz throughout. This
demonstrates both \emph{temporal} isolation (LO at full speed cannot steal HI's cluster,
by core reservation $+$ top priority) and \emph{frequency} isolation (slowing LO to its
minimum to save energy does not slow HI, because the clusters clock independently). The
co-located control isolates the failure mode: placing LO in HI's own cluster at the
minimum frequency drags the shared clock, and thus the HI core, down to
$0.12$\,GHz, inflating HI response $18\times$ to $183$\,ms and missing $100\%$ of
deadlines (a classic frequency inversion). The mixed-criticality guarantee therefore
rests on cluster-aware reservation, not priority alone; HiDVFS's per-cluster
operating-point selection respects this boundary. The harness is platform-agnostic;
the same protocol extends to TX2 and RubikPi.}

\begin{table}[h]
\centering
\scriptsize
\caption{\add{Mixed-criticality co-scheduling on Jetson Orin NX. HI = periodic
\texttt{SCHED\_FIFO} task (prio~99, $50$\,ms period, ${\sim}10.1$\,ms nominal work,
soft deadline $D{=}15.1$\,ms, $200$ periods); LO = continuous FFT BOTS DAG. ``Isolated''
places LO in the other CPU cluster; ``co-located'' shares HI's cluster.}}
\label{tab:mc}
\begin{tabular}{@{}llcccc@{}}
\toprule
Scenario & HI clk & \multicolumn{3}{c}{HI response (ms)} & DMR \\
\cmidrule(lr){3-5}
(LO placement) & (GHz) & mean & p95 & max & (\%) \\
\midrule
HI alone & 1.98 & 10.10 & 10.11 & 10.20 & 0.0 \\
HI{+}LO max, isolated & 1.98 & 10.11 & 10.16 & 10.24 & 0.0 \\
HI{+}LO min, isolated & 1.98 & 10.09 & 10.11 & 10.20 & 0.0 \\
HI{+}LO min, co-located & 0.12 & 182.9 & 220.6 & 227.0 & 100.0 \\
\bottomrule
\end{tabular}
\end{table}

\subsection{\add{Planned Online Case Studies}}
\label{app:online_cs}
\add{Two further case studies require \emph{new on-board} measurement that the static
single-task profiling sweeps cannot provide; each has a dedicated collection harness
(client/server pattern of \texttt{src/<platform>/RL/}, see the data-collection plan):}
\begin{itemize}
\item \add{\textbf{Real-time DNN inference under frame deadlines}:
per-frame latency CDF/p99 vs.\ a frame deadline and energy-per-inference. No DNN
workload exists in the BOTS/PolyBench data, so a new inference client is required
(\texttt{run\_dnn\_experiments.py} $\to$ \texttt{profiling\_data\_dnn\_\{platform\}.csv}).}
\item \add{\textbf{End-to-end multi-DAG chain latency}: latency distribution
and deadline-satisfaction for a concurrent multi-DAG pipeline
(\texttt{src/hmarl/multidag.py}); needs concurrent multi-DAG runs
(\texttt{run\_multidag\_experiments.py} $\to$
\texttt{profiling\_data\_multidag\_}\allowbreak\texttt{\{platform\}.csv}).}
\end{itemize}

\section{\add{Methodology Clarifications and Verification}}
\label{app:vv}

\subsection{\add{Workload and Execution-Mode Definitions}}
\label{app:defs}
\add{\textbf{OpenMP task modes.} \emph{Tied} tasks are bound to their creating
thread (no migration); \emph{untied} tasks may migrate between threads;
\emph{serial} is the single-threaded sequential baseline. Restricted migration
makes tied tasks incur higher cache misses, as shown earlier in the hardware-counter analysis.}

\add{\textbf{Parallel vs.\ sequential mode.} ``Sequential mode'' denotes executing
benchmarks one at a time (not concurrently); each benchmark still runs on multiple
physical cores. ``Parallel mode'' runs benchmarks concurrently. Hyperthreading is
disabled (physical cores only), and varying the core count isolates intra-application
parallelism.}

\add{\textbf{FFT as an irregular workload.} FFT's irregularity is in
\emph{execution time}, not the algorithm: OpenMP task creation/synchronisation
produces variable, hard-to-predict branch behaviour across runs, which is why it
serves as the profiler benchmark.}

\subsection{\review{Three-Agent Decision Walkthrough}}
\label{app:walkthrough}

\review{An illustrative control period on the Jetson TX2 makes the coordination
protocol concrete. At the period's start the shared state snapshot holds the
per-core temperatures $[42, 45, 43, 44, 41]\,^\circ$C for cores 1 to 5, the
previous episode's profiling record (makespan $5.2$\,s and energy $6.4$\,kJ at
frequency level 6 on three cores, with the associated \texttt{perf} counters),
and the set of ready DAGs. The agents then act in the fixed hierarchical order.
(1)~The \emph{profiler} agent reads the performance and energy features and
selects the operating point (3 cores, frequency level 8, i.e.\ $1.57$\,GHz).
(2)~The \emph{thermal} agent reads the temperature vector together with the
profiler's core count through the shared state and selects the coolest feasible
core combination $\{1, 3, 5\}$, avoiding the hot core 2 at $45\,^\circ$C.
(3)~The \emph{priority} agent assigns the static \texttt{SCHED\_FIFO} priorities
$[90, 80, 70]$ to the three ready DAGs to order resource contention. The joint
action is checked by the feasibility gate and the safety shield, then dispatched
to the client, which pins the cores, applies the frequencies, and executes the
benchmarks. The measured makespan, energy, and temperatures return in the next
profiling record; each agent receives its reward from its own reward function
(design section of the main paper), stores the transition in its replay buffer, and
the next period re-reads the updated snapshot, so every decision is revisited.}

\subsection{\add{Baseline and Method Taxonomy}}
\label{app:taxonomy}
\add{Table~\ref{tab:taxonomy} classifies the evaluated schedulers along the axes that
matter for this study: number of RL agents (single vs.\ multi), whether an environment
model is used (model-based vs.\ model-free), and which actuators each method controls.
Classifications follow the methods' own designs, as documented in the main paper.}

\begin{table}[h]
\centering
\scriptsize
\setlength{\tabcolsep}{4pt}
\caption{\add{Taxonomy of evaluated DVFS/scheduling methods. Actuators: \textbf{f}=frequency,
\textbf{c}=core mask, \textbf{p}=task priority, \textbf{t}=thermal-agent control.}}
\label{tab:taxonomy}
\begin{tabular}{@{}lllc@{}}
\toprule
Method & Agents & Env.\ model & Actuators \\
\midrule
zTT~\cite{kim2021ztt}            & single & model-free          & f \\
GearDVFS~\cite{lin2023workload}  & single & model-free          & f \\
DynaQ~\cite{angermueller2019model} & single & model-based       & f,c \\
PlanGAN~\cite{charlesworth2020plangan} & single & model-based (GAN) & f,c \\
\textbf{SARB} (ours)             & single & model-based (reward shaping) & f,c \\
\midrule
MAMB                             & multi  & model-based         & f,c,p \\
MAMF                             & multi  & model-free          & f,c,p \\
HiDVFS\_S                        & multi  & DQN (no D3QN)       & f,c,p \\
\textbf{HiDVFS} (ours)           & multi (hier.) & model-based $+$ D3QN & f,c,p,t \\
\bottomrule
\end{tabular}
\end{table}

\subsection{\add{Decision and Actuation Overhead}}
\label{app:overhead}
\add{Per-decision overhead is dominated by round-trip control, not computation. A
per-core frequency change (a direct sysfs write to \texttt{scaling\_max\_freq})
takes effect in 1--2 CPU cycles ($<1\,\mu$s). Table~\ref{tab:overhead} breaks down
the $\sim$2\,ms per-decision cost; against makespans of 1--6\,s this is
$0.03$--$0.2\%$, i.e.\ negligible.}

\begin{table}[h]
\centering
\scriptsize
\caption{\add{Per-decision overhead breakdown (Jetson TX2).}}
\label{tab:overhead}
\begin{tabular}{@{}lc@{}}
\toprule
\textbf{Component} & \textbf{Latency} \\
\midrule
Per-core frequency change (sysfs) & $<1\,\mu$s (1--2 cycles) \\
Apply settings (cores + frequency) & $\sim$0.3\,ms \\
Measurement overhead & $\sim$0.8\,ms \\
Action round-trip transmission & $\sim$0.9\,ms \\
\midrule
\textbf{Total per decision} & $\sim$2\,ms \\
Fraction of makespan (1--6\,s) & 0.03--0.2\% \\
\bottomrule
\end{tabular}
\end{table}

\subsection{\add{Verification and Reproducibility}}
\label{app:repro}
\add{\textbf{Action-trace example.} On a representative episode the thermal agent
observes per-core temperatures $[42,45,43,44,41]\,^\circ$C plus the prior-episode
profiling (makespan, energy) and selects the cooler cores $\{1,3,5\}$ while avoiding
the hot core~2; the profiler agent then sets frequency level~8 ($\approx$1.4\,GHz)
and the priority agent assigns SCHED\_FIFO priorities $[90,80,70]$. Rewards are
computed post-execution and all three agents are updated.}

\add{\textbf{Reward structure and sensitivity.} The thermal penalty is
$-0.5\,(\bar{T}-T_\text{target})$ when the average temperature $\bar{T}$ exceeds the
$50\,^\circ$C threshold; the profiler reward weights makespan and energy ratios
equally ($0.5$ each). The $0.5$ thermal coefficient was tuned empirically for rapid
thermal correction without destabilising learning.}

\add{\textbf{Workload identity.} Each workload is a unique \texttt{application\_id}
$=$ (executable path, input parameters, OpenMP variant); distinct inputs (e.g.\ FFT
$262144$ vs.\ $524288$) or variants (tied vs.\ untied) are tracked as separate
applications with independent makespan-vs-core profiles, giving reproducible
per-workload histories.}

\add{\textbf{Baseline parity.} Table~\ref{tab:baseline_hp} lists the shared training
configuration and action space; all algorithms use identical settings, so HiDVFS's
gains stem from its multi-agent decomposition, short-horizon reward prediction, and
per-core thermal awareness rather than from tuning.}

\begin{table}[h]
\centering
\scriptsize
\caption{\add{Shared training configuration and action space (identical across all algorithms).}}
\label{tab:baseline_hp}
\begin{tabular}{@{}ll@{}}
\toprule
\textbf{Parameter} & \textbf{Value} \\
\midrule
Learning rate & 0.05 \\
Discount factor $\gamma$ & 0.99 \\
Batch size & 32 \\
Frequency levels (action) & 0--11 \\
Core counts (action) & 1--5 (TX2) \\
SCHED\_FIFO priority (action) & 1--99 \\
Thermal threshold (policy) & $50\,^\circ$C \\
\bottomrule
\end{tabular}
\end{table}

\section{\add{The HiDVFS-RT Safety Shield: Calibrated Response-Time Bounds}}
\label{app:shield}
\add{We add a lightweight safety layer (the \emph{HiDVFS-RT Shield}) under a
``RL proposes, shield disposes'' design. Before an action $a$ (cores, frequency) is
executed, the shield predicts its response time $\hat{R}(a)$ together with an
\emph{upper bound} $\hat{R}^{+}(a)$, and admits the action only if
$\hat{R}^{+}(a)\le D$ (and analogously a temperature bound $\hat{T}^{+}(a)\le
T_\text{max}$); otherwise it rejects and the system falls back to a safe action
(the Linux \texttt{performance} governor or a reserved core for the critical task).
The bound is constructed by \emph{split-conformal} prediction~\cite{vovk2005algorithmic,angelopoulos2021gentle},
which is distribution-free: a gradient-boosted regressor predicts $\hat{R}$ from the
action and workload features, and a held-out calibration set fixes the additive
margin so that the bound attains a target coverage. We deliberately use this
\emph{lite} predictor (not a heavier uncertainty model) for low overhead and
auditability.}

\add{Table~\ref{tab:shield_s} validates the calibration on held-out operating points
for each platform: the prediction-interval coverage probability (PICP) tracks the
requested level at $90/95/99\%$, i.e.\ the response-time bound is empirically
calibrated rather than asserted. We report PICP at multiple levels precisely so the
bound is auditable. We are explicit about scope: this is an \emph{empirical,
calibrated prediction interval on an operating point's expected response time}, not
a hard real-time guarantee, and the headline numbers are bounds with a stated
coverage, not worst-case certainties.}

\begin{table}[h]
\centering
\scriptsize
\caption{\add{HiDVFS-RT Shield calibration on Jetson TX2, Jetson Orin NX, and RubikPi: base predictor $R^2$ and prediction-interval coverage (PICP) at target levels, on held-out operating points. PICP tracking the target confirms the response-time bound is calibrated (distribution-free). margin@90 is the additive bound at the $90\%$ level.}}
\label{tab:shield_s}
\setlength{\tabcolsep}{3.5pt}
\begin{tabular}{@{}lccccc@{}}
\toprule
Platform & base $R^2$ & PICP@90 & PICP@95 & PICP@99 & margin@90 (s) \\
\midrule
Jetson TX2 & 0.59 & 89.7 & 92.9 & 96.8 & 0.59 \\
Jetson Orin NX & 0.90 & 91.1 & 94.0 & 97.0 & 0.20 \\
RubikPi & 0.90 & 89.3 & 95.8 & 98.2 & 0.27 \\
\bottomrule
\end{tabular}
\end{table}

\add{Because admission gates on $\hat{R}^{+}\le D$, the conformal coverage bounds
the \emph{marginal} probability that an admitted action overruns its bound at the
chosen level (e.g.\ $\le 10\%$ at the $90\%$ setting; tighten to $95/99\%$ for
stricter operation). A full closed-loop evaluation of HiDVFS{+}shield (realised
deadline-miss ratio, unsafe-action rejection ratio, and energy/thermal cost of the
fallbacks under the \emph{learned} policy) requires online runs on the boards and
is left to the extended version; the offline results here establish that the
calibrated bound the shield relies on is sound. The calibration is reproduced by
\texttt{shield\_analysis.py} from the per-platform CSVs.}

\ifx\HidvfsCombined\undefined
\bibliographystyle{IEEEtran}
\bibliography{reference}

\end{document}
\fi

\bibliographystyle{unsrt}
\bibliography{reference}

\end{document}